%% file: main.tex
\documentclass[journal,compsoc]{IEEEtran}
\usepackage{cite}
\usepackage{amsmath,amssymb,amsfonts}
\usepackage{graphicx,color}
\usepackage{textcomp}
\usepackage{xcolor}
\usepackage{hyperref}
\hypersetup{hidelinks=true}
\usepackage{algorithm}
\usepackage{algorithmic}
\usepackage[font=scriptsize]{caption}
\def\BibTeX{{\rm B\kern-.05em{\sc i\kern-.025em b}\kern-.08em
    T\kern-.1667em\lower.7ex\hbox{E}\kern-.125emX}}
\AtBeginDocument{\definecolor{tmlcncolor}{cmyk}{0.93,0.59,0.15,0.02}\definecolor{NavyBlue}{RGB}{0,86,125}}

\usepackage{bbm}
\usepackage{bm}

\def\BibTeX{{\rm B\kern-.05em{\sc i\kern-.025em b}\kern-.08em
    T\kern-.1667em\lower.7ex\hbox{E}\kern-.125emX}}
\usepackage{mathtools}
\usepackage{booktabs}
\usepackage{multirow}
\usepackage{subfig}

\newcommand{\E}[1]{\mathbb{E}\left[ #1 \right]} 
\newcommand{\mc}[1]{\mathcal{#1}}   
\newcommand{\mb}[1]{\mathbf{#1}}    
\DeclareMathOperator*{\argmax}{arg\,max}    
\DeclareMathOperator*{\argmin}{arg\,min}    

\usepackage{glossaries}
\newacronym{ann}{ANN}{Artificial Neural Network}
\newacronym{ai}{AI}{Artificial Intelligence}
\newacronym{rl}{RL}{Reinforcement Learning}
\newacronym{fl}{FL}{Federated Learning}
\newacronym{dqn}{DQN}{Deep Q-Network}
\newacronym{cl}{CL}{Continual Learning}
\newacronym{ns}{NS}{Network Slicing}
\newacronym{mel}{MeL}{Meta Learning}
\newacronym{drl}{DRL}{Deep Reinforcement Learning}
\newacronym{qos}{QoS}{Quality of Service}
\newacronym{kpi}{KPI}{Key Performance Indicator}
\newacronym{iot}{IoT}{Internet of Things}
\newacronym{ue}{UE}{User Equipment}
\newacronym{mdp}{MDP}{Markov Decision Process}
\newacronym{fifo}{FIFO}{First-In First-Out}
\newacronym{m2m}{M2M}{Machine to Machine}
\newacronym{urllc}{URLLC}{Ultra-Reliable Low-Latency Communications}
\newacronym{embb}{eMBB}{enhanced Mobile Broadband}
\newacronym{ofdma}{OFDMA}{Orthogonal Frequency Division Multiple Access}
\newacronym{vr}{VR}{Virtual Reality}
\newacronym{mec}{MEC}{Mobile Edge Computing}
\newacronym{cdf}{CDF}{Cumulative Distribution Function}
\newacronym{fdma}{FDMA}{Frequency Division Multiple Access}
\newacronym{tdma}{TDMA}{Time Division Multiple Access}

\definecolor{color0}{HTML}{0000FF}
\definecolor{color1}{HTML}{A300D8}
\definecolor{color2}{HTML}{DC35AA}
\definecolor{color3}{HTML}{FA6772}
\definecolor{color4}{HTML}{FF9B00}

\definecolor{cornflowerblue121150222}{RGB}{121,150,222}
\definecolor{darkseagreen12518595}{RGB}{125,185,95}
\definecolor{darkslategray38}{RGB}{38,38,38}
\definecolor{darkslategray66}{RGB}{66,66,66}
\definecolor{lightgray204}{RGB}{204,204,204}
\definecolor{palevioletred227108143}{RGB}{227,108,143}
\definecolor{peru21513172}{RGB}{215,131,72}
\definecolor{sandybrown22319372}{RGB}{223,193,72}

\def \ffffwidth{0.225\linewidth}

\def \fwidth{0.95\columnwidth}
\def \fheight{0.5\columnwidth}

\usepackage[utf8]{inputenc}
\usepackage{pgfplots}
\DeclareUnicodeCharacter{2212}{−}
\usepgfplotslibrary{groupplots,dateplot}
\usetikzlibrary{patterns,shapes.arrows}
\pgfplotsset{compat=newest}

\begin{document}

\markboth{Fast Context Adaptation in Cost-Aware Continual Learning}{Lahmer \emph{et al.}}

\title{Fast Context Adaptation in Cost-Aware Continual Learning}

%
\title{Fast Context Adaptation in Cost-Aware Continual Learning}

\author{Seyyidahmed~Lahmer,~\IEEEmembership{Student Member,~IEEE,}
        Federico~Mason,~\IEEEmembership{Member,~IEEE,}
        Federico~Chiariotti,~\IEEEmembership{Member,~IEEE,}
        and~Andrea~Zanella,~\IEEEmembership{Senior~Member,~IEEE}
\thanks{This work was supported by the EU H2020 MSCA ITN project Greenedge (grant no. 953775) and by the European Union, under the Italian National Recovery and Resilience Plan (NRRP) of NextGenerationEU, as part of the REDIAL (SoE0000009) Young Researchers grant and the partnership on ``Telecommunications of the Future'' (PE0000001) - program ``RESTART.''}}

\IEEEtitleabstractindextext{%
\begin{abstract}
In the past few years, \gls{drl} has become a valuable solution to automatically learn efficient resource management strategies in complex networks with time-varying statistics. However, the increased complexity of 5G and Beyond networks requires correspondingly more complex learning agents and the learning process itself might end up competing with users for communication and computational resources. This creates friction: on the one hand, the learning process needs resources to quickly convergence to an \textit{effective} strategy; on the other hand, the learning process needs to be \textit{efficient}, i.e., take as few resources as possible from the user's data plane, so as not to throttle  users' \gls{qos}. 

In this paper, we investigate this trade-off and propose a dynamic strategy to balance the resources assigned to the data plane and those reserved for learning. With the proposed approach, a learning agent can quickly converge to an efficient resource allocation strategy and adapt to changes in the environment as for the \gls{cl} paradigm, while minimizing the impact on the users' \gls{qos}. Simulation results show that the proposed method outperforms static allocation methods with minimal learning overhead, almost reaching the performance of an ideal out-of-band \gls{cl} solution.
\end{abstract}

}

\maketitle

\section{Introduction}

\glsresetall
\IEEEPARstart{T}{he} role of \gls{ai} in communication networks has become more and more central with the transition from 4G to 5G, and learning is at the core of the 6G standardization process~\cite{letaief2019roadmap}. Mobile networks are no longer designed as rigid entities that the final users have to adapt to, rather are becoming customizable services evolving according to the users' needs~\cite{yang2015software}. The \gls{ns} paradigm supports this approach by enabling the definition of multiple logical networks overlaying the same physical infrastructure~\cite{afolabi2018network}, with each \emph{slice} devoted to a specific class of service. This allows applications with very different requirements to coexist and share spectrum resources. However, managing \gls{ns}, as well as other advanced application scenarios, requires judicious allocation of both transmission and computational resources to users, according to their \gls{qos} targets, in a fast-paced scenario~\cite{rlInNetworkingSurvey}, which is expected to become even more straining with 6G.

Hand-designed resource allocation strategies may not be up to this challenge, so that growing attention has been dedicated to machine-learning approaches. In particular, \gls{drl} is  considered a promising framework for deriving adaptable and robust strategies for network orchestration~\cite{rlInNetworkingSurvey} and resource allocation~\cite{sami2021ai}. 

\gls{drl}'s \emph{effectiveness} in dealing with complex scenarios is indeed well-established: with proper training, the \gls{drl} agents can find foresighted policies aiming for long-term objectives~\cite{sutton2018reinforcement}, significantly improving network performance. Such promising results, however, have been typically obtained in \textit{stationary} environments: if this assumption is not satisfied, the performance of pre-trained \gls{drl} agents may dramatically decrease when the network dynamic shifts away from the training environment. 

Approaches based on the \gls{cl} paradigm~\cite{de2021continual} are designed to deal with non-stationary systems. \gls{cl} enables the adaptation of a learning agent to a series of subsequent tasks that, in a network scenario, may represent different network configurations. We observe that combining \gls{cl} and \gls{drl} for managing network resources in non-stationary scenarios has a non-negligible cost in terms of energy, computation, and communication resources~\cite{neda2022survey}. These resources are necessarily subtracted to the \emph{data plane}, i.e., the part of the system that is responsible for transmitting, processing, and forwarding user data packets. Therefore, supporting the learning represents an overhead for the system, which can negatively impact users' \gls{qos}. We use the term \textit{cost of learning} to indicate the impact that the learning process can have on users' performance due to the resources it requires. Considering such a cost implies that the learning framework, in addition to being effective, must also be \textit{efficient}, that is, require as few resources as possible to achieve its goals. The cost of learning problem is particularly critical considering the ever-larger size of most recent \gls{drl} neural networks, and the growing demand for efficient systems, as for the green networking paradigm~\cite{chih2021ai}. We observe that \gls{mec}~\cite{mao2017survey} solutions does not solve the problem, but just move it at the network edge. In fact, while \gls{mec} allows computationally-expensive tasks (such as the training of \gls{drl} algorithms) to be carried out directly in dedicated edge nodes physically close to the data sources, still the limited transmission, computational and energetic resources of such nodes have to be shared between data and learning planes.

Finding a balance between the number of resources to be used for improving the system reactivity to variations and those to be allocated for serving the users may be a very difficult task. This is particularly critical in the case of \gls{cl} systems, in which agents must constantly adapt to new working conditions. As the very same network resources are also used for the training, a trade-off between the capability of \gls{drl} agent to learn new tasks and its performance during the current task arises. 

Note that, although at a first sight this problem may recall the well-known exploration-exploitation problem in learning systems, it is actually fundamentally different. In fact, the exploration-exploitation problem involves finding a balance between exploring new strategies, with the risk of temporarily losing some performance, and exploiting the currently learned strategy that, however, may be globally suboptimal, thus wasting part of the system capacity. In this setting, the resources required by the learning process are typically ignored: whatever action is taken, the outcome is assumed to be available to the learner, enriching its {experience}, without any cost. 
In this paper, instead, we look at the resources needed to transfer such information to the learner and turn it into experience, irrespective of whether it originates from an exploration or exploitation action.  From a theoretical perspective, the scenario we look at is hence a \gls{mel} one, in which the agent's actions determine the efficiency of the learning data aggregation and processing. 

The cost of learning is therefore a fundamental aspect to be considered in modern network design, and recent works have proposed learning-based frameworks that are computation-aware \cite{mason2022no}. Despite the high interest of the scientific community in this field, the cost of learning for \gls{drl} models is still a relatively unexplored subject in the networking literature, and even the most recent works on resource allocation and \gls{ns} ignore the true cost of combining \gls{drl} and \gls{cl} in modern networks \cite{mason2022using}, making the effectiveness of state of the art \gls{drl} solutions questionable.
%

In this work, we analyze the trade-off between \textit{effectiveness} and \textit{efficiency} in \gls{cl}, formally defining the resource allocation problem and presenting a heuristic solution to allocate resources to the data and learning planes. The proposed scheme effectively controls training in a \gls{cl} framework, maximizing the efficiency of the training (i.e., reducing the learning plane overhead) while still achieving effective resource allocation (i.e., the same \gls{qos} as the ideal approach that assumes learning does not consume users' resources) in a reasonable time. Although we applied our optimization framework to a networking scenario, it is actually more general and can be adapted to any learning-based allocation problem in which the allocated resources are also required for the agent training, such as \gls{mec} job scheduling.

The major contributions of our work are the following:
\begin{itemize}
    \item We define a theoretical model to analyze the trade-off between \textit{effectiveness} and \textit{efficiency} of learning-based resource allocation schemes;
    \item We propose a \gls{cl} strategy to enable the resource allocation scheme to adapt to sudden changes in traffic dynamics;
    \item We test the proposed model in a \gls{ns} use case, in which learning agent and system users compete the same network resources;
    \item We compare the benefits and drawbacks of the proposed approach against a static resource-sharing scheme between data and learning planes, and an ideal strategy that considers out-of-band resources for the agent training (or, equivalently, assumes the learning agent does not consume any user-plane resources). Our simulation results show that the proposed heuristic performs closely to the ideal (out-of-band) approach, minimizing the impact of learning plane traffic during the training. 
\end{itemize}
A partial and preliminary version of this work was presented in~\cite{lahmer2022cost}. In this paper, we extend that work by introducing the \gls{cl} approach, providing a much richer set of results, and deepening the discussion and analysis of our observations. 

The rest of this paper is organized as follows: first, Sec.~\ref{sec:related} reports the most significant related work. We then present the model for optimizing data and learning plane in Sec.~\ref{sec:system} and the \gls{ns} use case definition in Sec.~\ref{sec:slicing}. Successively, Sec.~\ref{sec:results} presents the simulation results. Finally, Sec.~\ref{sec:conc} concludes the paper and discusses some possible avenues for future work.

\section{Related Work}
\label{sec:related}

While the latest advances in \gls{ai} have made it possible to reach stunning performance levels in multiple fields, there is still a large gap between human cognition and \gls{ai} models in terms of adaptation. 
Most of the current learning models need to be retrained from scratch every time a new task has to be accomplished, with a high cost in terms of computational power and time.  
For this purpose, the scientific community has recently leveraged the \gls{cl} paradigm, which focuses on learning a series of subsequent data, associated with different tasks, without \emph{catastrophically forgetting} the past knowledge~\cite{li:2019}. Therefore, in \gls{cl} scenarios the goal is to adapt to a time-varying environment, working on one task at a time and assuming that future information is inaccessible. This model appeals to the resource allocation problem considered in this manuscript, since in realistic networks the type, number and requirements of the users keep changing over time, making the system non-stationary (though stationarity can be assumed during the \textit{coherence intervals}, i.e., the time periods during which the main system parameters do not change).

A baseline \gls{cl} solution may involve a pre-trained model that is iteratively adapted to new tasks (or to changes in the environment), e.g., taking advantage of {curriculum learning}, as done in~\cite{graves:2016}. 
{Replay-based} methods form a more recent class of \gls{cl} algorithms, which store past experience in memory or exploit a generative model to reproduce it, using this information as model input while training on new tasks~\cite{rolnick:2019, isele:2018}. 
{Regularization-based} methods, which introduce a penalty term in the model's loss function with the goal of avoiding performance degradation in past tasks~\cite{li:2017, ahn:2019}, form another class of solutions.
An extension of the aforementioned class is proposed in~\cite{zenke:2017}, where the authors estimate the importance of each learned parameter and prevent the modifications of such parameters that most affect performance in past tasks. 
Finally, {architecture-based} methods define an additional branch of the model for each task, freezing the previously learned parameters when training the model on new scenarios~\cite{mallya:2018, serra:2018}.
An example is provided in~\cite{schwarz:2018}, where the authors developed a model organized into two blocks: the first is retrained every time a new task arises, while the latter distills the knowledge acquired for future reuse. 

From a different perspective, \gls{cl} algorithms aim at defining a strategy to detect the best settings for training a learning model in new scenarios. 
This concept falls within the \gls{mel} paradigm, which 
focuses on convergence speed and stability~\cite{hospedales:2021}. 
\gls{mel} methods differ in the output of the learning optimization, which may include the weight initialization strategy~\cite{finn:2019}, the optimizer algorithm~\cite{wichrowska:2017}, the loss function~\cite{houthooft:2018}, the dimensions of the learning architecture~\cite{real:2019}, and other hyper-parameters. 

The methodology used for the \gls{mel} optimization task may be based on gradient descent~\cite{finn:2017}, evolutionary algorithms~\cite{stanley:2019}, or learning-based approaches. For instance, the authors of~\cite{xu:2018} develop a \gls{cl} model in which a \gls{drl} agent has to define the optimal settings of the task-specific block, balancing between validation accuracy and model complexity.
Besides, \gls{mel} methods differ in terms of optimization goals, which may either be fully based on the model performance on a validation set or consider more specific aspects, such as the adaptability of the solution to multiple tasks~\cite{snell:2017}, the greater importance of fast adaptation than asymptotic performance~\cite{antoniou:2018}, and the difference between online and offline learning scenarios~\cite{andrychowicz:2016}.

In particular, the use of \gls{mel} methods for optimizing \gls{drl} models is a relatively new field. 
In this scenario, the combination of \gls{cl} and \gls{mel} avoids the need for a centralized agent that can handle each possible state-action pair and enables the definition of multiple and more straightforward policies. The authors of~\cite{nagabandi:2018} show the benefits of \gls{mel} in a real-world scenario, analyzing a \gls{drl} robotic system that exploits a recurrent module to preserve past knowledge and speed up the training process.  Interestingly, when applying \gls{mel} combined with \gls{drl}, a fundamental parameter is the choice of the exploration policy by which the agent interacts with the environment, which is an absent aspect in classification tasks. For instance, the authors of~\cite{gupta:2018} develop a \gls{mel} model where prior information is used to define agnostic exploration policies enabling a better agent adaptation to multiple learning problems. 

In the context of \gls{cl} and \gls{mel} for 5G and 6G network management, drift detection is a critical task: if the environment changes abruptly, the \gls{mel} paradigm requires the learner to first become aware of the change, and delayed detection may lead to a violation of the service requirements~\cite{yang:2021}. Drifts can be classified as abrupt or gradual depending on the period during which the system performance degrades. 
The literature presents several drift detection algorithms, usually considering the prediction error of the learning model for estimating environment changes. 
It is possible to consider both heuristic~\cite{bifet:2009} or learning-based strategies~\cite{faria:2016}, with different advantages and drawbacks in terms of, e.g., false alarm probability and assumptions on the drift characteristics.  

Despite the numerous works investigating \gls{cl} and \gls{mel} in supervised and reinforcement learning scenarios, to the best of our knowledge none of them considers the trade-off between efficiency and effectiveness proposed in this manuscript.  
In the following sections, we will take advantage of solutions inspired by the literature, analyzing their impact in terms of both training efficiency and user performance. 
We will consider a baseline \gls{cl} system where the models trained for previous tasks are stored in a central memory, similarly to what is done in~\cite{graves:2016}.
Besides, we will consider a simple drift detection algorithm to monitor the environment statistics and trigger the retraining of the learning model. 
We chose relatively simple techniques for both drift detection and \gls{cl} in order to focus on the main aspect of our analysis, which is the trade-off between efficiency and effectiveness in resource allocation problems, as represented by the cost of learning.

\section{System Model}\label{sec:system}

Let us consider a generic resource allocation problem, which is modeled as an infinite horizon \gls{mdp} defined by the tuple $(\mc{S},\mc{A},\mb{P},R,\gamma)$: $\mc{S}$ represents the state space, $\mc{A}$ is the action space (which is potentially different for each state), 
$\mb{P}:\mc{S}\times\mc{A}\times\mc{S}\rightarrow[0,1]$ 
is the transition probability matrix, which depends on the current state, the action chosen by the agent, and the landing state,  $R:\mc{S}\times\mc{A}\times\mc{S}\rightarrow\mathbb{R}$ is the reward function, and $\gamma\in[0,1)$ is the discount factor. Time is divided in slots, and the slot index is denoted by $t \in \mathbb{Z^{+}}$. The ultimate objective of a \gls{drl} agent is to find the optimal policy $\pi^*:\mc{S}\rightarrow\mc{A}$, which maximizes the expected long-term reward:
\begin{equation}
    \pi^*=\argmax_{\pi:\mc{S}\rightarrow\mc{A}}\E{\sum_{t=0}^{\infty}\gamma^t R(\mb{s}_t,\pi(\mb{s}_t),\mb{s}_{t+1})}.
\end{equation}

Let us assume that, in each time slot $t$, the system can allocate $N$ resource blocks, which may represent communication bandwidth, computational cycles, or energy units, depending on the specific application: the type of resource may affect the definition of the specific \gls{mdp}, but is immaterial for our reasoning. In the following we hence generally refer to a \emph{request}, which can be a packet to be transmitted, a computing job to be executed, or an action to be taken, and we assume that each request requires exactly one resource block of some kind (transmission capacity, computational power, or energy).

\begin{figure*}[!t]
\centering
    \includegraphics[width=0.75\textwidth]{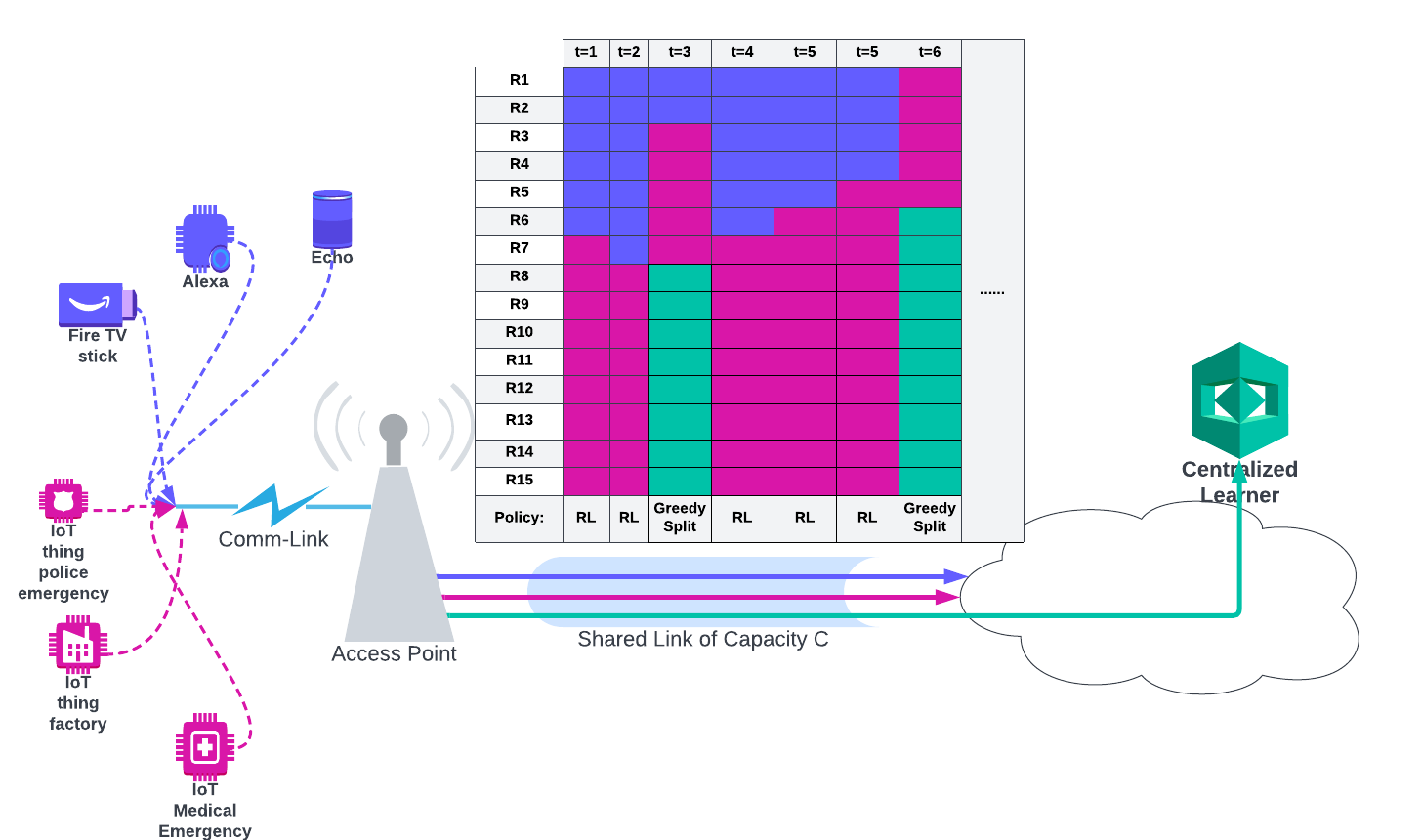}
    \label{fig:scenario}
    \caption{Schematic of the learning control loop in a communication scenario: nodes on the left-hand side represent users, belonging to two different slices (blue and magenta). The connection between the base station and the Internet is used to carry both the users' traffic (blue and magenta streams) and the data needed to train the learner in the cloud (green stream), according to the allocation scheme graphically shown above the link.}
    \label{fig:schematic}
\end{figure*}

The system resources are assumed to be partitioned into $M$ different \textit{slices}, where a slice may serve a single user, or a group of users with the same features. The action space then contains all possible resource allocation vectors that split the $N$ resources among the $M$ slices:
\begin{equation}
    \mc{A}=\left\{\mb{a}\in\{0,\ldots,N\}^M:\sum_{m=1}^M a_m= N\right\}.\label{eq:general_action_space}
\end{equation}

Furthermore, we assume that each slice is associated to a \gls{fifo} queue of requests: each queue has a limited size $Q$, after which the system starts dropping older requests for that slice to make room for newer ones. 

In this work, we focus on \glspl{kpi} tied to the latency with which the requests of the different slices are served. However, the approach can be generalized to consider other metrics. 

We hence indicate by $T_{m,i}$ the latency of the $i$-th request from slice $m$, which depends on the time it spends in the queue before being assigned a resource. Dropped or rejected requests have an infinite latency by definition. The $i$-th request from slice $m$ is generated at time $t_{m,i}$, and age $\Delta_{m,i}(t)$ is defined as:
\begin{equation}
    \Delta_{m,i}(t)=t-t_{m,i}.
\end{equation}
We can then define the reward function:
\begin{equation}
    R(\mb{s},\mb{a},\mb{s}')=\sum_{m=1}^M\sum_{i=1}^{a_m}f_m\left(\Delta_{q_m(i)}\right),\label{eq:general_reward}
\end{equation}
where $\Delta_{q_m(i)}$ is the age of the packet in position $i$ of the $m$-th queue at the current time $t$, and $f_m:\mathbb{N}\rightarrow[0,1]$ is a function mapping the latency of each request to slice $m$'s resulting \gls{qos}. With a slight abuse of notation, we define $f(\varnothing)=0$, where $\varnothing$ indicates that there is no packet in that position in the queue. We can distinguish between slices with \emph{hard} timing requirements, for which the \gls{qos} of a request is 1 if it is served within a maximum latency, and 0 if it exceeds that deadline; and \emph{soft} timing requirements, for which the \gls{qos} is a generic monotonically decreasing function of the latency. 

We can then distinguish between \textit{rejected} and \textit{dropped} packets, the first being packets that find a full queue and are immediately discarded, the second referring to  queued packets  whose age is higher than the deadline and, in case of hard timing requirements, are hence dropped before service since they would not contribute to the \gls{qos} of the slice and just waste resources. We remark that only slices with hard timing requirements can experience dropped packets, while packet rejection can occur in any slice. 

It should also be noted that dropped or rejected requests do not generate any rewards, as they are never included in the sum. The state of the system is then represented by the age of each request contained in each queue, so that in the most general case, $\mc{S}=\left(\{\varnothing\}\cup\mathbb{N}\right)^{M\times Q}$.

The objective of the learning agent is then to learn how to allocate resources among users, so as to maximize their \gls{qos} parameters; it should also be aware of the slices that have a higher risk of violating hard timing requirements and schedule resources to avoid missing deadlines. However, learning is also a computational process, and the \gls{drl} agent may take up some of the same resources that may be allocated to the users in order to improve its policy. As we highlighted in our previous work~\cite{mason2022no}, considering the cost of learning can lead to significantly different choices, limiting the amount and type of experience samples that are selected for training: this is also true regardless of the type of resource the learning requires.

However, even that work only considered static policies, which set up a separate virtual channel (either divided in time or in frequency) for the learning data, strictly separating the learning and data planes. Equivalently, an agent learning how to schedule tasks in an edge server could reserve a certain percentage of computation time to self-improvement, but the amount was decided in advance. This is clearly suboptimal: intuitively, the relative returns from policy self-improvement decrease over time, as the agent gradually converges to the optimal policy. After convergence, and as long as the environment statistics are stable, the value of further improvements to the policy is zero by definition. A dynamic policy for adapting the allocation between requests and learning should then take this into account. 

Furthermore, the current state of the system also needs to be taken into account: if delaying the queued requests further does not have a large impact on the \gls{qos}, the system can take away resources from the slices in order to improve the resource allocation policy, but if the impact is big, e.g., if some requests from a slice with hard timing requirements are already close to the deadline, they need to be prioritized, choosing immediate gains over potential future improvements.

This is particularly important for non-stationary environments, in which the coherence time of the \gls{mdp} statistics is finite: in this kind of system, the learning agent needs to adapt the allocation to the changing statistics of the environment, and cannot rely on offline training, but must keep learning from experience and adapt to the changes proactively.

\subsection{Learning Plane Resource Allocation}

One of the problems of including the learning plane in the resource allocation policy is the circularity of the policy: in order to learn when to allocate resources to policy improvement, a \gls{drl} agent needs to first learn when learning is important. As the policy evolves over time and learning becomes less of a priority, this makes the reward that the agent perceives dependent on the agent's own reduced resources demand, making the learning more difficult.

In order to avoid this problem, we set an external rule to regulate learning, so that the environment that the agent sees is stationary. We define a generic resource allocation vector space $\mc{Z}$ as follows:
\begin{equation}
    \mc{Z}=\left\{\mb{z}\in\{0,\ldots,N\}^M:\sum_{m=1}^M z_m\leq N\right\}.
\end{equation}
We remark here that $\mc{Z}$ is not the action space, but rather a superset of it, i.e., $\mc{A}\subseteq\mc{Z}$: the definition of the action space in~\eqref{eq:general_action_space}  indeed only considers allocations that assign all of the available resources to the users' slices, while $\mc{Z}$ also includes actions that allocate only part of the resources to the users: if $\sum_{m=1}^M z_m<N$, the remaining resources are allocated to the learning.

We can then divide the time slots in two categories, which we name \textit{\gls{drl}} and \textit{learning}: in \gls{drl} slots, all the resources are allocated to the users' slices according to the action chosen by the \gls{drl} agent, while in learning slots the resources are divided between the learning process and the users' slices, according to a simple, empirical strategy. We also remark that learning slots are not considered as experience samples for the \gls{drl} training, as the action $\mb{z}$ in these slots might not belong to the action space $\mc{A}$ considered in the \gls{drl} slots.

Fig.~\ref{fig:schematic} shows a basic schematic of the process in the communication use case: the two classes of users, corresponding to \gls{iot} (magenta) and human communications (blue), transmit over a shared link, and the resources in each time slot (which correspond to bandwidth and time resources in the uplink to the Cloud) are allocated following a dynamic division. Slots 3 and 6 in the figure are learning slots: a significant portion of the resources is allocated to the learning plane (green line). We remind the reader that this networking example, while being the main motivation for our study, is not the only application of the model, which may also be used to allocate scarce computational or energy resources to different users in a dynamic scenario.

For the sake of simplicity, we assume that each slot will be used as a learning slot with probability $\rho(t)$, which decreases linearly over time as the learned policy becomes more stable. The actual shape of $\rho(t)$ (i.e., the learning curve) shall be defined based on the coherence time of the scenario, i.e., the  number of slots $\tau$ over which the statistics of the environment will be approximately stationary. As explained later (see \eqref{eq:rhot}) here we choose a linear function, but other choices are possible. 
We now need to define an allocation strategy in learning slots.

\subsection{Greedy Allocation Strategy}

\begin{figure}[t!]
\centering
\label{fig:allocationPolicy}
\centerline{\includegraphics[width=3.4in]{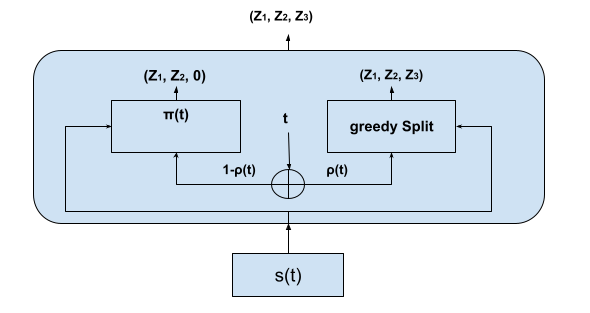}}
\caption{Schematic of the learning plane resource allocation policy.}
\label{fig:greedy}
\end{figure}
To define an strategy for splitting resources in the learning slots, we consider two contrasting objectives: minimizing the loss of \gls{qos} for users, and maximizing the number of experience samples that can be transferred to the learner. 

To capture the first aspect, we define a function $\hat{R}:\mc{S}\times\mc{Z}\rightarrow\mathbb{R}$ that represents an approximation of the instantaneous reward for each resource allocation, considering only the information available in the current state. If the \gls{qos} functions $\{f_m(\cdot)\}$ are known, we can consider the following function:
\begin{equation}
    \hat{R}(\mb{s},\mb{z})=\sum_{m=1}^M\left[\sum_{i=1}^{z_m}f_m\left(\Delta_{q_m(i)}\right)-\sum_{\mathclap{j=z_m+1}}^{L}f_m\left(\Delta_{q_m(i)}+1\right)\right].
    \label{eq:Rhat}
\end{equation}
Note that maximizing \eqref{eq:Rhat} may lead to suboptimal resource allocations, since $\hat{R}(\mb{s},\mb{z})$ does not account for the long-term reward. 

To model the second aspect, we consider that each \gls{drl} slot generates an experience sample, which requires $\ell$ packets (and as many resources) to be transferred to the learner. Due to memory limitations, we assume that the number of samples that can be buffered cannot exceed $E$. We hence define a second function $S(\mb{z},e)$ that captures the effectiveness of the allocation $\mb{z}$ in transferring the $e\in\{0,\ldots,E\}$ samples in the experience queue. In this paper, we define $S$ in terms of the number of experience samples that we manage to transmit in the learning slot, defined by~\eqref{eq:Sze} in our use case, but, once again, this is a reasonable but arbitrary choice and other options might be more suitable for other scenarios. 

The greedy strategy is then the solution to the following optimization problem: 
\begin{equation}
    \mb{z}^*(\mb{s},\mb{z},e)=\argmax_{\mb{z}}\left(M^{-1}\hat{R}(\mb{s},\mb{z})+E^{-1}S(\mb{z},e)\right).
\end{equation}
If $f_m$ is concave for all slices with a soft timing requirement, we can exploit the \gls{fifo} nature of the queue to provide a simple iterative solution, starting from the empty assignment and gradually assigning resources to either one of the slices or the learning process, depending on the value of the utility function. 
Fig.~\ref{fig:greedy} shows a schematic of the full learning plane resource allocation strategy in a simple case with $M=2$: at each time step, the node randomly selects either the \gls{drl} agent or (with probability $\rho(t)$) the greedy allocation, which reserves some resources for the learning plane.

\subsection{Continual Learning}\label{ssec:continual}
We assume the environment can be characterized by a set of parameters $\bm{\omega}$, which determine its stochastic dynamic. From time to time, this parameters can instantaneously change, making the environment non-stationary (in the \gls{drl} jargon, changing the task) and requiring the strategy to be updated in order to pursue the new task.  

To cope with such a non-stationary context we proposed a \gls{cl} strategy similar to the work in~\cite{graves:2016}, but including considerations on the cost of learning. When a context-change is detected, say from  $\bm{\omega}$ to  $\bm{\omega}^\prime$, we consider its significance by means of a distance function $\eta(\bm{\omega},\bm{\omega}^\prime)$: if it is larger than a threshold $\eta_{\text{thr}}$, we consider the environment to be novel enough to warrant a retraining. On the other hand, if the change is smaller than the threshold, we implictly assume that the policy is close enough to the optimum that maintaining it is cheaper than running a new training phase.

We define the environment index $k\in\mathbb{N}$, which starts from 0 and is incremented at every significant change in the environment. Our solution maintains a record of the past environments and the respective learned policies, so that as the environment shifts into the new context $\bm{\omega}_{k+1}=\bm{\omega}^\prime$, we can find the closest past environment:
\begin{equation}
    j^*=\argmin_{j\in\{0,\ldots,k\}}\eta(\bm{\omega}_j,\bm{\omega}^\prime).
\end{equation}
If the previously experienced environment is close enough to the new one, i.e., $\eta(\bm{\omega}_{j^*},\bm{\omega}^\prime)<\eta_{\text{thr}}$, we can apply the stored policy directly, relying on a short training phase with increased exploration rate and training probability $\rho(t)$ to adapt to the small change. If no environment in the memory is close enough, training needs to begin from scratch, with slower and more expensive training phase.

\section{Network Slicing Use Case}\label{sec:slicing}

\setlength{\textfloatsep}{15pt}
\begin{table}[t]
    \centering
    \caption{Use case and learning parameters.}
    \footnotesize
    {\renewcommand{\arraystretch}{1.25}
    \begin{tabular}{@{}rlcc@{}}
    \toprule
    \multicolumn{2}{c}{Parameter} & Symbol & Value \\ \midrule
    \multicolumn{4}{c}{\textbf{Communication system}}\\ \midrule
    \multicolumn{2}{c}{Number of subchannels} & $N$ & 15\\
    \multicolumn{2}{c}{Slot time duration} & $\tau$ & 1~ms\\
    \multicolumn{2}{c}{Packet queue length} & $Q$ & 1500 \\
    \multicolumn{2}{c}{Packet size} & $L$ & 512~B \\
    \multicolumn{2}{c}{Link capacity} & $C$ & 7.68~MB/s \\
     \midrule
     \multicolumn{4}{c}{\textbf{Learning plane}}\\ \midrule
     \multicolumn{2}{c}{Discount factor} & $\gamma$ & 0.95\\
     \multicolumn{2}{c}{Learning queue length} & $E$ & 1500\\
     \multicolumn{2}{c}{Packets required for each sample} & $\ell$ & 3\\
     \multicolumn{2}{c}{Initial learning slot probability} & $\rho_0$ & 0.2\\
     \multicolumn{2}{c}{Final learning slot probability} & $\rho_f$ & 0.01\\
     \multicolumn{2}{c}{Learning slot probability decay} & $\sigma$ & $8\times10^{-4}$\\
     \multicolumn{2}{c}{Learning slot decay pace} & $H$ & 1000\\
     \multicolumn{2}{c}{Queue pressure parameter} & $\chi_1$ & 1400\\

     \bottomrule
    \end{tabular}}
    \label{tab:params}
\end{table}

\begin{table*}[t]
    \centering
    \caption{Traffic Model Parameters}
    \footnotesize
    {\renewcommand{\arraystretch}{1.25}
    \resizebox{0.99\linewidth}{!}{
    \begin{tabular}{@{}ccccccccccccc@{}}
    \toprule
    \multicolumn{1}{c}{Env. Index} & \multicolumn{2}{c}{Policies} &\multicolumn{4}{c}{\textbf{Slice 1 Params}} & \multicolumn{6}{c}{\textbf{Slice 2 Params}}  \\
    \midrule
      &  $\omega_{start}$ & $\omega_{end}$ &$U_1$ & $R_1$ & $\mathbf{O}^{(1)}$ & $\E{U_1}$ & $U_2$  & $R_2$ & $\mathbf{O}^{(2)}$ &  $\E{U_2}$&$T_{\text{soft}}^{(2)}$ & $T_{\max}^{(2)}$ \\
    \midrule
      $\bm{\omega}_{0}$&$\theta \sim \mathcal{N}(0,0.1)$&$\theta_{\omega_{0}}$&28&512kB/s&$\begin{pmatrix} 0.617 & 0.382\\
     0.544 & 0.455\\
     \end{pmatrix}$&11.561&5&512kB/s&$\begin{pmatrix} 0.156 & 0.843\\
     0.763 & 0.236\\
     \end{pmatrix}$&2.625&50ms& 70ms\\
     
      $\bm{\omega}_{12}$&$\theta_{\omega_{1}}$&$\theta_{\omega_{12}}$&15&512kB/s&$\begin{pmatrix} 0.158 & 0.841\\
     0.218 & 0.781\\
     \end{pmatrix}$&11.908&30&512kB/s&$\begin{pmatrix} 0.925 & 0.074\\
     0.672 & 0.327\\
     \end{pmatrix}$&2.976&50ms& 70ms\\

     $\bm{\omega}_{102}$&$\theta_{\omega_{23}}$&$\theta_{\omega_{102}}$&20&512kB/s&$\begin{pmatrix} 0.797 & 0.202\\
     0.316 & 0.683\\
     \end{pmatrix}$&7.810&83&512kB/s&$\begin{pmatrix} 0.949 & 0.050\\
     0.547 & 0.452\\
     \end{pmatrix}$&6.944&50ms& 70ms\\

     $\bm{\omega}_{110}$&$\theta_{\omega_{75}}$&$\theta_{\omega_{110}}$&9&512kB/s&$\begin{pmatrix} 0.696 & 0.303\\
     0.156 & 0.843\\
     \end{pmatrix}$&5.937&37&512kB/s&$\begin{pmatrix} 0.804 & 0.195\\
     0.620 & 0.379\\
     \end{pmatrix}$&8.870&50ms& 70ms\\
     \bottomrule
    \end{tabular}}
    }
    \label{tab:params-env}
\end{table*}
To substantiate the approach on a practical but easy to analyze use case, we consider the resource allocation problem in a simple network slicing scenario. We assume a common communication link is used to transmit both the data packets generated by the users, which belong to two different network slices, and the pieces of information used to feed the learner. Time is divided in slots of constant duration $\tau$, and in each slot the transmission channel can carry $N$ orthogonal and identical resource blocks. The scenario fits the general model presented in the previous section, as the communication resources are shared between the data and learning planes. The full parameters for the scenario, which we will describe in this section, are given in Tab.~\ref{tab:params}.

\subsection{Communication System Model}
We consider two slices, corresponding to the two types of data sources:
\begin{itemize}
    \item Slice 1 is for bulky file transfer, for which we do not set any strict latency constraints. However, we want the system to have the highest possible reliability to ease the burden on the higher layers. As such, $f_1(T)=1$ for all finite values of $T$, but the \gls{qos} is 0 if $T$ is infinite (i.e., if the packet is dropped);
    \item Slice 2 is intended for interactive traffic, such as video conferencing or \gls{vr} traffic, with a strict latency deadline: packets need to be transmitted with a maximum latency $T_{\text{soft}}^{(2)}$. For the sake of simplicity, we assume that, after $ T_{\text{soft}}^{(2)}$, the utility decreases linearly with time, dropping to 0 if the latency is higher than $T_{\max}^{(2)}\geq T_{\text{soft}}^{(2)}$, i.e., $f_2(x) = 1$ if $0\leq x \leq T_{\text{soft}}^{(2)}$, and $f_2(x) = \max\left(0,1-(x-T_{\text{soft}}^{(2)})/(T_{\max}^{(2)}-T_{\text{soft}}^{(2)})\right)$ if $x>T_{\text{soft}}^{(2)}$.
\end{itemize}
We remark that, although these \gls{qos} functions are reasonable, they may not be the most appropriate to represent the considered slices. Since the purpose of this study is to gain insights on the cost of learning in dynamic systems, more than proposing a quantitative performance analysis of the use-case, we prefer these neatly-shaped functions that allow for a qualitative performance analysis while easing the interpretability of the results.  

The number of active users in each slice is variable, making traffic non-deterministic. 
We consider a maximum number of active users $U_m\in\mathbb{N}$ for each slice $m\in\{1,2\}$. Each user follows a on-off model, which can be modeled as a Gilbert-Elliott binary Markov chain with transition probability matrix $\mb{O}^{(m)}$. In state 0, the user does not transmit, while in state 1, it transmits packets of size $L$ with a constant bitrate $R_m$.

The aggregate traffic generated by slice $m$ is then represented by the number $u_m(t)$ of active users at time $t$, multiplied by $R_m$. We can then define a Markov chain over $u_m\in\{0,\ldots,U_m\}$, with the following transition probabilities:
\begin{equation}
\begin{aligned}
P(u_m(t+1)=v | u_m(t) = u)=\sum_{\mathclap{w=\max(0,u+v-U_m)}}^{\mathclap{\min(u,v)}}({O}^{(m)}_{11})^w({O}^{(m)}_{10})^{u-w}\\
\times\binom{u}{w}\binom{U_m-u}{v-w}({O}^{(m)}_{01})^{v-w}({O}^{(m)}_{00})^{U_m-u-v+w}.
\end{aligned}
\end{equation}
The expected traffic $G_m$ from slice $m$ can be computed as:
\begin{equation}
    \E{G_m}=\frac{O^{(m)}_{01}U_mR_m}{O^{(m)}_{01}+O^{(m)}_{10}}.
\end{equation}
On the other hand, the total channel capacity is simply:
\begin{equation}
    C=\frac{NL}{\tau}\,.
\end{equation}
With the values in Tab.~\ref{tab:params}, we obtain $C=7.68\text{~MB/s}$. 

Note that, based on the definition, slice 1 can only experience rejected packets, while slice 2 can have both rejected and dropped packets (if their age exceeds the deadline $T_{\max}^{(2)}$). 

\subsection{Learning Plane}
In this part we define the two components of the learning plane, i.e., the \gls{drl} agent, which will assign resources during the \gls{drl} slots, and the greedy split approach, which manages resource allocation in the learning slots. 
\paragraph*{\gls{drl} agent settings}
We use a \gls{dqn}~\cite{mnih2015human} for the agent, as the problem is simple enough not to require more advanced architectures. 

We consider a simplified state: for each slice $m\in\{1,2\}$, the input to the network is given by the following values:
\begin{itemize}
    \item The number $q_m\in\{0,\ldots,Q\}$ of packets in the queue;
    \item The minimum latency $T^{\min}_m$ for packets transmitted in the previous slot;
    \item The maximum latency $T^{\max}_m$ for packets transmitted in the previous slot;
    \item The average latency $T^{\text{avg}}_m$ for packets transmitted in the previous slot;
    \item The number $d_m$ of discarded (dropped or rejected) packets in the previous slot;
    \item The current number $a_m$ of resource blocks allocated to the slice.
\end{itemize}
The values for each queue are contained in the tuple $\mb{s}^{(m)}=(q_m,T^{\min}_m,T^{\max}_m,T^{\text{avg}}_m,d_m,a_m)$, to which we add another parameter $\xi^{(m)}$, i.e., the difference in the utility for slice $m$ if packets are not transmitted in the next slot. For slice 1, i.e., the latency-insensitive one, this corresponds to the expected number of rejected packets; for the second slice, it is only applicable if some packets are close to or over the soft deadline $T_{\text{soft}}^{(2)}$.
If the head-of-line packets are close to $T_{\text{max}}^{(2)}$, this can even lead to packet drops. We can define it as follows:
\begin{equation}
    \xi^{(2)}=\sum_{i=1}^{q_m}f_2\left(\Delta_{q_2(i)}\right)-f_2\left(\Delta_{q_2(i)}+1\right).
\end{equation}
As the first slice does not have latency requirements, there are no equivalent parameters for it. All the input values are normalized to fit in the range between 0 and 1. 

The input to the \gls{dqn} is then given by $\mb{s}^{(m)},\xi^{(m)}$ which corresponds to a total of 13 values; the training parameters are defined in Tab.~\ref{tab:params}.

For what concerns the action space, we denote by $\mb{a}_t=[a_1,a_2]$  the resource allocation vector during slot $t$. At each step, the \gls{drl} makes an action $\delta_t$ to change the resource allocation as: 
\begin{equation}
   \mb{a}_{t+1}=\mb{a}_t+\delta_t. 
\end{equation}
For the sake of simplicity and interpretability of the results, we admit only actions $\delta_t\in\{(1,-1),(0,0),(-1,1)\}$ that change the allocation to each slice of at most 1 resource block per step. The outputs of the \gls{dqn} correspond to the estimated long-term value of selecting each $\delta_t$, so the network only has 3 output values. 

 The full network architecture is given in Table~\ref{tab:agent}\footnote{The complete implementation of the DQN agent and dynamic resource allocation is available at~\url{https://github.com/slahmer97/costoflearning}}.

\setlength{\textfloatsep}{15pt}
\begin{table}[t]
    \centering
    \caption{\gls{dqn} architecture.}
    \footnotesize
    {\renewcommand{\arraystretch}{1.25}
    \begin{tabular}{@{}ccc@{}}
    \toprule
     \multicolumn{2}{c}{\textbf{Layer size}}           &\multirow{2}{*}{\textbf{Activation function}}\\
    Input & Output &\\
     \midrule
     13 & 64 & ReLU\\
     64 & 32 & ReLU\\
     32 & 3 & Linear\\
     \bottomrule
    \end{tabular}}
    \label{tab:agent}
\end{table}

\paragraph*{Greedy split algorithm settings} We set the size of the experience sample queue $E=1500$, and implement an early rejection policy. When a sample is generated, its rejection probability is equal to $\frac{e}{E}$, i.e., to the current pressure on the queue. Consequently, samples that find a full queue are always rejected, but sometimes samples that could fit in the queue are dropped in favor of new experiences, avoiding too many correlated samples filling the queue.

The probability of selecting a slot as a learning slot decays linearly, starting from an initial value $\rho_0$ and gradually decaying to a value $\rho_f$: every $H$ steps the learning rate is decreased by a constant value $\sigma$:
\begin{equation}
    \rho(t)=\max\left(\rho_f,\rho_0-\left\lfloor\frac{t}{H}\right\rfloor\sigma\right).
    \label{eq:rhot}
\end{equation}

Finally, we consider the greedy allocation in the learning slots. As the first slice has no latency requirements, we consider allocating resources to it greedily only when the number of packets in the queue is higher than a threshold $\chi_1$: in this way, we avoid packet rejections, but also leave more resources for learning plane and latency-sensitive packets.

We can then define the following estimated rewards:
\begin{align}
    \hat{R}_1(\mb{s},\mb{z})&=\min(0,z_1-\min(q_1-\chi_1,N));\\
    \hat{R}_2(\mb{s},\mb{z})&=\min(0,z_2-\min(\xi_2,N));\\
    S(\mb{z},e)&=-\left(\min(e,N)-\left(N-\sum_{m=1}^2 z_m\right)\right). \label{eq:Sze}
\end{align}
The minimum operation ensures that resources will not be allocated to a slice once the queue pressure is below the limit $\xi_1$ or all packets with a close deadline are served, respectively.
We can define the following problem:
\begin{equation}
\begin{aligned}
    \mb{z}^*(\mb{s},\mb{z},e)=\argmax_{\mb{z}\in\mc{Z}} S(\mb{z},e)+\sum_{m=1}^2\hat{R}_m(\mb{s},\mb{z}).
\end{aligned}
\end{equation}
As the problem can easily be converted to an integer linear problem, we can easily solve it through iterative methods. 


\subsection{Continual Learning}

In our environment, the statistics of the traffic change periodically every 500 seconds: the parameter vector $\bm{\omega}$ includes $\mathbf{O}^{(1)}$, $\mathbf{O}^{(2)}$, $U_{1}$, or $U_{2}$, and as a result, corresponding changes are made to the transition matrix $\mb{P}$. The policy for a given set of environmental parameters $\bm{\omega}$ is defined by the set of corresponding trained weights vectors in the \gls{drl} neural network, $\theta_{\bm{\omega}}$.


In the slicing task, the average traffic for each slice is enough to characterize a new environment, and we can identify each context with the vector $\bm{\omega}=(\E{U_1},\E{U_2})$. Additionally, we define the threshold $\eta$ as the point at which we trigger this event. In other words, we only initiate a change event if the distance between two environments is greater than $\eta$. While the average may not accurately represent the environment due to potential variance changes with a constant average, in our specific scenario, the average proved sufficient in maintaining a system performance near the ideal one (as described in the following section). The implementation of a robust method for detecting changes in the environment is critical, as the failure to identify a true change or the detection of a false change can lead to a degradation in system performance. However, in this study, we focus on the efficiency-vs-effectiveness tradeoff of the learning and defer the study of advanced and reliable context-recognition problem to future research.

\begin{figure*}[!ht]
	\centering
	\subfloat[%
	Instantaneous number of active flows.
    \label{sfig:arrival0}
	]{\input{extension-tikz-figures/arrival-0}}
 \vfill
	\subfloat[%
	Smoothed number of active flows.
    \label{sfig:arrival1}
	]{\input{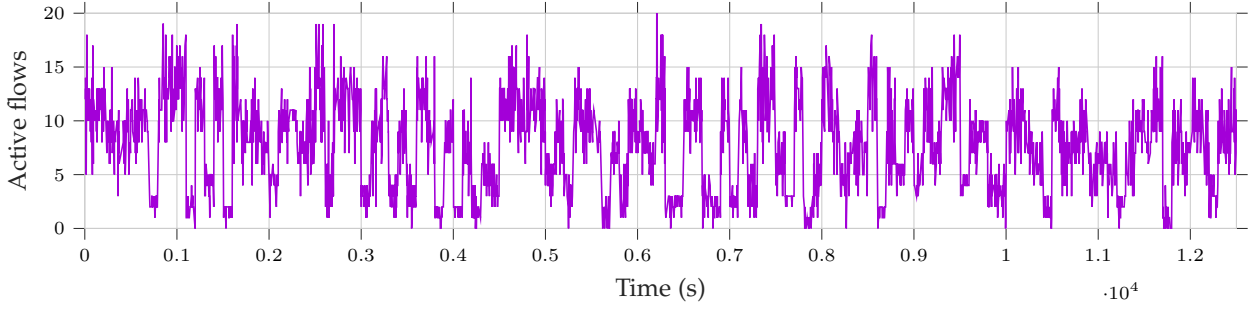}
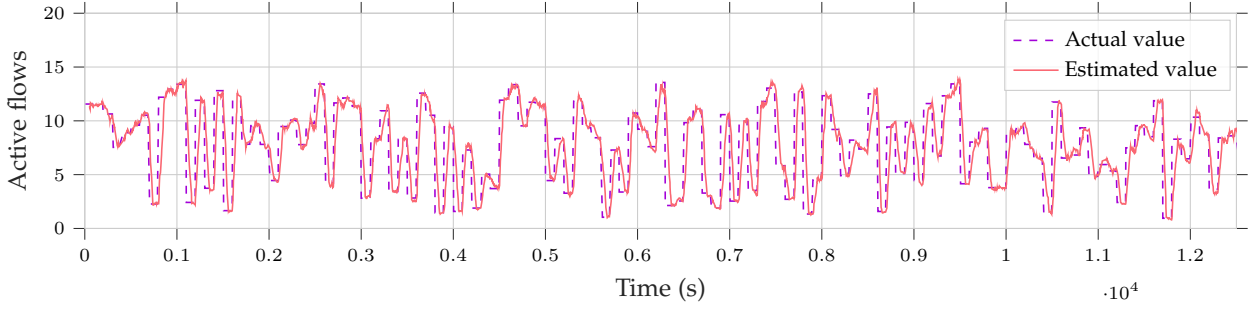}

	\caption{Number of active flows over time for each coherence period in the first slice.}\vspace{-0.3cm}
	\label{fig:arrival-shape}
\end{figure*}

Following the strategy we outlined in Sec.~\ref{sec:system}-\ref{ssec:continual}, the weights of the neural network are chosen from the closest environment observed in the past. We can also make an additional consideration: if the offered traffic is decreasing for both slices, the previous policy will still obtain good results, as the new environment is substantially easier than the previous one. We then define a strict minority relation between vectors, so that $\mb{x}\prec\mb{y}$ if the two vectors $\mb{x}$ and $\mb{y}$ are the same length and each element of $\mb{x}$ is smaller than the corresponding element of $\mb{y}$:
\begin{equation}
    \mb{x}\prec\mb{y}\Leftrightarrow |\mb{x}|=|\mb{y}| \wedge x_i<y_i,\ \forall i\in\{1,\ldots,|\mb{x}|\}.
\end{equation}
We also employ the Euclidean distance to define $\eta(\bm{\omega},\bm{\omega}')$ between two environments, so the centralized agent is updated as follows:
\begin{multline}
\mb{\theta}_{\bm{\omega}_{k+1}}=
\begin{cases}
    $$
    \mb{\theta}_{\bm{\omega}_k},  & \mbox{if } \bm{\omega}_{k+1} \prec \bm{\omega}_k;\\
    \mb{\theta}_{\bm{\omega}_{j^*}}, & \mbox{if } \lVert\bm{\omega}_{k+1}-\bm{\omega}_{j^*}\rVert_2 <\eta_{\text{thr}};\\
    \mb{\theta}\sim\mc{N}(0, 0.1), & \mbox{otherwise},
    $$
    \end{cases}
\end{multline}
where $||\mb{x}||_2$ is the $\ell_2$ norm of vector $\mb{x}$. After the new weight vector is selected, the algorithm temporarily increases both the training slot probability $\rho(t)$ and the exploration rate of the \gls{drl} agent, so that the new policy can be adapted to the new task.

\section{Simulation Settings and Results}\label{sec:results}
In this section, we present numerical findings that demonstrate the efficacy of the dynamic learning plane resource allocation policy in a non-stationary environment. To assess the proposed framework, we run the resource allocation for $64000$ seconds, corresponding to 128 coherence periods lasting 500 seconds each. As each allocation step corresponds to 1~ms, this means that the environment is statistically stable for $5\times10^{5}$ steps, then abruptly transitions to a different behavior. The changes in the environment are produced using Algorithm~\ref{algo:env_gen}. In the algorithm, we denote the probability of a user belonging to slice $m$ being active as $\text{on}_m$, and the uniform distribution between $a$ and $b$ as $\mc{U}(a,b)$. The full parameters of the simulation model are given in Tables~\ref{tab:params} and~\ref{tab:params-env}.

We consider four different benchmarks for the proposed scheme:
\begin{itemize}
    \item \emph{Out-of-band}: this scheme represents the ideal case in which training data is transmitted over a side channel with infinite capacity. This aligns with the common assumption in the literature of free training, and represents an upper bound for performance;
    \item \emph{\gls{fdma}}: here we assume 1 resource block in each slot is reserved to the learning plane, while the other 14 resource blocks can be freely allocated to the users' slices; 
    \item \emph{\gls{tdma}}: we consider a time division between the learning and data planes, in which all available resources are allocated to the learning plane once every $T_{\ell}$ slots. We consider two cases, with $T_{\ell}=10$ and $T_{\ell}=100$.
\end{itemize}
In the following, we will also consider a normalized reward, equal to 1 if all packets are delivered with utility 1 (i.e., before $T_{\text{soft}}^{(2)}$ if they belong to slice 2) and 0 if all packets are dropped or rejected.

Fig.~\ref{sfig:arrival0} shows the total number of active flows (i.e., of users transmitting data in the slot) in the first slice during the entire simulation time, while Fig.~\ref{sfig:arrival1} reports a smoothed time average, along with the tracked number of active flows. We can clearly see that the smoothed average is tracked relatively well, so that any significant drift is detected promptly and dealt with by the \gls{cl} scheme.

\begin{algorithm}[t]
 \caption{Environment Parameter Update}
 \begin{algorithmic}[1]
 \renewcommand{\algorithmicensure}{\textbf{Output:}}
 \ENSURE   $\mb{O}^{(1)}$, $\mb{O}^{(2)}$, $\mb{U_{1}}$, $\mb{U_{2}}$
  \WHILE {$\frac{\mathbb{E}[G_m]}{C}\notin[0.75,1.1]$}
      \FOR{$\mb{i} \in [1,2]$}
          \STATE $\mb{O}^{(i)}[1][1] = \mc{U}(0.05, 0.95)$
          \STATE $\mb{O}^{(i)}[1][0] = 1.0 - \mb{O}^{(i)}[1][1]$
         
          \STATE $\mb{O}^{(i)}[0][0] = \mc{U}(0.05, 0.95)$
          \STATE $\mb{O}^{(i)}[0][1] = 1.0 - \mb{O}^{(i)}[0][0]$
      \ENDFOR
  \STATE $\mb{on_0} = \frac{\mb{O}^{(0)}[0][1]}{(\mb{O}^{(0)}[0][1] + \mb{O}^{(0)}[1][0])}$
  \STATE $\mb{on_1} = \frac{\mb{O}^{(1)}[0][1]}{(\mb{O}^{(1)}[0][1] + \mb{O}^{(1)}[1][0])}$

  \STATE $\mb{U_{1}} =$ \texttt{random}$\left(2, \left\lfloor\frac{14}{\mb{on_0}}\right\rfloor, 1\right)$
  \STATE $\mb{U_{2}}  = \left\lfloor\frac{\max(\left\lfloor15 - \mb{U_{1}}\mb{on_0}\right\rfloor, 1) }{\mb{on_1}}\right\rfloor$

  \STATE Compute the resulting $\mathbb{E}[G_m]$
 \ENDWHILE
 \RETURN $\mb{O}^{(1)}$, $\mb{O}^{(2)}$, $\mb{U_{1}}$, $\mb{U_{2}}$
 \end{algorithmic} 
 \label{algo:env_gen}
 \end{algorithm}

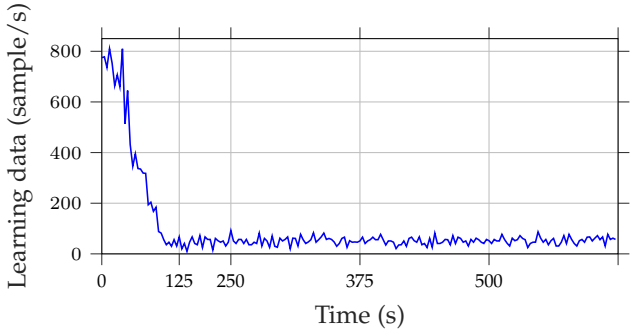
\begin{figure}[t]
\centering
\input{tikz-figures/averageForwardedExperiences.tex}
\caption{Average number of forwarded learning samples per second.}\vspace{-0.3cm}
\label{fig:forwardedExperiences}
\end{figure}

\begin{figure}[t]
\centering
\input{tikz-figures/overheadPerGreedyAction.tex}
\caption{Empirical CDF of the reward loss during learning slots.}\vspace{-0.45cm}
\label{fig:greedy-overhead}
\end{figure}
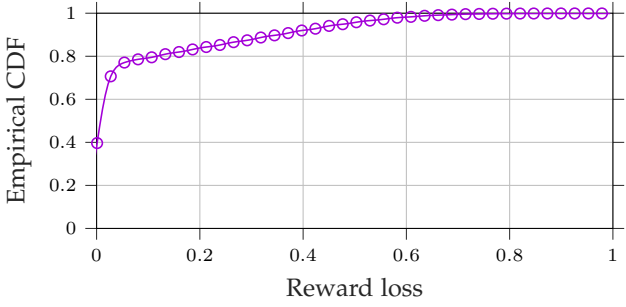

We can also consider the effect of the shared resources on both the learning and data planes: Fig.~\ref{fig:forwardedExperiences} shows how many experience samples are forwarded to the Cloud during the training process. Following the linear decay of $\rho(t)$, the number of new experience samples transmitted for training is initially very high, but decreases over about 70 seconds to reach the minimum, which is between 40 and 50 samples per second. This rate is high enough to guarantee that changes in the environment statistics are captured, but does not impact the final performance, as we will show in the following. Furthermore, we can analyze the impact of learning slots on the instantaneous reward by looking at the empirical \gls{cdf} of the reward penalty from using the greedy allocation, shown in Fig.~\ref{fig:greedy-overhead}: the reward loss is 0 in 40\% of cases, and below 0.1 in 80\% of cases. This means that the greedy allocation can still guarantee good performance in most cases, and as such, is a robust strategy for the learning slots.

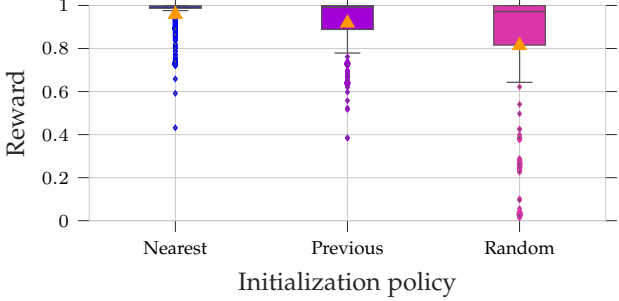
\begin{figure}[!t]
\centering
\input{extension-tikz-figures/aggregated-perf-20}
\caption{Distribution of the performance with different initialization strategies after drift is detected.}\vspace{-0.3cm}
\label{fig:aggregated-perf}
\end{figure}

\begin{figure*}[t!]
\centering
\input{extension-tikz-figures/shared_legend}
\vspace{-0.5cm}
\end{figure*}

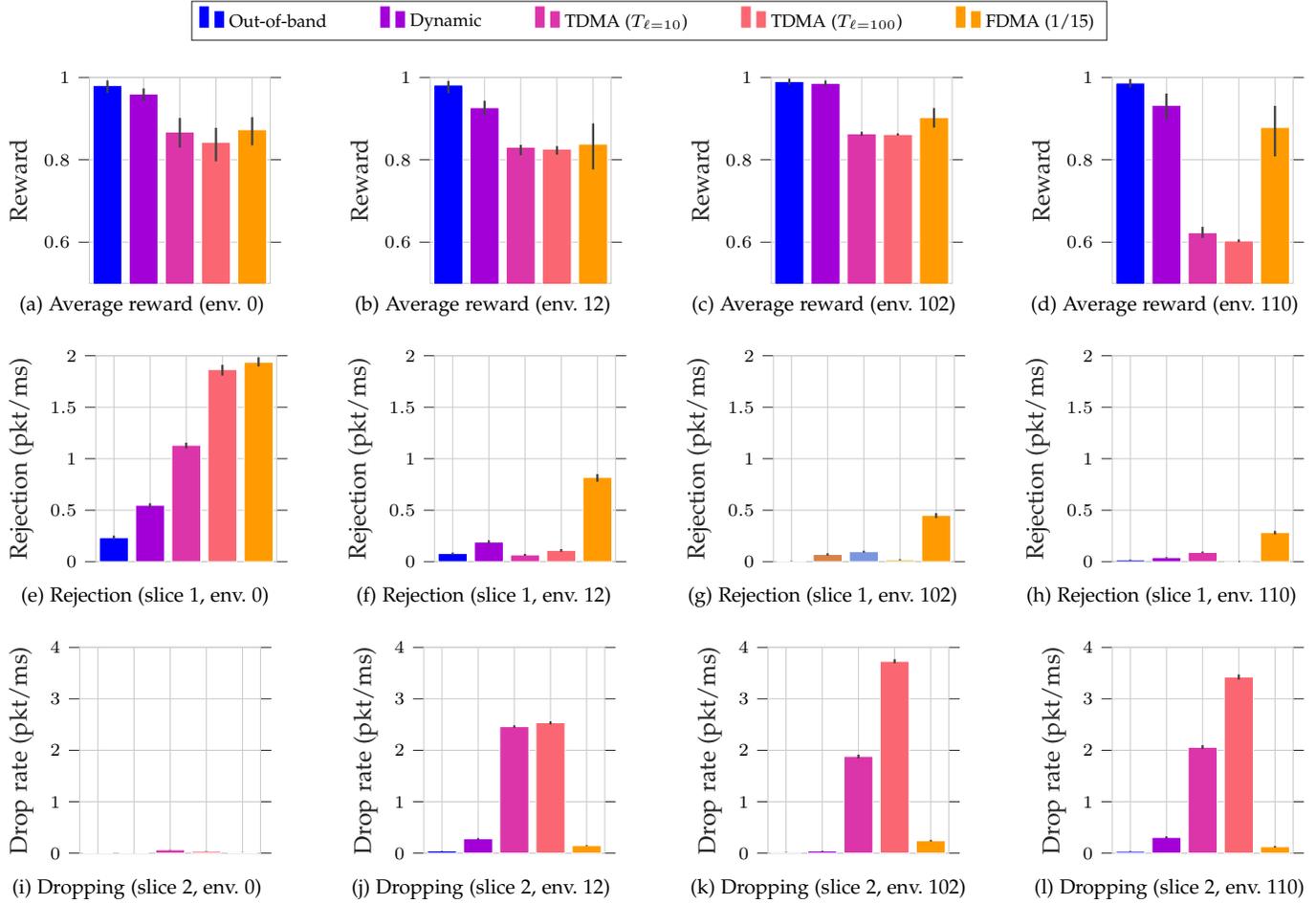
\begin{figure*}[ht!]
    \centering
\subfloat[%
  Average reward (env. 0) \label{fig:perft00}
]{\input{extension-tikz-figures/perf-task-00}}\hfill
\subfloat[%
  Average reward (env. 12) \label{fig:perft12}%
]{\input{extension-tikz-figures/perf-task-12}}\hfill
\subfloat[%
  Average reward (env. 102) \label{fig:perft102}
]{\input{extension-tikz-figures/perf-task-102}}\hfill
\subfloat[%
 Average reward (env. 110) \label{fig:perft110}%
]{\input{extension-tikz-figures/perf-task-110}}\\
\subfloat[%
  Rejection (slice 1, env. 0) \label{fig:ploss00}%
]{\input{extension-tikz-figures/packet-loss-task-00}}\hfill
\subfloat[%
  Rejection (slice 1, env. 12) \label{fig:ploss12}%
]{\input{extension-tikz-figures/packet-loss-task-12}}\hfill
\subfloat[%
  Rejection (slice 1, env. 102) \label{fig:ploss102}%
]{\input{extension-tikz-figures/packet-loss-task-102}}\hfill
\subfloat[%
  Rejection (slice 1, env. 110) \label{fig:ploss110}%
]{\input{extension-tikz-figures/packet-loss-task-110}}\\
\subfloat[%
Dropping (slice 2, env. 0) \label{fig:pdead00}%
]{\input{extension-tikz-figures/packet-dead-task-00}}\hfill
\subfloat[%
Dropping (slice 2, env. 12) \label{fig:pdead12}%
]{\input{extension-tikz-figures/packet-dead-task-12}}\hfill
\subfloat[%
Dropping (slice 2, env. 102) \label{fig:pdead102}%
]{\input{extension-tikz-figures/packet-dead-task-102}}\hfill
\subfloat[%
Dropping (slice 2, env. 110) \label{fig:pdead110}%
]{\input{extension-tikz-figures/packet-dead-task-110}}
\caption{Performance of the schemes in four different sampled environments, measured by the average normalized reward, the packet rejection rate for slice 1, and the packet drop rate for slice 2.}\vspace{-0.3cm}
\label{fig:performance}
\end{figure*}

We have explored the impact of three different initialization strategies on system performance: the proposed strategy, which uses the nearest recorded environment, provides a significant boost over keeping the previous environment or randomly selecting a memorized one, as Fig.~\ref{fig:aggregated-perf} shows: the nearest environment selection improves \gls{cl} by starting the \gls{dqn} with weights that are already close to the correct ones, reducing the mistakes in the initial phases of the retraining (as shown by the limited number of outliers in the boxplot).

We also sampled four different coherence periods for the system (i.e., periods of time during which the context does not change), whose parameters are reported in Tab.~\ref{tab:params-env}. The reported index corresponds to the time of their appearance in the simulation. In all the selected environments, the load is greater than 0.945, i.e., the offered traffic is very close to the channel capacity, and in env. 12 and 110, the load is around 0.99. Fig.~\ref{fig:performance} shows the average reward for these environments, along with the packet drop rate for slice 1 (i.e., packets exceeding $T_{\max}^{(2)}$, which are dropped from the queue as their utility is 0) and the packet rejection rate for slice 2 (i.e., packets which find a full buffer and are discarded directly). The indices of the periods represent an incremental number of different environment seen by the \gls{cl} agent: the first, with index 0, is the first to be seen, while there are 12 other periods between 0 and 12, and so on. Each period has the same duration, i.e., 500 seconds. 
 
As a first observation, we note that the ideal out-of-band policy, which neglects the cost of learning, clearly outperforms those that reserve some resources to the learning plane, which confirms that the cost of learning is not negligible and needs to be accounted for when designing the resource allocation strategies, as we have done in our "Dynamic" scheme. 
The bar plots in Fig.~\ref{fig:perft00}-\subref*{fig:perft110}, in fact, show that our scheme can outperform the static \gls{fdma} and \gls{tdma} resource allocation strategies, almost reaching the same performance as the ideal out-of-band system. The only cases with an appreciable performance gap between our scheme and the out-of-band system are env. 12 and env. 110: as we remarked above, these are the most challenging ones, with a total load close to or over 99\% of the nominal link capacity. In these limit cases, any learning policy that requires resources for the training will unavoidably determine the violation of the \gls{qos} requirements for some users, which further highlights the importance of the cost of learning in the system design.

\begin{figure*}[!t]
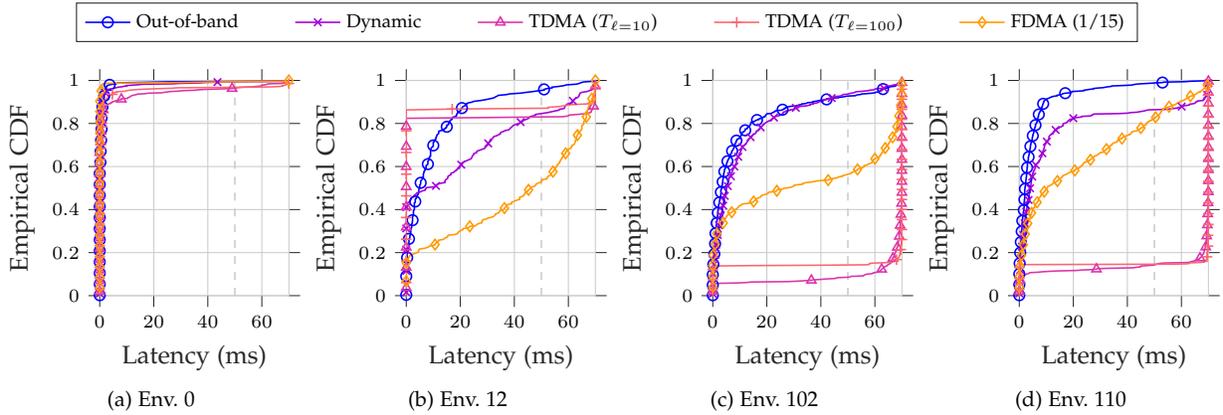

    \centering
\subfloat[%
  Env. 0 \label{fig:latency00}%
]{\input{extension-tikz-figures/latency1-task-00}}
\subfloat[%
  Env. 12 \label{fig:latency12}%
]{\input{extension-tikz-figures/latency1-task-12}}
\subfloat[%
  Env. 102 \label{fig:latency102}%
]{\input{extension-tikz-figures/latency1-task-102}}
\subfloat[%
  Env. 110 \label{fig:latency110}%
]{\input{extension-tikz-figures/latency1-task-110}}
    \caption{Empirical CDF of the latency for slice 2 in the four selected environments.}\vspace{-0.45cm}
    \label{fig:latency}
\end{figure*}

The relative simplicity of the system model we considered makes it possible to analyze in depth the choices made by the different schemes. From the bar plots in Fig.~\ref{fig:ploss00} to Fig.~\ref{fig:pdead110} we can observe that the \gls{fdma} and \gls{tdma} schemes tend to drop or reject a significant number of packets in all environments, while the ideal and dynamic ones manage to limit the number of unserved packets for both slices. Interestingly, even the ideal scheme drops a significant number of packets from slice 1 in env. 0, but performance is still high. We can explain this by considering Fig.~\ref{fig:latency}, which shows the empirical \gls{cdf} of the latency for packets in slice 2. Fig.~\ref{fig:latency00} clearly shows that almost all packets have utility 1, i.e., are delivered before $T_{\text{soft}}^{(2)}$: in this case, all schemes tend to privilege slice 2, filling the queue in slice 1 more often. We should also consider that the learning agents start from scratch in environment 0, i.e., they have no pre-trained weights to start from, and we should expect a relatively large number of mistakes.

We can also see that the ideal and dynamic schemes have matching latency profiles in env. 102, as Fig.~\ref{fig:latency102} shows, while the \gls{fdma} and \gls{tdma} schemes tend to transmit more packets with a latency close to $T_{\max}^{(2)}$. In environments 12 and 110, shown in Fig.~\ref{fig:latency12} and Fig.~\ref{fig:latency110}, respectively, the dynamic scheme drops more packets than the ideal one, and has a higher overall latency, but still outperforms the static allocation schemes. Interestingly, the two \gls{tdma} schemes tend to have better latency performance than the dynamic scheme in env.~12, but cannot improve the utility: the fraction of packets with a latency higher than $T_{\text{soft}}^{(2)}$ is the same for all three schemes, and the two \gls{tdma} ones drop a large number of packets, causing a significant performance difference. In this case, serving most packets from slice 2 as soon as they arrive is not advantageous, as it leads to worse performance overall for \gls{tdma}.

\begin{figure*}[t!]
\centering
\input{extension-tikz-figures/shared_legend-line}
\vspace{-0.5cm}
\end{figure*}

\section{Conclusions and Future Directions}\label{sec:conc}
In this work, we have designed a dynamic resource allocation policy, which can mediate between the learning and data planes, controlling the trade-off between effectiveness and efficiency of \gls{drl} models. Unlike most works in the learning-based networking literature, we specifically consider the cost of learning, i.e., the resources required by the training process of a \gls{drl} agent, and show that our dynamic policy can outperform static schemes and, after a short transition phase, match the performance of an ideal system with an out-of-band learning plane. Furthermore, the adaptability of the scheme is demonstrated by applying it in a \gls{cl} setting with environment changes, to which the dynamic scheme adapts extremely quickly.

Possible extensions of the work certainly include the adaptation of the scheme to more complex scenarios, with a larger number of resources and slices and more stringent \gls{qos} requirements,  as those for \gls{urllc}. However, even more interesting would be addressing some theoretical questions, such as the interplay between the cost of learning and active learning, which requires to select the most valuable samples to be transmitted in order to accelerate the training, particularly when  resources in the learning plane are scarce. As mentioned, furthermore, the detection of context changes that trigger a retraining of the network is another open challenge. Finally, of particular interest is the design of meta-learning schemes that can learn when the resource allocation scheme needs to be retrained, balancing the potential performance improvement that could be brought about by a retrained policy and the cost to learn it, relative to the expected system coherence time. 

\bibliographystyle{IEEEtran}
\bibliography{biblio.bib}

\newpage

\newpage

\vfill\pagebreak

\end{document}

%% file: extension-tikz-figures/arrival-0.tex
\begin{tikzpicture}
	
	\definecolor{darkorange25512714}{RGB}{255,127,14}
	\definecolor{darkslategray38}{RGB}{38,38,38}
	\definecolor{forestgreen4416044}{RGB}{44,160,44}
	\definecolor{lightgray204}{RGB}{204,204,204}
	\definecolor{steelblue31119180}{RGB}{31,119,180}
	\pgfplotsset{every tick label/.append style={font=\scriptsize}}
	\begin{axis}[
		width=\fwidth*2,
		height=\fheight,
		axis line style={lightgray204},
		tick align=outside,
		x grid style={lightgray204},
        xlabel=\textcolor{darkslategray38}{Time (s)},
		ylabel=\textcolor{darkslategray38}{Active flows},
		xmajorgrids,
		xmin=0.0, xmax=12500,
		xtick style={color=darkslategray38},
		y grid style={lightgray204},
		ymajorgrids,
		ymin=0, ymax=20,
		ytick style={color=darkslategray38}
		]
		\addplot [semithick, color1]
		table[x expr=\thisrowno{0}/1000, y index=1] {%
			7399 13
			7578 12
			8806 14
			12093 12
			15205 5
			22598 18
			27449 10
			36682 13
			37377 12
			41172 10
			50082 13
			58194 10
			60849 11
			61023 9
			61705 12
			61946 11
			67207 10
			71715 8
			76866 14
			82373 5
			82940 13
			87224 12
			87235 17
			90145 12
			92894 6
			93533 12
			109708 10
			122080 11
			126514 9
			131540 13
			132727 8
			142082 13
			142674 12
			143426 11
			147829 11
			152751 12
			153054 13
			153395 10
			160667 11
			163928 13
			163973 11
			168035 12
			176224 12
			185473 12
			186027 13
			186176 8
			186349 13
			187335 13
			194005 11
			200014 11
			201881 9
			206170 11
			206450 15
			207740 10
			216910 13
			228837 10
			239307 7
			243470 9
			245629 10
			246873 8
			251087 10
			255250 9
			264498 8
			267475 6
			274084 11
			274440 6
			277418 11
			290993 11
			291295 11
			292697 15
			293245 8
			295391 13
			298694 7
			308671 9
			308820 9
			311028 10
			315826 9
			321221 5
			321711 9
			322771 8
			350496 10
			355649 9
			356133 8
			360228 10
			360751 3
			363096 8
			367722 11
			368118 7
			369387 9
			371478 6
			417225 8
			418013 10
			426171 8
			428944 6
			431894 10
			436904 10
			438477 5
			442448 6
			460903 13
			461651 9
			471449 9
			471779 12
			478326 5
			488361 8
			496037 7
			497409 11
			506443 11
			515744 12
			515780 10
			517643 10
			519312 9
			538149 8
			539727 10
			544485 14
			549493 7
			553582 10
			555372 7
			556190 6
			561121 11
			565398 7
			570495 8
			579093 11
			579922 14
			584035 13
			587409 8
			592531 10
			603065 12
			612275 10
			614162 11
			614247 11
			615725 12
			619324 8
			633819 11
			635913 11
			647338 11
			650561 12
			651709 11
			652578 9
			654434 13
			656317 11
			668079 11
			676114 10
			678838 10
			679219 9
			685674 9
			703194 3
			706705 2
			707740 2
			712216 2
			712888 2
			725773 2
			726543 2
			727373 3
			728987 3
			737894 2
			738533 2
			742023 3
			746802 2
			754534 3
			760588 1
			772530 3
			773473 3
			775308 3
			775855 3
			776540 2
			777648 2
			778611 3
			779017 2
			780559 3
			782638 3
			785709 2
			787082 2
			787446 2
			801770 10
			805692 8
			807045 12
			807588 9
			808973 14
			813056 11
			821224 11
			833134 11
			835040 11
			850628 19
			851450 19
			857373 13
			859060 12
			866315 14
			872739 8
			878284 18
			884307 12
			887388 8
			888871 11
			888895 12
			902373 13
			906041 12
			916350 15
			917205 16
			921326 14
			924806 16
			928557 13
			935051 12
			936295 12
			936949 9
			937454 14
			938245 14
			943694 12
			949265 10
			949802 18
			950316 10
			951907 10
			952274 11
			953631 13
			953671 10
			963531 10
			975904 11
			976968 15
			977023 15
			988149 13
			991149 14
			992680 11
			995496 14
			998237 17
			999197 13
			1000649 14
			1001133 13
			1005141 13
			1005228 15
			1006684 12
			1007697 16
			1007837 14
			1009594 13
			1013916 13
			1016478 14
			1017194 12
			1018188 16
			1025979 13
			1031696 17
			1041816 13
			1055038 14
			1055853 15
			1056923 9
			1059690 16
			1061970 12
			1063013 13
			1066187 13
			1068473 10
			1070448 14
			1071359 10
			1087448 17
			1097590 18
			1101447 1
			1105903 1
			1106651 3
			1116779 2
			1117774 3
			1119828 2
			1122573 3
			1123326 1
			1134029 1
			1142409 2
			1145209 2
			1148005 2
			1157876 3
			1158897 4
			1160780 2
			1165454 2
			1172671 3
			1173753 3
			1174605 2
			1179171 2
			1196800 0
			1197973 0
			1200017 11
			1200372 13
			1210398 12
			1224922 14
			1225038 13
			1225360 9
			1239647 12
			1248065 14
			1249262 11
			1250575 15
			1250736 12
			1259711 14
			1259973 14
			1261241 9
			1263695 12
			1274140 9
			1280108 10
			1295561 13
			1297578 14
			1302093 3
			1303784 5
			1304746 4
			1308209 3
			1312576 5
			1313228 4
			1315155 6
			1324336 5
			1325654 4
			1327862 4
			1335178 4
			1341777 5
			1346818 3
			1352740 3
			1356558 1
			1357216 4
			1359255 5
			1370773 5
			1372373 4
			1380145 5
			1381076 4
			1383533 5
			1387023 4
			1393092 4
			1399522 2
			1399534 3
			1404767 17
			1410771 12
			1415347 13
			1419576 12
			1420308 16
			1422390 10
			1426845 13
			1430165 14
			1442413 14
			1443395 8
			1444467 11
			1450692 9
			1452045 11
			1455163 11
			1458651 13
			1466224 11
			1466497 9
			1475473 11
			1478134 14
			1485960 11
			1487514 13
			1490014 12
			1491881 17
			1495628 14
			1499903 12
			1506910 1
			1509529 2
			1512146 2
			1518147 2
			1528305 1
			1532633 0
			1533895 2
			1534563 1
			1535261 2
			1536873 2
			1545743 2
			1547373 2
			1558986 2
			1559961 2
			1563392 2
			1564440 1
			1567279 1
			1569785 2
			1573557 2
			1576181 2
			1581949 2
			1586099 1
			1588126 2
			1590966 2
			1598671 1
			1599067 2
			1599698 2
			1604415 18
			1605826 12
			1608516 8
			1608531 14
			1611232 12
			1613337 17
			1617473 13
			1620904 16
			1621055 6
			1622023 13
			1625539 10
			1643237 14
			1646133 17
			1650141 12
			1650539 19
			1650751 9
			1651290 8
			1653096 13
			1655512 13
			1656551 14
			1657722 16
			1660975 10
			1692033 12
			1707225 7
			1707865 7
			1709679 6
			1711187 10
			1715448 4
			1716359 6
			1719933 6
			1721894 10
			1724342 10
			1736057 8
			1739353 8
			1742590 7
			1745660 12
			1747291 10
			1750903 9
			1751651 8
			1753336 11
			1756151 9
			1756981 11
			1758952 9
			1765511 8
			1790209 8
			1802480 10
			1803897 8
			1807643 10
			1810454 11
			1810492 12
			1816898 4
			1819605 11
			1822313 12
			1823268 13
			1829716 8
			1829727 11
			1832803 10
			1833534 11
			1834485 9
			1835033 10
			1845017 11
			1845372 14
			1848337 12
			1849901 10
			1860198 10
			1860495 6
			1862879 11
			1863735 6
			1870360 10
			1872924 11
			1874035 6
			1882531 9
			1884647 12
			1886030 11
			1897094 9
			1904973 8
			1905725 9
			1913417 8
			1917172 10
			1922072 5
			1923819 6
			1929729 6
			1935077 7
			1941559 9
			1947570 7
			1948950 5
			1953200 8
			1956154 8
			1962336 7
			1966641 8
			1982154 7
			1983666 10
			1989220 8
			2011851 1
			2012059 5
			2014106 5
			2014884 4
			2015600 4
			2018367 4
			2018447 4
			2022059 4
			2040965 6
			2056061 4
			2059483 5
			2059941 5
			2079349 2
			2087408 5
			2093419 4
			2099611 4
			2100002 9
			2108185 11
			2115493 9
			2115991 7
			2133893 12
			2136993 13
			2137314 9
			2140332 8
			2142157 8
			2148119 11
			2150777 11
			2150948 9
			2166819 10
			2169062 12
			2169776 10
			2172913 9
			2182235 10
			2184346 10
			2185187 11
			2193509 8
			2195732 10
			2197571 9
			2199114 10
			2199679 12
			2199768 6
			2200720 10
			2212824 9
			2213222 10
			2213963 10
			2229696 11
			2231836 11
			2240176 11
			2250479 11
			2256904 11
			2257275 10
			2263205 11
			2265116 11
			2271392 11
			2273756 11
			2277480 11
			2280787 10
			2289049 11
			2292932 9
			2293705 10
			2293979 10
			2304123 10
			2304224 8
			2314703 6
			2321205 7
			2326332 7
			2328195 6
			2336447 9
			2338564 9
			2340588 7
			2341442 3
			2342584 8
			2343847 8
			2347720 8
			2353802 9
			2357248 8
			2361650 7
			2364820 7
			2367856 8
			2367944 10
			2370005 8
			2371492 8
			2372863 8
			2377286 8
			2379241 8
			2382135 6
			2382759 8
			2384788 8
			2390018 8
			2391204 8
			2395127 8
			2397029 6
			2404854 10
			2413934 12
			2415762 6
			2421591 4
			2424768 12
			2425153 7
			2428095 8
			2429782 9
			2431686 10
			2436060 10
			2437374 8
			2441646 15
			2448549 10
			2449689 13
			2451516 5
			2462965 8
			2465044 6
			2466155 6
			2467149 9
			2472576 10
			2473055 11
			2478381 9
			2482397 7
			2483408 6
			2491342 8
			2500393 15
			2504040 8
			2507435 13
			2509785 11
			2509977 19
			2525101 13
			2526539 14
			2531773 13
			2539275 19
			2543348 13
			2551425 14
			2551953 11
			2554920 13
			2566549 8
			2572698 17
			2573132 12
			2574420 11
			2574643 8
			2583218 11
			2584494 19
			2587381 15
			2588634 10
			2588847 9
			2599573 15
			2613933 2
			2615962 4
			2624191 3
			2624595 6
			2628693 4
			2630998 4
			2631727 1
			2634806 5
			2636752 3
			2636822 6
			2642611 5
			2645688 10
			2649882 1
			2662575 5
			2667197 2
			2677053 4
			2677984 5
			2682134 7
			2682859 2
			2683175 4
			2683467 4
			2685208 4
			2690940 8
			2692230 3
			2692842 3
			2695379 5
			2696700 4
			2697968 3
			2703566 19
			2703906 15
			2707397 7
			2714579 13
			2715003 13
			2715154 11
			2732366 12
			2749133 14
			2749145 12
			2760302 10
			2772724 12
			2773318 11
			2775843 13
			2777837 16
			2780194 13
			2783761 9
			2784243 10
			2787408 14
			2792461 13
			2793079 9
			2797020 12
			2800586 10
			2806556 13
			2811916 15
			2812229 11
			2812365 10
			2817970 12
			2820117 12
			2820564 7
			2822897 11
			2828222 12
			2836682 14
			2850680 13
			2856522 13
			2869696 7
			2876702 14
			2884859 15
			2903729 12
			2909970 12
			2910622 12
			2920430 9
			2924534 15
			2930449 6
			2935237 13
			2949155 10
			2950194 10
			2950355 11
			2955266 11
			2955363 9
			2955962 14
			2957102 11
			2961250 13
			2964543 12
			2968715 12
			2973587 13
			2979957 10
			2986902 9
			2987318 10
			2987344 9
			2993579 11
			2993774 13
			3000514 3
			3000540 4
			3001798 2
			3004031 4
			3004744 3
			3009178 3
			3010396 3
			3017367 4
			3018367 3
			3025085 3
			3031853 3
			3032985 2
			3039367 4
			3042991 3
			3043344 3
			3052952 1
			3054184 3
			3056277 2
			3060194 4
			3060678 1
			3062462 2
			3062835 4
			3072584 3
			3085014 3
			3086666 2
			3087836 4
			3090751 5
			3095427 4
			3095475 3
			3096847 3
			3105464 5
			3116156 10
			3118087 7
			3122115 8
			3122914 6
			3125522 10
			3127121 6
			3128927 3
			3129345 8
			3130436 7
			3141344 9
			3142866 8
			3144304 7
			3147225 11
			3150804 10
			3155793 13
			3162080 2
			3164327 13
			3169789 8
			3179106 9
			3197611 10
			3208441 8
			3210576 9
			3212962 10
			3216510 9
			3217806 12
			3218772 12
			3219700 12
			3220204 6
			3226475 7
			3227036 9
			3227490 11
			3229744 13
			3236100 16
			3248415 13
			3251354 13
			3252098 13
			3252959 12
			3277177 16
			3278161 12
			3280767 12
			3283541 10
			3287347 10
			3293390 11
			3293655 9
			3294820 9
			3300669 3
			3303719 4
			3304304 4
			3306921 4
			3307617 5
			3312631 4
			3318014 2
			3320091 4
			3321249 4
			3325227 3
			3329464 2
			3334419 2
			3337624 4
			3341909 4
			3345909 4
			3349790 3
			3357223 4
			3357712 1
			3366903 4
			3369916 4
			3372071 5
			3376076 1
			3380185 2
			3383867 2
			3384350 5
			3388180 3
			3390863 4
			3393217 4
			3397254 4
			3402526 9
			3402683 7
			3410290 8
			3415802 10
			3420755 8
			3422900 8
			3424242 5
			3432286 9
			3434290 9
			3444362 11
			3450771 10
			3451030 9
			3451686 6
			3452235 8
			3453473 8
			3453733 6
			3464575 13
			3467194 7
			3467388 7
			3470731 7
			3484070 11
			3495339 4
			3497363 6
			3501427 1
			3502927 3
			3504733 3
			3505323 3
			3506838 3
			3511463 2
			3523865 5
			3524096 2
			3530707 3
			3531237 2
			3532161 6
			3535354 2
			3541029 5
			3543527 3
			3546225 2
			3547687 0
			3559641 3
			3560770 4
			3561318 3
			3569834 3
			3570207 2
			3576902 2
			3579446 4
			3592376 2
			3592533 2
			3592636 3
			3593850 3
			3595264 3
			3607164 15
			3609147 11
			3609542 10
			3610441 16
			3610566 11
			3614113 13
			3621130 14
			3643411 11
			3652379 13
			3652536 11
			3655777 12
			3656022 8
			3661589 14
			3662334 11
			3663333 12
			3672528 14
			3686816 11
			3687789 13
			3689177 12
			3694922 13
			3703287 10
			3705914 16
			3706702 11
			3709089 11
			3715530 8
			3715867 11
			3718709 14
			3718965 9
			3720183 9
			3721203 9
			3727219 5
			3740504 11
			3741193 8
			3746549 9
			3747146 10
			3747768 10
			3754573 9
			3759764 8
			3783713 9
			3791936 10
			3793032 7
			3794226 6
			3795126 16
			3797000 7
			3797369 5
			3801689 1
			3805301 1
			3807563 2
			3809727 2
			3811560 2
			3813284 1
			3815970 2
			3822268 2
			3827657 2
			3828080 1
			3828380 2
			3828835 1
			3831303 1
			3840501 1
			3842676 2
			3843845 1
			3844320 2
			3850973 1
			3857811 1
			3858596 0
			3867355 0
			3872813 2
			3873406 1
			3878021 2
			3888546 2
			3891701 2
			3900976 6
			3906698 8
			3913177 7
			3914066 10
			3917865 10
			3930875 7
			3937355 9
			3946276 10
			3951538 7
			3951901 9
			3952551 10
			3953081 7
			3956987 10
			3957339 10
			3963156 9
			3965451 10
			3967912 7
			3969012 11
			3971829 11
			3974997 9
			3975633 8
			3989411 9
			3996460 8
			3999639 12
			4003223 2
			4006443 2
			4012065 2
			4019100 1
			4023588 2
			4027318 2
			4031862 2
			4038502 2
			4040069 2
			4042230 2
			4046228 2
			4053832 1
			4055122 2
			4057285 2
			4059292 2
			4062508 2
			4064140 2
			4069525 2
			4073160 1
			4073861 2
			4081514 2
			4095890 2
			4096350 1
			4102598 7
			4104907 10
			4105787 11
			4108094 5
			4109434 5
			4113441 9
			4114936 6
			4115516 8
			4117455 11
			4118636 11
			4121462 8
			4126713 6
			4129795 9
			4137637 9
			4139971 6
			4146899 7
			4154155 6
			4157908 6
			4159894 10
			4162424 8
			4162674 9
			4165014 5
			4165281 3
			4172331 8
			4173449 5
			4177142 8
			4179014 7
			4182932 6
			4183144 6
			4183786 8
			4190402 8
			4190684 14
			4191942 5
			4191965 9
			4192933 8
			4201600 1
			4203677 3
			4212599 4
			4213111 1
			4216298 2
			4217657 1
			4219417 1
			4219455 2
			4220836 2
			4225487 4
			4228795 1
			4228831 1
			4232298 1
			4234436 1
			4236383 3
			4237300 4
			4240422 0
			4240541 0
			4244445 1
			4244759 3
			4246000 0
			4250652 0
			4252235 1
			4252463 2
			4267419 1
			4267706 2
			4274658 2
			4294897 1
			4298509 2
			4309339 6
			4319347 6
			4319875 7
			4326619 6
			4329716 5
			4330419 4
			4331418 5
			4338370 8
			4341416 7
			4343240 4
			4346817 5
			4355433 5
			4357746 6
			4360781 3
			4363742 5
			4366862 5
			4370991 4
			4372551 4
			4375411 5
			4377166 6
			4380940 8
			4386386 5
			4387002 6
			4387571 5
			4392149 3
			4393824 7
			4402150 4
			4411221 4
			4413317 4
			4419581 6
			4421354 2
			4421811 6
			4426080 2
			4428360 3
			4430620 3
			4431347 6
			4435969 3
			4440238 4
			4441956 6
			4442713 5
			4443354 6
			4453538 2
			4454790 3
			4457125 3
			4463875 3
			4469999 5
			4471286 3
			4475805 4
			4479655 2
			4483992 3
			4487395 3
			4491115 3
			4494990 3
			4495950 3
			4502058 12
			4503341 13
			4504238 13
			4507899 11
			4510687 14
			4515174 11
			4519203 12
			4520366 10
			4523599 13
			4525240 12
			4527287 11
			4530534 11
			4533975 12
			4534573 10
			4535550 13
			4536822 11
			4539854 13
			4554299 13
			4557216 10
			4559595 12
			4562328 9
			4564717 12
			4567735 12
			4568799 10
			4572750 12
			4578784 11
			4579334 11
			4583988 11
			4588016 12
			4590796 13
			4592826 11
			4594983 13
			4595377 11
			4596708 10
			4601597 14
			4601623 16
			4601857 13
			4606106 12
			4607051 15
			4612053 10
			4612268 15
			4612984 11
			4613737 8
			4616449 11
			4618175 16
			4618394 15
			4622301 16
			4625940 11
			4627230 13
			4627703 11
			4630379 12
			4631700 9
			4635020 14
			4637475 13
			4637507 17
			4638906 10
			4643244 15
			4647081 15
			4648206 10
			4650003 13
			4652432 9
			4668799 15
			4670064 13
			4670490 11
			4671504 11
			4684133 17
			4684145 14
			4685304 13
			4695302 13
			4706883 7
			4707724 9
			4712121 11
			4713463 11
			4715194 13
			4722378 12
			4723948 9
			4732020 8
			4734469 11
			4737685 9
			4739087 12
			4740670 11
			4756293 9
			4763222 14
			4782133 6
			4785591 9
			4786783 9
			4799950 9
			4800725 18
			4803570 12
			4804478 11
			4806301 12
			4806741 15
			4807850 9
			4810072 11
			4810844 13
			4814275 13
			4817909 16
			4819859 8
			4838334 11
			4842338 13
			4848789 12
			4851236 11
			4865449 11
			4867574 13
			4873442 13
			4874236 10
			4874963 12
			4884155 13
			4887297 9
			4900001 10
			4914146 14
			4918433 14
			4920627 11
			4928774 13
			4935514 15
			4937461 11
			4944928 10
			4945396 12
			4955204 5
			4957123 12
			4957304 13
			4974603 11
			4977272 8
			4977991 9
			4979780 13
			4987952 14
			4994130 11
			4995678 12
			4997462 8
			4998121 9
			5000612 4
			5000719 5
			5005651 6
			5018203 5
			5020834 7
			5023280 5
			5023311 4
			5024447 4
			5030427 4
			5030475 6
			5031847 4
			5032869 4
			5033711 7
			5045520 5
			5051156 3
			5051720 4
			5053584 3
			5057069 6
			5057115 5
			5060236 5
			5060448 5
			5060666 2
			5064345 6
			5071586 4
			5076344 3
			5082120 4
			5082225 6
			5090793 5
			5094617 6
			5103746 6
			5104789 9
			5105733 7
			5117263 5
			5120624 12
			5132611 8
			5139431 9
			5141785 8
			5145576 10
			5154700 11
			5155204 9
			5156271 5
			5170346 9
			5173084 10
			5183415 6
			5187040 10
			5187098 12
			5187959 12
			5193596 11
			5198360 5
			5198906 7
			5202000 3
			5217598 5
			5231311 5
			5235669 3
			5247913 2
			5250592 0
			5253056 4
			5254350 5
			5255091 4
			5257345 4
			5260227 4
			5263811 4
			5264352 4
			5264464 1
			5271074 3
			5271146 4
			5271368 4
			5272596 4
			5272624 3
			5275535 4
			5276909 4
			5278409 3
			5280909 3
			5283845 2
			5284209 4
			5287819 2
			5292712 3
			5306741 9
			5315093 10
			5318612 12
			5319972 14
			5323161 11
			5325863 13
			5328217 7
			5340829 11
			5342841 10
			5345290 13
			5345642 13
			5346697 11
			5348741 12
			5352019 13
			5355755 12
			5359242 15
			5360043 12
			5360920 15
			5364529 12
			5367286 10
			5385771 9
			5386027 13
			5387577 10
			5388881 9
			5396186 10
			5399575 11
			5402039 11
			5402194 10
			5403715 11
			5410901 7
			5415659 10
			5416657 8
			5419070 6
			5426508 9
			5435628 12
			5438496 9
			5439733 11
			5445112 6
			5452218 13
			5458813 9
			5463401 9
			5477537 13
			5477959 8
			5478437 11
			5478527 10
			5480345 13
			5482410 9
			5483525 8
			5484285 10
			5491127 11
			5504834 9
			5512081 8
			5512841 9
			5514446 9
			5515059 10
			5524427 11
			5554118 9
			5567810 7
			5573661 8
			5587875 10
			5589681 8
			5590777 9
			5596589 8
			5618376 1
			5621816 0
			5626426 1
			5629922 2
			5641702 0
			5648813 1
			5650530 2
			5653709 2
			5653965 2
			5655383 3
			5671297 0
			5685556 3
			5690805 0
			5694683 0
			5713632 7
			5713981 7
			5727517 6
			5728032 7
			5729226 8
			5729774 7
			5732000 9
			5736689 8
			5740434 6
			5742184 7
			5742520 5
			5749822 9
			5751647 9
			5752472 9
			5754785 6
			5756130 7
			5775501 7
			5777676 6
			5777742 8
			5779609 8
			5793596 7
			5794602 7
			5796032 10
			5798348 7
			5800400 6
			5806887 4
			5817485 1
			5817975 3
			5818235 5
			5818835 3
			5845130 1
			5847533 3
			5851109 6
			5852898 3
			5855584 5
			5862571 4
			5864268 2
			5868930 1
			5869125 3
			5873236 5
			5874472 6
			5876426 2
			5884644 5
			5886987 4
			5887342 4
			5887499 2
			5887769 4
			5887896 2
			5889428 7
			5889485 2
			5890782 4
			5893716 4
			5894845 4
			5896087 5
			5898040 4
			5900737 12
			5908981 9
			5912943 10
			5914448 13
			5915252 9
			5915583 9
			5919619 13
			5920351 9
			5922040 7
			5925034 12
			5926692 8
			5928611 12
			5931253 12
			5937670 10
			5939518 8
			5940390 12
			5941218 7
			5948710 9
			5952295 11
			5953173 11
			5962266 9
			5966907 12
			5971117 12
			5975955 8
			5978784 11
			5980001 13
			5981653 10
			5986357 10
			5986720 7
			5989679 9
			5999088 9
			6001741 7
			6005207 9
			6013535 8
			6014669 9
			6019706 8
			6022597 8
			6022620 10
			6024581 9
			6029853 10
			6031977 9
			6032596 12
			6041042 11
			6041105 11
			6046142 9
			6047460 10
			6048653 9
			6052307 8
			6052602 5
			6054793 11
			6058279 7
			6072327 8
			6075563 9
			6076615 7
			6086621 7
			6091059 9
			6097606 7
			6099343 9
			6114434 9
			6114726 8
			6116600 7
			6119992 7
			6122731 6
			6126167 7
			6127020 5
			6134414 6
			6137511 8
			6139981 7
			6141522 6
			6146999 4
			6157608 7
			6163929 7
			6169173 4
			6170654 8
			6171526 6
			6189972 6
			6197186 9
			6200076 12
			6200903 12
			6203989 14
			6207283 14
			6207445 14
			6208434 20
			6212712 20
			6215375 9
			6225021 11
			6227243 14
			6231703 12
			6247409 12
			6253919 18
			6255406 15
			6256755 11
			6266112 11
			6267464 18
			6268429 7
			6268504 8
			6271959 18
			6274218 11
			6279371 14
			6279481 14
			6279711 8
			6289017 9
			6289058 14
			6294111 16
			6294665 6
			6296221 13
			6300585 2
			6306398 1
			6336382 3
			6345044 3
			6351750 1
			6352752 3
			6356612 2
			6359321 0
			6367878 2
			6371840 2
			6374622 3
			6381430 2
			6382976 2
			6385108 2
			6385702 2
			6393204 1
			6401529 2
			6402875 3
			6408806 3
			6409036 3
			6410243 2
			6413595 2
			6417677 3
			6422727 3
			6425781 3
			6428926 3
			6432671 3
			6432890 3
			6433555 2
			6433657 3
			6446741 3
			6448827 3
			6450794 3
			6457653 2
			6473415 3
			6476354 1
			6476879 3
			6480797 3
			6483596 2
			6486001 3
			6490768 3
			6503936 9
			6505767 11
			6508541 12
			6514420 9
			6516652 10
			6523972 10
			6525669 9
			6526009 12
			6530423 11
			6531921 11
			6532617 11
			6540822 15
			6543763 12
			6547345 9
			6551283 11
			6553811 15
			6554464 11
			6564304 12
			6565535 13
			6565794 9
			6570909 7
			6574790 7
			6579562 11
			6597071 10
			6599007 11
			6604212 10
			6605093 10
			6605185 10
			6615863 11
			6617835 12
			6618357 11
			6620345 9
			6622771 11
			6627683 12
			6632841 8
			6633995 14
			6634746 11
			6635290 10
			6635642 13
			6640270 9
			6643997 11
			6644524 12
			6647900 10
			6661404 11
			6667488 14
			6672752 11
			6675278 11
			6676027 12
			6676686 8
			6677577 8
			6685585 11
			6685662 9
			6686186 14
			6687051 11
			6692039 10
			6694669 10
			6695731 11
			6701731 4
			6705659 3
			6706657 4
			6707172 0
			6714478 2
			6716508 2
			6725628 5
			6725761 1
			6726170 2
			6728496 3
			6729733 5
			6730616 4
			6745322 3
			6755964 2
			6761278 3
			6767394 1
			6767537 3
			6768527 3
			6772118 4
			6773525 5
			6788044 4
			6789367 3
			6802081 2
			6807955 3
			6816043 2
			6818398 0
			6818850 2
			6820264 2
			6833637 1
			6837202 1
			6839113 3
			6841306 2
			6842986 3
			6843212 2
			6844564 3
			6859192 2
			6860387 2
			6863661 1
			6865788 3
			6868411 1
			6886589 2
			6888333 2
			6895508 2
			6899323 1
			6901809 9
			6902825 5
			6912789 11
			6914105 12
			6914177 14
			6919922 8
			6930914 7
			6931702 11
			6934082 9
			6939809 14
			6943965 11
			6945641 12
			6946203 10
			6952219 10
			6961297 10
			6963413 10
			6972146 10
			6977639 9
			6980805 8
			6982496 14
			6984683 12
			6985395 11
			6994609 10
			7003981 2
			7004091 3
			7017506 2
			7019774 2
			7022093 3
			7030880 4
			7036560 1
			7043922 3
			7046130 3
			7048896 2
			7052657 2
			7064416 3
			7065367 2
			7068845 2
			7076352 2
			7078984 4
			7086032 1
			7089292 4
			7103190 10
			7115472 12
			7118188 11
			7119778 12
			7127240 15
			7130038 11
			7145828 9
			7146492 6
			7146760 11
			7154268 8
			7156675 11
			7158930 8
			7166426 9
			7167097 9
			7176987 7
			7179485 11
			7180782 10
			7188040 9
			7201839 2
			7202943 5
			7209619 3
			7210351 5
			7215338 3
			7218427 3
			7218611 1
			7222392 4
			7229416 5
			7230390 6
			7242295 5
			7244524 2
			7250965 4
			7252266 6
			7257985 3
			7266189 1
			7270001 4
			7270422 7
			7270615 2
			7271904 3
			7276401 4
			7276720 2
			7288844 1
			7290542 4
			7292559 2
			7296915 2
			7299317 5
			7302906 8
			7303535 9
			7303588 10
			7304642 13
			7309706 11
			7312192 14
			7312597 10
			7313002 13
			7317557 14
			7319149 13
			7322596 15
			7331105 18
			7334917 11
			7336017 9
			7342602 10
			7342710 19
			7347193 12
			7350287 12
			7356409 8
			7361254 15
			7361969 9
			7363311 6
			7366676 11
			7376621 18
			7383897 16
			7387241 10
			7393440 11
			7399392 12
			7403090 16
			7404362 15
			7404726 11
			7405270 9
			7416401 9
			7419264 14
			7422692 11
			7427511 8
			7444188 13
			7446409 15
			7447985 12
			7452437 12
			7461526 12
			7463418 17
			7475403 11
			7483445 18
			7487186 13
			7502895 4
			7504070 6
			7509932 7
			7516773 3
			7520900 7
			7522547 5
			7523405 5
			7524760 8
			7539273 9
			7542648 5
			7555243 6
			7565480 8
			7569178 5
			7571569 6
			7571592 4
			7574591 3
			7581075 6
			7582715 7
			7584683 5
			7584930 9
			7587327 6
			7587626 7
			7591087 7
			7594367 6
			7599711 9
			7602754 3
			7603300 3
			7608420 2
			7610378 3
			7611784 3
			7612465 2
			7614607 3
			7615080 3
			7615227 3
			7617079 3
			7618402 2
			7618454 3
			7622828 3
			7624577 3
			7626594 3
			7630218 3
			7635342 3
			7643264 3
			7647132 3
			7647525 3
			7666108 3
			7669468 3
			7677584 2
			7680757 3
			7685714 3
			7689173 3
			7690312 3
			7698675 3
			7702883 10
			7703543 11
			7708607 13
			7710215 14
			7710951 15
			7712240 11
			7715489 11
			7718659 16
			7719626 9
			7723227 15
			7724385 14
			7726098 15
			7726719 14
			7728715 14
			7728818 14
			7730366 13
			7747671 13
			7751923 11
			7759091 12
			7767572 10
			7776546 15
			7784130 14
			7786709 11
			7787874 14
			7800544 3
			7810197 2
			7812635 1
			7814698 0
			7818710 2
			7819803 0
			7822004 1
			7823666 1
			7829548 0
			7841090 1
			7857332 0
			7858686 1
			7861387 0
			7864725 1
			7876855 1
			7878692 2
			7879270 1
			7883451 1
			7883698 0
			7890700 2
			7903577 3
			7909492 1
			7911762 4
			7912358 6
			7914586 2
			7918720 7
			7921682 4
			7921751 3
			7924275 4
			7924664 5
			7930979 4
			7937189 5
			7944341 2
			7945830 7
			7947607 3
			7947650 2
			7951070 7
			7952210 3
			7952511 7
			7954030 1
			7959313 7
			7960568 3
			7964105 3
			7969525 4
			7971640 6
			7976455 4
			7976621 4
			7977064 4
			7986286 6
			7987132 5
			7989979 8
			7992771 4
			7996338 6
			8004316 12
			8007409 10
			8007636 14
			8013726 15
			8014445 14
			8014644 12
			8014987 11
			8017441 10
			8030767 9
			8031188 15
			8033080 15
			8046329 14
			8047096 17
			8071186 11
			8078742 9
			8078939 16
			8098487 14
			8100590 9
			8104459 15
			8104471 14
			8113490 10
			8114755 12
			8117171 9
			8123070 8
			8132743 9
			8134009 5
			8140422 11
			8153613 10
			8163870 9
			8164686 5
			8165501 6
			8165806 13
			8170376 12
			8180260 10
			8186706 8
			8187656 9
			8191620 8
			8204376 3
			8212542 3
			8215597 6
			8226224 6
			8227004 5
			8231617 4
			8239356 5
			8251468 3
			8255147 8
			8265501 0
			8275369 5
			8279585 5
			8280443 7
			8282489 7
			8291184 7
			8294398 3
			8302138 7
			8308627 6
			8311405 8
			8316040 6
			8320987 7
			8332990 7
			8341485 8
			8351662 9
			8353079 5
			8370939 10
			8374867 4
			8380510 8
			8380992 11
			8385710 6
			8387841 9
			8390839 7
			8391117 8
			8395388 3
			8402595 11
			8404091 5
			8408450 5
			8412572 9
			8426738 9
			8429130 8
			8429766 10
			8444360 5
			8457635 5
			8461582 7
			8463710 8
			8464803 7
			8470401 8
			8474548 8
			8487088 8
			8491482 4
			8491745 3
			8494044 5
			8495371 9
			8506387 15
			8525257 11
			8525562 13
			8532126 17
			8541055 18
			8550155 9
			8555735 14
			8569664 16
			8580282 10
			8582100 14
			8590830 10
			8592607 16
			8593016 12
			8596070 12
			8598029 10
			8609105 0
			8621454 2
			8625846 1
			8626016 1
			8627948 1
			8632132 2
			8637637 2
			8643669 2
			8649316 2
			8651748 2
			8653598 2
			8655290 1
			8655988 1
			8658726 1
			8662212 1
			8669349 2
			8675767 1
			8676188 2
			8678080 1
			8679811 1
			8685413 2
			8691626 2
			8692096 2
			8705868 9
			8712007 6
			8713537 7
			8714311 15
			8728092 11
			8728792 4
			8729244 12
			8736028 15
			8739681 5
			8742832 7
			8745590 8
			8749667 11
			8753754 13
			8755635 4
			8762171 11
			8762663 8
			8763208 10
			8764177 10
			8766271 10
			8769263 6
			8779009 6
			8784620 6
			8787187 12
			8794145 12
			8794172 14
			8795334 8
			8801458 6
			8802511 6
			8807491 6
			8813446 3
			8819582 5
			8819652 5
			8821996 6
			8822230 6
			8825260 5
			8826095 5
			8836620 5
			8841383 6
			8844085 5
			8853244 4
			8853823 6
			8854750 5
			8859745 5
			8860054 4
			8860597 5
			8867405 6
			8871109 5
			8874396 5
			8880897 6
			8888008 4
			8889054 5
			8898503 5
			8906995 11
			8917089 12
			8919945 11
			8924585 9
			8925073 13
			8925443 12
			8926340 11
			8936942 7
			8939398 12
			8950475 9
			8952774 11
			8961031 10
			8964787 11
			8977990 8
			8980798 12
			8986485 11
			8987721 7
			8990507 13
			9003219 4
			9006646 4
			9027805 3
			9034725 4
			9035818 3
			9040485 5
			9041438 4
			9041993 7
			9044962 5
			9045878 3
			9052150 4
			9057231 4
			9058797 5
			9060702 4
			9085734 7
			9089631 6
			9095384 4
			9106874 11
			9107130 13
			9110285 7
			9116787 12
			9118175 9
			9118294 10
			9124159 14
			9134349 11
			9135336 13
			9139705 13
			9143055 10
			9145980 8
			9151183 10
			9153060 11
			9158663 10
			9161346 12
			9163077 8
			9163326 14
			9170598 8
			9173780 11
			9176872 13
			9176893 12
			9180797 6
			9189426 13
			9189482 9
			9189713 9
			9194399 13
			9198486 15
			9203581 6
			9205644 6
			9214890 7
			9218384 7
			9223414 6
			9223639 6
			9223777 8
			9230615 4
			9243146 6
			9246907 8
			9250639 7
			9255395 7
			9259867 8
			9260158 7
			9263207 9
			9271162 6
			9273377 10
			9278879 7
			9279918 9
			9282639 5
			9283237 8
			9283563 6
			9285519 7
			9290116 6
			9292280 7
			9293345 8
			9295022 9
			9309294 13
			9309436 15
			9314138 14
			9315047 12
			9315940 11
			9317881 12
			9319320 13
			9325322 14
			9344681 12
			9346097 13
			9346766 13
			9347220 12
			9351578 12
			9351582 10
			9352263 11
			9359046 12
			9361979 13
			9366525 13
			9368865 12
			9370938 15
			9375002 14
			9381986 11
			9388767 10
			9390544 11
			9394492 14
			9403321 10
			9403642 15
			9404215 14
			9408754 16
			9418702 15
			9433845 8
			9437058 18
			9442565 13
			9445406 13
			9451937 9
			9456340 14
			9473476 16
			9478972 15
			9479244 14
			9479442 15
			9490316 16
			9490839 14
			9492933 15
			9497318 18
			9506134 4
			9506757 3
			9519283 4
			9524900 4
			9528264 3
			9538370 5
			9540135 4
			9546259 5
			9550542 4
			9560879 4
			9566038 5
			9567977 4
			9572151 4
			9576632 4
			9576828 4
			9576938 5
			9578250 4
			9597228 4
			9608431 12
			9609291 8
			9609431 6
			9620917 10
			9625433 7
			9629737 9
			9640016 7
			9641152 10
			9663664 9
			9664540 7
			9667862 7
			9668787 11
			9668876 6
			9670212 11
			9674651 8
			9676498 9
			9677495 5
			9694160 8
			9694562 9
			9699170 8
			9706167 9
			9711665 9
			9711791 10
			9715413 10
			9715698 10
			9719172 9
			9719273 8
			9727079 9
			9727478 8
			9733688 10
			9743829 10
			9744919 9
			9746833 9
			9756018 10
			9758133 8
			9759779 8
			9759910 8
			9761355 9
			9776275 8
			9779999 10
			9783009 9
			9786450 9
			9791520 9
			9798149 7
			9798918 10
			9801387 2
			9802402 3
			9805925 3
			9812131 2
			9812957 2
			9815100 4
			9818161 4
			9819142 2
			9819211 4
			9823064 5
			9827942 3
			9835297 7
			9839359 3
			9843040 2
			9843768 3
			9844734 5
			9845186 3
			9865765 4
			9874942 4
			9878499 3
			9879652 3
			9880438 2
			9881488 5
			9885199 4
			9891541 5
			9898950 5
			9902211 4
			9902497 2
			9902974 4
			9908353 1
			9919773 3
			9923564 3
			9923922 6
			9925982 4
			9927493 3
			9930831 4
			9948054 3
			9959395 2
			9960987 4
			9964123 1
			9966426 3
			9977100 6
			9981309 3
			9983965 5
			9986388 5
			9990143 3
			9994702 0
			9997230 1
			10004142 10
			10006471 9
			10031274 11
			10038891 6
			10040520 12
			10044969 9
			10046570 12
			10048353 7
			10054849 11
			10055483 9
			10068666 10
			10069531 15
			10073229 11
			10080911 11
			10081579 9
			10088263 12
			10094319 7
			10098241 6
			10099720 5
			10115934 13
			10119024 9
			10124346 9
			10130612 7
			10132499 15
			10134689 7
			10152010 12
			10154530 11
			10158126 10
			10161462 13
			10163248 6
			10165842 12
			10166883 10
			10183805 13
			10185204 10
			10186818 10
			10203354 11
			10204970 7
			10215341 10
			10222477 8
			10241068 7
			10241832 7
			10245903 8
			10247038 7
			10248127 11
			10254406 9
			10255688 5
			10257607 6
			10261344 9
			10263006 6
			10278360 9
			10281007 7
			10284067 7
			10288794 8
			10291603 5
			10296396 8
			10302079 4
			10330337 8
			10331651 7
			10333883 7
			10334318 5
			10334539 8
			10335988 6
			10339869 5
			10346626 6
			10347370 5
			10359544 6
			10366500 6
			10367885 7
			10369799 7
			10371538 8
			10373588 4
			10375228 8
			10376268 7
			10376877 7
			10386202 4
			10391547 8
			10392299 9
			10409371 1
			10412057 2
			10414137 2
			10426461 3
			10426691 2
			10432994 1
			10439076 2
			10446374 1
			10447301 3
			10457627 3
			10458340 2
			10458506 1
			10461777 2
			10464404 1
			10467849 2
			10481332 1
			10486003 1
			10487188 3
			10488804 0
			10492843 0
			10495151 1
			10495444 3
			10500475 11
			10505405 11
			10508566 10
			10513599 10
			10515371 13
			10519556 10
			10526836 12
			10537828 9
			10559562 13
			10560518 14
			10561190 10
			10571521 15
			10571738 13
			10573831 8
			10574795 11
			10575816 10
			10578542 13
			10583704 7
			10594241 10
			10594580 10
			10600800 9
			10601093 7
			10602425 5
			10603769 7
			10604342 10
			10610011 4
			10611555 3
			10614997 6
			10622393 10
			10623392 7
			10623410 8
			10628053 4
			10628847 9
			10632627 8
			10639486 4
			10647487 8
			10651184 4
			10659046 9
			10659754 5
			10660941 3
			10661098 6
			10663881 10
			10671698 6
			10673313 7
			10686923 8
			10687194 7
			10698755 6
			10709718 7
			10709841 9
			10711338 3
			10713335 9
			10713400 7
			10714653 9
			10722431 9
			10726724 8
			10733085 7
			10733207 10
			10747270 6
			10751007 7
			10755543 9
			10758140 6
			10760175 9
			10761363 9
			10777819 8
			10781518 7
			10788396 8
			10792842 9
			10794208 6
			10794955 7
			10799533 7
			10800839 10
			10803102 8
			10806425 12
			10808672 9
			10810399 8
			10811442 12
			10811974 9
			10824249 13
			10829142 11
			10833541 5
			10836781 8
			10839447 11
			10839504 9
			10847284 7
			10847303 11
			10860044 9
			10860811 9
			10862377 8
			10864065 12
			10869396 11
			10882606 9
			10900303 6
			10906891 7
			10908868 5
			10909404 7
			10913464 3
			10916148 7
			10917255 7
			10932582 3
			10933962 2
			10952021 6
			10962260 5
			10962980 4
			10967026 3
			10970818 7
			10976438 7
			10984754 8
			10984977 3
			10989009 5
			10993515 6
			11000992 7
			11012251 6
			11017326 9
			11019735 6
			11024631 7
			11025229 5
			11026699 8
			11040197 8
			11043680 7
			11043865 4
			11048243 4
			11070336 8
			11080091 6
			11080602 8
			11082264 8
			11086183 8
			11091285 6
			11093663 9
			11097289 6
			11098326 7
			11106174 1
			11108780 5
			11115134 4
			11116579 6
			11131580 5
			11133486 6
			11151481 3
			11158025 4
			11166579 8
			11177057 5
			11180360 5
			11181907 4
			11183470 5
			11195203 9
			11196067 8
			11206162 2
			11210561 2
			11213879 3
			11214918 3
			11228345 0
			11231553 3
			11234059 1
			11243201 1
			11244294 3
			11256266 2
			11259210 3
			11260322 2
			11278629 2
			11281766 2
			11288343 3
			11296979 3
			11300265 8
			11314593 5
			11322006 5
			11329492 4
			11339715 10
			11344815 8
			11351715 8
			11353702 6
			11353928 4
			11363814 9
			11365430 7
			11367283 7
			11368328 10
			11374547 3
			11375137 10
			11377565 4
			11380820 12
			11385054 3
			11387561 4
			11390095 5
			11396004 7
			11405287 7
			11410937 11
			11421622 9
			11426588 10
			11431701 11
			11432318 11
			11434635 8
			11436443 10
			11438327 9
			11439072 10
			11448319 11
			11452844 9
			11452969 10
			11453040 10
			11465828 10
			11465930 9
			11472596 9
			11475135 7
			11475367 9
			11481259 12
			11495366 10
			11495589 12
			11496804 9
			11502977 9
			11511632 8
			11519244 8
			11528042 8
			11530550 7
			11532228 8
			11538683 7
			11542414 11
			11543431 6
			11565949 7
			11566532 8
			11570974 8
			11576152 8
			11578463 11
			11598664 9
			11598853 9
			11601580 13
			11602934 10
			11603787 13
			11606050 13
			11606779 15
			11613169 9
			11628694 12
			11629501 12
			11631025 16
			11632469 8
			11639434 13
			11641865 13
			11644455 10
			11645606 10
			11651134 12
			11653665 12
			11666871 9
			11673961 13
			11676401 15
			11682920 9
			11684820 11
			11689627 14
			11691111 12
			11697022 16
			11701490 2
			11701499 1
			11701685 1
			11706031 2
			11706532 1
			11711307 1
			11711760 0
			11715303 0
			11722098 0
			11728113 0
			11728705 4
			11734222 1
			11734653 0
			11734723 2
			11739022 2
			11739724 2
			11740718 1
			11750805 1
			11755374 0
			11757270 2
			11757384 0
			11759501 0
			11759764 2
			11760939 1
			11773633 1
			11781381 0
			11796575 0
			11800211 5
			11800732 8
			11809606 9
			11813701 11
			11816722 4
			11822942 7
			11823173 6
			11824141 12
			11828491 6
			11831232 7
			11832683 11
			11839040 6
			11839100 9
			11843513 6
			11859736 7
			11863894 11
			11867382 4
			11878297 7
			11881137 6
			11881747 8
			11883312 9
			11884163 9
			11898173 12
			11899412 9
			11905204 9
			11907432 6
			11908035 7
			11913548 7
			11915926 5
			11919169 6
			11919216 7
			11919338 6
			11919448 6
			11919474 7
			11921444 7
			11922149 8
			11922315 7
			11929694 5
			11930227 7
			11937504 5
			11943028 3
			11949258 7
			11953612 8
			11955337 5
			11955771 9
			11960416 4
			11961281 5
			11972032 8
			11987647 7
			11988385 8
			11995980 7
			12007984 8
			12020812 9
			12026456 14
			12028034 13
			12030255 12
			12044869 8
			12050857 14
			12057183 9
			12062335 10
			12067126 11
			12071976 10
			12072163 11
			12075980 7
			12094046 12
			12098456 10
			12098593 9
			12099616 12
			12100636 10
			12102149 11
			12102283 10
			12103683 9
			12104591 11
			12110530 7
			12119909 9
			12121065 10
			12124631 12
			12148986 7
			12155702 5
			12173837 11
			12175850 8
			12176571 4
			12177453 9
			12182649 10
			12186996 10
			12189053 11
			12191856 7
			12195074 8
			12196136 5
			12196629 5
			12197790 8
			12198842 6
			12205120 1
			12210503 3
			12215838 5
			12218250 1
			12222089 1
			12225177 3
			12231317 5
			12233503 4
			12238532 3
			12240062 4
			12247986 3
			12249735 3
			12253476 4
			12253875 5
			12254482 3
			12262776 3
			12264625 3
			12266766 3
			12276919 3
			12277338 1
			12283178 2
			12284877 3
			12289295 3
			12291054 2
			12292326 2
			12298531 4
			12299779 5
			12304257 10
			12307402 6
			12310408 9
			12314075 8
			12314730 9
			12316398 7
			12317226 12
			12322932 9
			12322980 7
			12329794 7
			12339652 8
			12340663 13
			12344148 8
			12347005 10
			12353880 9
			12359336 8
			12361422 7
			12361809 8
			12364344 8
			12365166 9
			12367324 7
			12369610 11
			12377201 9
			12378103 7
			12388681 6
			12391744 7
			12395086 7
			12395263 9
			12397252 7
			12398155 11
			12400144 8
			12404330 10
			12406730 9
			12414894 8
			12415175 12
			12422001 5
			12424755 9
			12425908 6
			12426602 7
			12430665 6
			12431714 7
			12435918 7
			12447284 15
			12447491 10
			12452837 9
			12468107 9
			12468529 9
			12475236 13
			12476452 9
			12479507 8
			12480364 14
			12485744 5
			12496390 7
			12496502 10
			12498910 9
			12499156 11
			12499647 11
			12519740 7
			12521948 8
			12543553 8
			12545514 7
			12545614 4
			12553741 5
			12558348 5
			12561906 7
			12565672 8
			12567206 7
			12568475 7
			12570601 7
			12573892 7
			12574588 6
			12577743 8
			12579379 7
			12582329 8
			12583172 7
			12584860 7
			12585222 6
			12592178 7
			12595589 8
			12598409 7
			12600438 13
			12606750 8
			12614002 9
			12620325 10
			12620372 8
			12620851 9
			12625304 7
			12625846 10
			12625913 5
			12629339 9
			12630561 11
			12638743 11
			12641572 9
			12641643 8
			12644608 12
			12648645 7
			12659382 7
			12664459 6
			12665841 5
			12669724 6
			12673793 8
			12675489 13
			12676759 10
			12694078 12
			12696532 14
			12697545 8
			12698729 9
			12703646 11
			12705243 14
			12705650 10
			12709025 16
			12729993 14
			12731352 14
			12731747 13
			12736171 15
			12738635 9
			12742312 13
			12748529 10
			12751978 12
			12755875 16
			12758254 11
			12799032 11
			12805592 9
			12809012 9
			12811339 10
			12817173 7
			12821032 6
			12823395 5
			12826687 9
			12826797 7
			12830722 9
			12834225 9
			12846549 8
			12848145 6
			12848238 11
			12853142 9
			12857979 7
			12861433 9
			12866354 9
			12870551 10
			12872817 10
			12873491 9
			12873728 8
			12878603 12
			12878937 12
			12881177 7
			12885951 10
			12888018 7
			12889446 9
			12890027 6
			12894410 9
			12894653 8
			12899453 9
		};

	\end{axis}
	
\end{tikzpicture}

%% file: tikz-figures/averageForwardedExperiences.tex
\begin{tikzpicture}

\definecolor{darkslategray38}{RGB}{38,38,38}
\pgfplotsset{every tick label/.append style={font=\scriptsize}}
\pgfplotsset{scaled x ticks=false}
\pgfkeys{/pgf/number format/.cd,1000 sep={\,}}

\begin{axis}[
width=\fwidth,
height=\fheight,
tick align=outside,
xlabel=\textcolor{darkslategray38}{Time (s)},
xmajorgrids,
xmajorticks=true,
xmin=0, xmax=2000,
xticklabels={0,125,250,375,500},
xtick={0,300,500,1000,1500,2000},
xtick style={color=darkslategray38},
ylabel=\textcolor{darkslategray38}{Learning data (sample/s)},
ymajorgrids,
ymajorticks=true,
ymin=0, ymax=850,
ytick style={color=darkslategray38}
]
\addplot [color0,semithick]
table {%
0 775
10 778
20 736
30 810
40 753
50 664
60 707
70 660
80 810
90 513
100 645
110 432
120 346
130 395
140 337
150 335
160 319
170 318
180 194
190 204
200 168
210 184
220 88
230 82
240 56
250 36
260 46
270 30
280 56
290 31
300 67
310 20
320 41
330 10
340 46
350 67
360 41
370 36
380 72
390 25
400 67
410 57
420 57
430 15
440 61
450 52
460 46
470 51
480 31
490 46
500 92
510 51
520 41
530 57
540 57
550 41
560 57
570 36
580 36
590 46
600 41
610 82
620 31
630 61
640 51
650 26
660 71
670 30
680 26
690 62
700 51
710 57
720 67
730 20
740 62
750 61
760 31
770 77
780 56
790 41
800 47
810 56
820 83
830 46
840 56
850 67
860 82
870 57
880 61
890 57
900 47
910 30
920 41
930 61
940 66
950 26
960 52
970 46
980 47
990 46
1000 51
1010 66
1020 41
1030 51
1040 56
1050 66
1060 56
1070 56
1080 77
1090 56
1100 35
1110 51
1120 51
1130 46
1140 21
1150 35
1160 36
1170 51
1180 30
1190 61
1200 66
1210 52
1220 46
1230 67
1240 36
1250 41
1260 25
1270 61
1280 26
1290 82
1300 41
1310 41
1320 46
1330 57
1340 56
1350 36
1360 72
1370 41
1380 66
1390 61
1400 46
1410 51
1420 31
1430 57
1440 46
1450 62
1460 56
1470 46
1480 41
1490 57
1500 52
1510 41
1520 57
1530 51
1540 51
1550 77
1560 57
1570 42
1580 31
1590 62
1600 52
1610 56
1620 72
1630 61
1640 56
1650 25
1660 41
1670 46
1680 46
1690 87
1700 62
1710 46
1720 57
1730 36
1740 52
1750 61
1760 31
1770 31
1780 46
1790 72
1800 41
1810 77
1820 57
1830 41
1840 31
1850 52
1860 57
1870 46
1880 66
1890 67
1900 51
1910 62
1920 72
1930 56
1940 67
1950 31
1960 77
1970 57
1980 62
1990 57
};
\end{axis}

\end{tikzpicture}

%% file: tikz-figures/overheadPerGreedyAction.tex
\begin{tikzpicture}

\definecolor{darkslategray38}{RGB}{38,38,38}
\pgfplotsset{every tick label/.append style={font=\scriptsize}}

\begin{axis}[
width=\fwidth,
height=\fheight,
tick align=outside,
xlabel=\textcolor{darkslategray38}{Reward loss},
xmajorgrids,
xmajorticks=true,
xmin=0, xmax=1,
xtick style={color=darkslategray38},
ylabel=\textcolor{darkslategray38}{Empirical CDF},
ymajorgrids,
ymajorticks=true,
ymin=0, ymax=1,
ytick style={color=darkslategray38}
]
\addplot [semithick, color1, mark=o,mark repeat=5]
table {%
-0.0571733563488781 0
-0.0518876944760251 0.00144288538966454
-0.046602032603172 0.0044985191209931
-0.041316370730319 0.0104932813528343
-0.0360307088574659 0.0213887255255372
-0.0307450469846129 0.0397338753587905
-0.0254593851117598 0.0683494165182956
-0.0201737232389068 0.109700626084916
-0.0148880613660537 0.165058794858702
-0.00960239949320068 0.233715907535937
-0.00431673762034763 0.312603053593691
0.000968924252505415 0.396579775283692
0.00625458612535846 0.479408150025411
0.0115402479982115 0.555117885707128
0.0168259098710646 0.619280089974833
0.0221115717439176 0.669751448217996
0.0273972336167706 0.706699780197422
0.0326828954896237 0.732028106602767
0.0379685573624767 0.748514185281699
0.0432542192353298 0.759001377105755
0.0485398811081828 0.76585473216322
0.0538255429810359 0.770733859974999
0.0591112048538889 0.774616327848509
0.064396866726742 0.77796108723705
0.069682528599595 0.780912920860594
0.0749681904724481 0.783483321142658
0.0802538523453011 0.785676997010047
0.0855395142181541 0.787557490224697
0.0908251760910072 0.789259989352021
0.0961108379638602 0.790966565612754
0.101396499836713 0.792861058206875
0.106682161709566 0.795080039700368
0.111967823582419 0.797674382846037
0.117253485455272 0.800592976123681
0.122539147328125 0.803694843132581
0.127824809200979 0.806787770031562
0.133110471073832 0.809682516381758
0.138396132946685 0.812245769945883
0.143681794819538 0.814435447281634
0.148967456692391 0.816308629357597
0.154253118565244 0.818001863454445
0.159538780438097 0.819691451088298
0.16482444231095 0.821545359326133
0.170110104183803 0.823678898131558
0.175395766056656 0.826124744681026
0.180681427929509 0.828824908445234
0.185967089802362 0.831647421915957
0.191252751675215 0.834423919305766
0.196538413548068 0.836997571158371
0.201824075420921 0.839267085328363
0.207109737293774 0.841213844883725
0.212395399166627 0.842905249926334
0.21768106103948 0.844475009288227
0.222966722912333 0.846087172155026
0.228252384785186 0.847893569244603
0.233538046658039 0.849994663853567
0.238823708530893 0.852412860506925
0.244109370403746 0.855085343659243
0.249395032276599 0.857879694419262
0.254680694149452 0.860629491308779
0.259966356022305 0.863180409180042
0.265252017895158 0.865433135086866
0.270537679768011 0.867370252847304
0.275823341640864 0.869059913314795
0.281109003513717 0.870636677904951
0.28639466538657 0.872265944092688
0.291680327259423 0.874101254564087
0.296965989132276 0.876244295015378
0.302251651005129 0.878716852186755
0.307537312877982 0.88145250116386
0.312822974750835 0.884312087600395
0.318108636623688 0.887120568621174
0.323394298496541 0.88971529497413
0.328679960369394 0.891991078936469
0.333965622242247 0.893928316544762
0.3392512841151 0.895596672923172
0.344536945987953 0.897134965382149
0.349822607860806 0.898714091375002
0.355108269733659 0.900492668184641
0.360393931606513 0.902575534638955
0.365679593479366 0.904984858015883
0.370965255352219 0.90765209660525
0.376250917225072 0.910434980041461
0.381536579097925 0.913156505380501
0.386822240970778 0.915655020731107
0.392107902843631 0.917829900059272
0.397393564716484 0.919668899718288
0.402679226589337 0.921250147492913
0.40796488846219 0.922720002926915
0.413250550335043 0.924254038073619
0.418536212207896 0.926011069569153
0.423821874080749 0.928090810268054
0.429107535953602 0.930505479592442
0.434393197826455 0.933173943879158
0.439678859699308 0.93594203255132
0.444964521572161 0.938624619341582
0.450250183445014 0.941056922405724
0.455535845317867 0.943138693374678
0.46082150719072 0.944857919213238
0.466107169063574 0.946288675997372
0.471392830936427 0.947566428053935
0.47667849280928 0.948849586175361
0.481964154682133 0.95027773827611
0.487249816554986 0.951936237194189
0.492535478427839 0.953835277518765
0.497821140300692 0.955909043646995
0.503106802173545 0.958035943397097
0.508392464046398 0.960074608550786
0.513678125919251 0.961904699744989
0.518963787792104 0.963459755895067
0.524249449664957 0.964742639839192
0.52953511153781 0.965820740108878
0.534820773410663 0.966804506068546
0.540106435283516 0.967816585504483
0.545392097156369 0.968959645804262
0.550677759029222 0.970290259251935
0.555963420902075 0.971804967874159
0.561249082774928 0.973442433491305
0.566534744647781 0.975101677610895
0.571820406520634 0.976671345554765
0.577106068393487 0.978060874092268
0.582391730266341 0.97922378285752
0.587677392139194 0.980166628183141
0.592963054012047 0.980942635648805
0.5982487158849 0.981633827876428
0.603534377757753 0.982327787561768
0.608820039630606 0.983095176454309
0.614105701503459 0.983972995012853
0.619391363376312 0.984957151030874
0.624677025249165 0.986006061721263
0.629962687122018 0.987054338760993
0.635248348994871 0.988032513604174
0.640534010867724 0.988886634746621
0.645819672740577 0.989591743057248
0.65110533461343 0.990155789116183
0.656390996486283 0.990614141034528
0.661676658359136 0.991017654572665
0.666962320231989 0.991418376103754
0.672247982104842 0.991856588267383
0.677533643977695 0.992351940016883
0.682819305850548 0.992900359618469
0.688104967723401 0.993477261596516
0.693390629596255 0.994046033297407
0.698676291469108 0.994569198769335
0.703961953341961 0.995018761929256
0.709247615214814 0.995382654969247
0.714533277087667 0.99566585158058
0.71981893896052 0.995886700618985
0.725104600833373 0.996070422948443
0.730390262706226 0.996242053118394
0.735675924579079 0.996420639575633
0.740961586451932 0.996615778091671
0.746247248324785 0.996826893612378
0.751532910197638 0.997045125926142
0.756818572070491 0.997257122595148
0.762104233943344 0.997449560685321
0.767389895816197 0.997613038931736
0.77267555768905 0.997744278343392
0.777961219561903 0.997846247948355
0.783246881434756 0.997926561305051
0.788532543307609 0.997994937513687
0.793818205180462 0.998060577405941
0.799103867053315 0.998130096927859
0.804389528926168 0.998206383149806
0.809675190799022 0.99828849935439
0.814960852671875 0.998372553624744
0.820246514544728 0.998453231042778
0.825532176417581 0.998525514266421
0.830817838290434 0.99858607636764
0.836103500163287 0.99863397716787
0.84138916203614 0.998670572886934
0.846674823908993 0.998698820685732
0.851960485781846 0.998722304440732
0.857246147654699 0.998744299410286
0.862531809527552 0.998767091610822
0.867817471400405 0.998791651863005
0.873103133273258 0.998817673896528
0.878388795146111 0.998843916088112
0.883674457018964 0.998868725620984
0.888960118891817 0.99889058276384
0.89424578076467 0.998908507573744
0.899531442637523 0.99892223222902
0.904817104510376 0.998932135702063
0.910102766383229 0.998939018402106
0.915388428256082 0.998943827658952
0.920674090128936 0.998947426989389
0.925959752001788 0.998950456671958
0.931245413874642 0.998953289426072
0.936531075747495 0.998956059064983
0.941816737620348 0.998958731720134
0.947102399493201 0.998961190874585
0.952388061366054 0.998963312967752
0.957673723238907 0.998965018080067
0.96295938511176 0.998966289964085
0.968245046984613 0.998967169620777
0.973530708857466 0.998967733429822
0.978816370730319 0.998968068252627
0.984102032603172 0.998968252467877
0.989387694476025 0.998968346364127
0.994673356348878 0.998968390702125
};
\end{axis}

\end{tikzpicture}

%% file: extension-tikz-figures/aggregated-perf-20.tex
\begin{tikzpicture}

\definecolor{cornflowerblue121150222}{RGB}{121,150,222}
\definecolor{darkseagreen12518595}{RGB}{125,185,95}
\definecolor{darkslategray38}{RGB}{38,38,38}
\definecolor{dimgray84}{RGB}{84,84,84}
\definecolor{forestgreen4416044}{RGB}{44,160,44}
\definecolor{lightgray204}{RGB}{204,204,204}
\definecolor{peru21513172}{RGB}{215,131,72}
	\pgfplotsset{every tick label/.append style={font=\scriptsize}}

\begin{axis}[
width=\fwidth,
height=\fheight,
axis line style={lightgray204},
legend style={fill opacity=0.8, draw opacity=1, text opacity=1, draw=lightgray204},
tick align=outside,
x grid style={lightgray204},
xlabel=\textcolor{darkslategray38}{Initialization policy},
xmajorticks=true,
xmin=-0.5, xmax=2.5,
xtick style={color=darkslategray38},
xtick={0,1,2},
xticklabels={Nearest,Previous,Random},
y grid style={lightgray204},
ylabel=\textcolor{darkslategray38}{Reward},
xmajorgrids,
ymajorgrids,
ymajorticks=true,
ymin=0, ymax=1,
ytick style={color=darkslategray38}
]
\path [draw=dimgray84, fill=color0, line width=0.48pt]
(axis cs:-0.15,0.987105654761905)
--(axis cs:0.15,0.987105654761905)
--(axis cs:0.15,0.998150974025974)
--(axis cs:-0.15,0.998150974025974)
--(axis cs:-0.15,0.987105654761905)
--cycle;
\path [draw=dimgray84, fill=color1, line width=0.48pt]
(axis cs:0.85,0.888696428571428)
--(axis cs:1.15,0.888696428571428)
--(axis cs:1.15,0.99953125)
--(axis cs:0.85,0.99953125)
--(axis cs:0.85,0.888696428571428)
--cycle;
\path [draw=dimgray84, fill=color2, line width=0.48pt]
(axis cs:1.85,0.815961180124222)
--(axis cs:2.15,0.815961180124222)
--(axis cs:2.15,1)
--(axis cs:1.85,1)
--(axis cs:1.85,0.815961180124222)
--cycle;
\addplot [line width=0.48pt, dimgray84, forget plot]
table {%
0 0.987105654761905
0 0.97625
};
\addplot [line width=0.48pt, dimgray84, forget plot]
table {%
0 0.998150974025974
0 1
};
\addplot [line width=0.48pt, dimgray84, forget plot]
table {%
-0.075 0.97625
0.075 0.97625
};
\addplot [line width=0.48pt, dimgray84, forget plot]
table {%
-0.075 1
0.075 1
};
\addplot [color0, mark=diamond*, mark size=1, mark options={solid,fill=dimgray84}, only marks, forget plot]
table {%
0 0.958781263569258
0 0.905568714719931
0 0.932669018671298
0 0.910664350846722
0 0.962593139383413
0 0.920891445940078
0 0.975676617455493
0 0.922380699088146
0 0.893102692140686
0 0.915214502822406
0 0.898081089882762
0 0.924384824142423
0 0.946352366478506
0 0.968760855405992
0 0.903006404689535
0 0.893126139817629
0 0.926628636561008
0 0.957723838471559
0 0.896811007381676
0 0.894420646982197
0 0.897550586191924
0 0.950744246634824
0 0.926174337820235
0 0.896543313069909
0 0.96191608771168
0 0.949000976986539
0 0.908510095527573
0 0.891376682587929
0 0.888302105948762
0 0.893178897090751
0 0.887912288319583
0 0.898128962223187
0 0.9085563395571
0 0.896612027789839
0 0.895207121146331
0 0.899403603994789
0 0.8925552540165
0 0.966340859748155
0 0.975878202344768
0 0.951457121146331
0 0.964983499782892
0 0.939363981762918
0 0.934197025618758
0 0.943150781589231
0 0.888573056882327
0 0.932513026487191
0 0.901269322622666
0 0.914098458532349
0 0.891837494572297
0 0.974596396005211
0 0.941672600955276
0 0.889450716456795
0 0.926216673903604
0 0.961161202778984
0 0.901601823708207
0 0.89105623100304
0 0.908075010855406
0 0.940641228831958
0 0.915251302648719
0 0.89384683022145
0 0.914140468953539
0 0.896923035171516
0 0.894304385584021
0 0.927864524533218
0 0.912754125054277
0 0.932804168475901
0 0.889549392097264
0 0.903164025184542
0 0.944053191489362
0 0.905125814155449
0 0.962396765089014
0 0.951030829353018
0 0.894290056448111
0 0.895049826313504
0 0.940084415584416
0 0.931714285714286
0 0.934681818181818
0 0.965337662337662
0 0.964324675324676
0 0.770535714285714
0 0.910014705882353
0 0.886659963985594
0 0.786471488595438
0 0.742334033613446
0 0.915714285714286
0 0.856071428571429
0 0.84086224489796
0 0.771509303721488
0 0.865136854741897
0 0.806607142857143
0 0.915714285714286
0 0.911607142857143
0 0.887029411764706
0 0.806935474189676
0 0.910663265306122
0 0.957722388955582
0 0.748516806722689
0 0.860899459783913
0 0.926607142857143
0 0.923295918367347
0 0.838474789915966
0 0.868543817527011
0 0.79875
0 0.848214285714286
0 0.917321428571429
0 0.883113445378151
0 0.839627551020408
0 0.886043517406963
0 0.758341536614646
0 0.772647058823529
0 0.754666566626651
0 0.757759903961584
0 0.73523019207683
0 0.733257803121249
0 0.729482893157263
0 0.722141956782713
0 0.722885954381753
0 0.730583433373349
0 0.732732292917167
0 0.723485294117647
0 0.72801830732293
0 0.725188775510204
0 0.728478391356543
0 0.730165966386555
0 0.728353541416567
0 0.727015306122449
0 0.726757803121249
0 0.727474489795919
0 0.728620348139256
0 0.735111044417768
0 0.728668967587035
0 0.731171968787515
0 0.731787515006002
0 0.72986524609844
0 0.725955582232893
0 0.733312725090036
0 0.729153961584634
0 0.728701380552221
0 0.730254201680673
0 0.732171068427372
0 0.727683073229292
0 0.731254801920769
0 0.729127551020408
0 0.727282112845138
0 0.729171968787515
0 0.733421968787515
0 0.729945378151261
0 0.734265006002401
0 0.727947779111645
0 0.725408163265306
0 0.72381362545018
0 0.729210084033614
0 0.730950780312125
0 0.721411164465786
0 0.726515606242497
0 0.72818487394958
0 0.733677971188476
0 0.727716986794718
0 0.731671668667467
0 0.728292016806723
0 0.729988295318127
0 0.729885954381753
0 0.729804021608643
0 0.730286914765907
0 0.731978091236495
0 0.729721488595438
0 0.726463085234094
0 0.728834333733494
0 0.726488295318127
0 0.734764105642257
0 0.72072268907563
0 0.734695678271309
0 0.724488595438175
0 0.739960984393758
0 0.725931572629052
0 0.729870648259304
0 0.727572328931572
0 0.731738895558224
0 0.728080432172869
0 0.731211584633854
0 0.727950480192077
0 0.726481692677071
0 0.736318427370949
0 0.724324429771909
0 0.728920168067227
0 0.731461584633853
0 0.724405162064826
0 0.725748199279712
0 0.727358643457383
0 0.72865156062425
0 0.735787515006003
0 0.729111344537815
0 0.737165666266507
0 0.732894957983194
0 0.732976290516207
0 0.730289615846339
0 0.724194777911164
0 0.73089555822329
0 0.724798619447779
0 0.72880762304922
0 0.733154561824731
0 0.729503601440576
0 0.728060324129652
0 0.728229291716687
0 0.726856242496999
0 0.731442076830732
0 0.732894657863145
0 0.729922569027612
0 0.734248199279712
0 0.727406062424971
0 0.727997599039616
0 0.726322328931573
0 0.725791416566627
0 0.722743397358944
0 0.734889255702281
0 0.725397058823529
0 0.724906662665066
0 0.7309849939976
0 0.72803031212485
0 0.729968487394959
0 0.729144357743097
0 0.724038115246098
0 0.729295018007203
0 0.730880552220888
0 0.732785714285715
0 0.731938175270109
0 0.728203781512605
0 0.734405762304922
0 0.728561224489796
0 0.730034513805522
0 0.726573829531813
0 0.728057022809124
0 0.732152761104442
0 0.722379651860744
0 0.729472989195679
0 0.732193277310925
0 0.728356842737095
0 0.730312424969988
0 0.728270108043217
0 0.726044417767107
0 0.730936974789916
0 0.724885354141657
0 0.732832833133253
0 0.728337334933973
0 0.727642857142858
0 0.732357142857143
0 0.727413265306123
0 0.729945678271309
0 0.731133853541417
0 0.731794717887155
0 0.731112244897959
0 0.733266506602641
0 0.733109843937575
0 0.730009603841537
0 0.725099339735895
0 0.7344318727491
0 0.72952581032413
0 0.727914165666267
0 0.732198079231693
0 0.726538115246098
0 0.730478391356543
0 0.727276710684273
0 0.728790816326531
0 0.727005702280912
0 0.733169567827131
0 0.725914165666266
0 0.920178571428571
0 0.971
0 0.898418367346939
0 0.855459183673469
0 0.928571428571429
0 0.973826530612245
0 0.971683673469388
0 0.919948979591837
0 0.932091836734694
0 0.746210618622449
0 0.873115593112245
0 0.96719387755102
0 0.914436224489796
0 0.948642857142857
0 0.959190476190476
0 0.961928571428571
0 0.961785714285714
0 0.964166666666666
0 0.964095238095238
0 0.961809523809524
0 0.965380952380952
0 0.966309523809524
0 0.964833333333333
0 0.969404761904762
0 0.97047619047619
0 0.973785714285714
0 0.935590816326531
0 0.902045918367347
0 0.894594387755102
0 0.897230612244898
0 0.898476020408163
0 0.90014081632653
0 0.899256632653061
0 0.896
0 0.900202040816326
0 0.8947
0 0.891906632653061
0 0.898425
0 0.896897448979592
0 0.894428571428571
0 0.894976530612245
0 0.893386734693878
0 0.89145
0 0.892828571428571
0 0.895984183673469
0 0.891135714285714
0 0.892576020408163
0 0.893641326530612
0 0.895308163265306
0 0.890632142857143
0 0.894442346938775
0 0.896302040816326
0 0.893691836734694
0 0.891879591836735
0 0.892946428571429
0 0.893218367346939
0 0.898642857142857
0 0.891148469387755
0 0.895551530612245
0 0.893232142857143
0 0.895426530612245
0 0.895629081632653
0 0.889803571428571
0 0.89892193877551
0 0.888392857142857
0 0.893151530612245
0 0.892966326530612
0 0.895488265306122
0 0.896073469387755
0 0.895234183673469
0 0.89498112244898
0 0.890782653061224
0 0.891510714285714
0 0.894751020408163
0 0.896670408163265
0 0.896247448979592
0 0.895920918367347
0 0.89323112244898
0 0.897413265306122
0 0.896385204081633
0 0.892588775510204
0 0.897147959183674
0 0.898624489795918
0 0.891427551020408
0 0.896342346938775
0 0.891870408163265
0 0.892532142857143
0 0.891399489795918
0 0.890570918367347
0 0.897608163265306
0 0.891684183673469
0 0.890552040816326
0 0.895297959183673
0 0.892970408163265
0 0.893601020408163
0 0.895679591836735
0 0.893733673469388
0 0.893494897959184
0 0.895202551020408
0 0.893435714285714
0 0.892038775510204
0 0.890459693877551
0 0.892383163265306
0 0.895986224489796
0 0.896091836734694
0 0.895598979591837
0 0.895816836734694
0 0.896132653061224
0 0.89581887755102
0 0.893292346938776
0 0.897352551020408
0 0.895844897959184
0 0.891545918367347
0 0.893386734693877
0 0.895489795918367
0 0.891682142857143
0 0.890823979591837
0 0.895424489795918
0 0.892728571428571
0 0.894948979591837
0 0.893214795918368
0 0.89490612244898
0 0.892994897959184
0 0.897357142857143
0 0.889845918367347
0 0.891014795918368
0 0.891440816326531
0 0.893787755102041
0 0.894537244897959
0 0.892527040816327
0 0.893598469387755
0 0.893923469387755
0 0.895108673469388
0 0.895589285714286
0 0.892665816326531
0 0.892112755102041
0 0.890872959183673
0 0.890658163265306
0 0.892054591836735
0 0.887467857142857
0 0.894397448979592
0 0.891345918367347
0 0.888386734693877
0 0.895859183673469
0 0.896745918367347
0 0.894324489795918
0 0.89378112244898
0 0.89369693877551
0 0.899029591836735
0 0.893857142857143
0 0.898719897959184
0 0.895962755102041
0 0.891386224489796
0 0.895325510204082
0 0.900043367346939
0 0.8965
0 0.893489795918367
0 0.896702040816327
0 0.89664693877551
0 0.894977040816327
0 0.894689795918367
0 0.889814795918368
0 0.895147448979592
0 0.894102551020408
0 0.888169897959184
0 0.890392346938775
0 0.893060714285714
0 0.894130612244898
0 0.891036734693877
0 0.89390612244898
0 0.893113265306123
0 0.893583673469388
0 0.891837755102041
0 0.892988265306123
0 0.892491326530612
0 0.887911734693877
0 0.891057142857143
0 0.893270918367347
0 0.895659693877551
0 0.894333163265306
0 0.658772959183674
0 0.814167602040816
0 0.968214285714286
0 0.786696428571429
0 0.801875
0 0.836964285714286
0 0.843660714285714
0 0.854375
0 0.869375
0 0.873214285714286
0 0.861607142857143
0 0.878660714285714
0 0.884464285714286
0 0.953839285714286
0 0.961964285714286
0 0.960982142857143
0 0.964017857142857
0 0.431749489795919
0 0.821085969387755
0 0.951071428571429
0 0.875892857142857
0 0.860625
0 0.876160714285714
0 0.931517857142857
0 0.879464285714286
0 0.888482142857143
0 0.937678571428571
0 0.591426020408164
0 0.890839795918367
0 0.973660714285714
0 0.738823979591836
0 0.955
0 0.935625
0 0.949107142857143
0 0.864425465838511
};
\addplot [line width=0.48pt, dimgray84, forget plot]
table {%
1 0.888696428571428
1 0.779017857142857
};
\addplot [line width=0.48pt, dimgray84, forget plot]
table {%
1 0.99953125
1 1
};
\addplot [line width=0.48pt, dimgray84, forget plot]
table {%
0.925 0.779017857142857
1.075 0.779017857142857
};
\addplot [line width=0.48pt, dimgray84, forget plot]
table {%
0.925 1
1.075 1
};
\addplot [color1, mark=diamond*, mark size=1, mark options={solid,fill=dimgray84}, only marks, forget plot]
table {%
1 0.761304321728692
1 0.729536914765906
1 0.732270408163266
1 0.728875150060024
1 0.726459783913565
1 0.72862274909964
1 0.731358043217287
1 0.730582833133253
1 0.727509303721489
1 0.724399759903961
1 0.728583133253301
1 0.726239795918367
1 0.728981992797119
1 0.727336734693877
1 0.72899849939976
1 0.726141956782713
1 0.729060024009604
1 0.7324393757503
1 0.727439975990396
1 0.729981992797119
1 0.730612244897959
1 0.732754801920768
1 0.729204081632653
1 0.727668667466987
1 0.731641956782713
1 0.734877551020409
1 0.729519507803122
1 0.729625450180072
1 0.730548019207683
1 0.727774609843938
1 0.73308943577431
1 0.732227190876351
1 0.729916566626651
1 0.72378181272509
1 0.725756302521009
1 0.735504801920768
1 0.732605342136855
1 0.72733193277311
1 0.732858043217288
1 0.730777310924369
1 0.735714285714285
1 0.728319327731093
1 0.728529711884754
1 0.729370048019208
1 0.726974789915967
1 0.725294117647059
1 0.730063025210084
1 0.730399159663866
1 0.727331932773109
1 0.72063025210084
1 0.721911764705883
1 0.729915966386555
1 0.731890756302521
1 0.722710084033614
1 0.727478991596639
1 0.724852941176471
1 0.728298319327731
1 0.730021008403362
1 0.72827731092437
1 0.726827731092437
1 0.726659663865547
1 0.727352941176471
1 0.728529411764706
1 0.735
1 0.728634453781513
1 0.731113445378151
1 0.731764705882353
1 0.72983193277311
1 0.725924369747899
1 0.73327731092437
1 0.729138655462185
1 0.728676470588236
1 0.730210084033614
1 0.732142857142858
1 0.727668067226891
1 0.731239495798319
1 0.729117647058824
1 0.727268907563025
1 0.729159663865546
1 0.733403361344538
1 0.729936974789916
1 0.734243697478991
1 0.727941176470589
1 0.725399159663866
1 0.723802521008404
1 0.729201680672269
1 0.730945378151261
1 0.72140756302521
1 0.726512605042017
1 0.728172268907563
1 0.733655462184874
1 0.727710084033614
1 0.731659663865546
1 0.72827731092437
1 0.729978991596639
1 0.729873949579832
1 0.729789915966387
1 0.730273109243698
1 0.731953781512605
1 0.729705882352941
1 0.726449579831933
1 0.728823529411765
1 0.726512605042017
1 0.734705882352941
1 0.720714285714286
1 0.73468487394958
1 0.724474789915966
1 0.739936974789916
1 0.725924369747899
1 0.729852941176471
1 0.727563025210084
1 0.731722689075631
1 0.728067226890756
1 0.731197478991597
1 0.727941176470588
1 0.726470588235294
1 0.736302521008404
1 0.724306722689076
1 0.728907563025211
1 0.731449579831932
1 0.724390756302521
1 0.725735294117647
1 0.727352941176471
1 0.728634453781513
1 0.735756302521009
1 0.729096638655462
1 0.737142857142858
1 0.732878151260504
1 0.73296218487395
1 0.730273109243698
1 0.724180672268907
1 0.730882352941177
1 0.724789915966387
1 0.728781512605042
1 0.733130252100841
1 0.729495798319328
1 0.728046218487395
1 0.728214285714286
1 0.726848739495799
1 0.731428571428571
1 0.732878151260504
1 0.729936974789916
1 0.734201680672269
1 0.727394957983194
1 0.727983193277311
1 0.726302521008404
1 0.72577731092437
1 0.722731092436975
1 0.734873949579832
1 0.725378151260504
1 0.724894957983193
1 0.730966386554623
1 0.728025210084034
1 0.729936974789916
1 0.729117647058824
1 0.724033613445378
1 0.729285714285715
1 0.730861344537815
1 0.732773109243698
1 0.731911764705883
1 0.728193277310924
1 0.734390756302521
1 0.728550420168067
1 0.730021008403361
1 0.72655462184874
1 0.728046218487395
1 0.732142857142858
1 0.722373949579832
1 0.729453781512605
1 0.732163865546219
1 0.728340336134454
1 0.730294117647059
1 0.728256302521009
1 0.726029411764706
1 0.730924369747899
1 0.724873949579832
1 0.73281512605042
1 0.728298319327731
1 0.727626050420169
1 0.732331932773109
1 0.727394957983193
1 0.729936974789916
1 0.731113445378152
1 0.731785714285714
1 0.73109243697479
1 0.733235294117647
1 0.733088235294118
1 0.73
1 0.725084033613446
1 0.734411764705883
1 0.729516806722689
1 0.727899159663866
1 0.73218487394958
1 0.726533613445378
1 0.73046218487395
1 0.727310924369748
1 0.728739495798319
1 0.726995798319328
1 0.733151260504202
1 0.725903361344537
1 0.663390306122449
1 0.671916613520408
1 0.636879464285715
1 0.699333227040816
1 0.691389827806122
1 0.649890625
1 0.63982350127551
1 0.645862723214286
1 0.634441964285714
1 0.668081951530612
1 0.693525031887755
1 0.676922672193877
1 0.670054049744898
1 0.646648596938776
1 0.643447225765306
1 0.640531887755102
1 0.64012005739796
1 0.639877072704082
1 0.643230070153061
1 0.644271843112245
1 0.642856983418367
1 0.641078125
1 0.643355389030612
1 0.642466198979591
1 0.639869897959184
1 0.635178571428571
1 0.636651785714286
1 0.638193399234694
1 0.641003507653061
1 0.639712531887755
1 0.642138073979592
1 0.640254464285714
1 0.641192442602041
1 0.639697704081633
1 0.639911989795918
1 0.642836894132653
1 0.636890943877551
1 0.639008450255102
1 0.6419921875
1 0.642174426020408
1 0.64495487882653
1 0.642676179846938
1 0.646916932397959
1 0.645138073979591
1 0.63665943877551
1 0.6399296875
1 0.636111288265306
1 0.63682350127551
1 0.639428890306122
1 0.639416454081632
1 0.636127551020408
1 0.636487244897959
1 0.632185267857143
1 0.643302774234694
1 0.634581632653061
1 0.63719850127551
1 0.639899553571429
1 0.643041135204081
1 0.640066964285714
1 0.64046875
1 0.641306281887755
1 0.643180803571429
1 0.638847257653061
1 0.637110012755102
1 0.641661192602041
1 0.638508131377551
1 0.646185586734694
1 0.640418526785714
1 0.635862085459184
1 0.647567442602041
1 0.642213169642857
1 0.633574776785714
1 0.641942920918367
1 0.641678890306122
1 0.640571588010204
1 0.641977359693878
1 0.642716358418367
1 0.646930803571429
1 0.634944674744898
1 0.636620216836735
1 0.642645567602041
1 0.639832589285714
1 0.635345982142857
1 0.644520567602041
1 0.640737085459184
1 0.638058035714286
1 0.640100446428571
1 0.633973214285714
1 0.639029017857143
1 0.639430803571429
1 0.636116071428571
1 0.639765625
1 0.637857142857143
1 0.642075892857143
1 0.635747767857143
1 0.637120535714286
1 0.642276785714286
1 0.634944196428571
1 0.637991071428571
1 0.636350446428571
1 0.639899553571429
1 0.638962053571429
1 0.637555803571429
1 0.638761160714286
1 0.641506696428571
1 0.640334821428571
1 0.637421875
1 0.643482142857143
1 0.637689732142857
1 0.631696428571429
1 0.633973214285714
1 0.638125
1 0.638426339285714
1 0.643180803571429
1 0.641841517857143
1 0.638292410714286
1 0.630390625
1 0.641171875
1 0.642779017857143
1 0.640167410714286
1 0.641272321428571
1 0.638359375
1 0.644386160714286
1 0.643716517857143
1 0.642845982142857
1 0.641171875
1 0.640100446428571
1 0.64046875
1 0.638861607142857
1 0.637890625
1 0.638694196428572
1 0.637689732142857
1 0.646763392857143
1 0.645658482142857
1 0.642075892857143
1 0.642310267857143
1 0.642712053571429
1 0.636785714285714
1 0.6428125
1 0.637321428571429
1 0.640803571428571
1 0.640904017857143
1 0.640033482142857
1 0.641071428571429
1 0.639698660714286
1 0.637388392857143
1 0.638627232142857
1 0.639665178571429
1 0.641540178571429
1 0.642745535714286
1 0.639363839285714
1 0.637924107142857
1 0.638962053571429
1 0.634944196428571
1 0.637287946428571
1 0.633537946428571
1 0.637455357142857
1 0.638258928571429
1 0.636919642857143
1 0.634776785714286
1 0.639095982142857
1 0.639464285714286
1 0.635345982142857
1 0.645993303571429
1 0.639732142857143
1 0.640602678571428
1 0.641808035714286
1 0.642243303571429
1 0.637890625
1 0.640770089285714
1 0.637689732142857
1 0.643247767857143
1 0.641439732142857
1 0.643113839285714
1 0.637756696428571
1 0.642310267857143
1 0.641573660714286
1 0.633973214285714
1 0.640234375
1 0.638426339285714
1 0.637723214285714
1 0.639296875
1 0.641037946428571
1 0.632868303571429
1 0.641540178571429
1 0.642444196428571
1 0.642845982142857
1 0.641573660714286
1 0.637957589285714
1 0.640770089285714
1 0.640636160714286
1 0.639229910714286
1 0.641774553571429
1 0.637723214285714
1 0.642310267857143
1 0.642879464285714
1 0.630055803571429
1 0.516518622448979
1 0.383500765306122
1 0.385179846938776
1 0.630829336734694
1 0.619574744897959
1 0.653473214285714
1 0.684054846938776
1 0.597205612244898
1 0.523303061224489
1 0.558037244897958
1 0.670023724489796
};
\addplot [line width=0.48pt, dimgray84, forget plot]
table {%
2 0.815961180124222
2 0.6430234375
};
\addplot [line width=0.48pt, dimgray84, forget plot]
table {%
2 1
2 1
};
\addplot [line width=0.48pt, dimgray84, forget plot]
table {%
1.925 0.6430234375
2.075 0.6430234375
};
\addplot [line width=0.48pt, dimgray84, forget plot]
table {%
1.925 1
2.075 1
};
\addplot [color2, mark=diamond*, mark size=1, mark options={solid,fill=dimgray84}, only marks, forget plot]
table {%
2 0.622196562835661
2 0.101935131195332
2 0.0562966472303227
2 0.0204715743440218
2 0.0266836734693863
2 0.0354081632653034
2 0.0332653061224473
2 0.0276086005830883
2 0.0369715743440214
2 0.0955153061224476
2 0.0275758017492714
2 0.0289052478134096
2 0.0375553935860042
2 0.0384693877551004
2 0.0233163265306111
2 0.0314285714285695
2 0.0288265306122422
2 0.0291348396501437
2 0.0370962099125337
2 0.0332696793002903
2 0.0393921282798803
2 0.0279103498542255
2 0.0279103498542252
2 0.0314285714285701
2 0.0251574344023319
2 0.0248469387755083
2 0.0193367346938766
2 0.0245408163265285
2 0.0145918367346916
2 0.0280743440233243
2 0.0135400874635557
2 0.0326574344023309
2 0.0363265306122431
2 0.0346690962099092
2 0.0306763848396463
2 0.0276042274052445
2 0.0326661807580152
2 0.0251530612244886
2 0.0262485422740518
2 0.0349489795918342
2 0.0332653061224473
2 0.0204103498542259
2 0.0378615160349837
2 0.0370940233236124
2 0.0210247813411066
2 0.021479591836734
2 0.0233206997084537
2 0.0253083090378994
2 0.0409227405247795
2 0.0351020408163241
2 0.0259205539358583
2 0.0242346938775505
2 0.0262266763848387
2 0.0201042274052468
2 0.0294475218658876
2 0.0257696793002908
2 0.0312820699708435
2 0.0351217201166161
2 0.0237973760932924
2 0.0338884839650129
2 0.0340021865889205
2 0.0179744897959174
2 0.0238061224489787
2 0.0274555393585989
2 0.0317806122448955
2 0.0333046647230305
2 0.03122303206997
2 0.0249825072886292
2 0.0328192419825061
2 0.0262332361516016
2 0.0274664723032064
2 0.0222776967930005
2 0.0282645772594747
2 0.0337069970845479
2 0.0282733236151584
2 0.0287543731778406
2 0.0376188046647213
2 0.0271603498542254
2 0.0332653061224479
2 0.0193476676384821
2 0.0260779883381906
2 0.0328104956268214
2 0.0334380466472277
2 0.0309846938775497
2 0.0291348396501449
2 0.0374023323615146
2 0.0370962099125354
2 0.0347981049562671
2 0.0378571428571414
2 0.0305102040816315
2 0.0343367346938767
2 0.028367346938773
2 0.0256122448979571
2 0.0369562682215727
2 0.0297492711370241
2 0.0297448979591833
2 0.029598396501455
2 0.0378658892128254
2 0.0216348396501441
2 0.0225685131195318
2 0.0312973760932925
2 0.0395911078717186
2 0.030658892128279
2 0.0275561224489791
2 0.0258134110787159
2 0.0295809037900863
2 0.0344919825072871
2 0.0354628279883359
2 0.0325174927113689
2 0.033870991253643
2 0.0269482507288621
2 0.0251355685131175
2 0.0225750728862955
2 0.0302718658892117
2 0.0299744897959171
2 0.0235371720116603
2 0.0236158892128276
2 0.0361669096209898
2 0.0283039358600565
2 0.0281268221574338
2 0.025973032069968
2 0.0346909620991237
2 0.033359329446062
2 0.0311618075801737
2 0.0223586005830876
2 0.0356224489795902
2 0.0237689504373171
2 0.0304249271137006
2 0.0195204081632638
2 0.0280065597667634
2 0.0345728862973763
2 0.0331559766763839
2 0.0248797376093285
2 0.0292310495626823
2 0.0282120991253638
2 0.0300816326530595
2 0.0363855685131181
2 0.0247288629737599
2 0.0320320699708437
2 0.0272062682215727
2 0.0438221574344005
2 0.034612244897959
2 0.0405619533527667
2 0.0395080174927106
2 0.0345357142857132
2 0.0350604956268223
2 0.0315860058309025
2 0.0242740524781329
2 0.0244686588921276
2 0.0321523323615146
2 0.0412900874635558
2 0.0256362973760929
2 0.0263797376093275
2 0.0341355685131176
2 0.0279365889212821
2 0.0236618075801733
2 0.0360772594752162
2 0.0288746355685107
2 0.0386793002915438
2 0.0245517492711362
2 0.0357208454810486
2 0.0245779883381908
2 0.0321392128279874
2 0.0306982507288605
2 0.0276260932944589
2 0.0303965014577243
2 0.0225706997084546
2 0.0274686588921277
2 0.0326814868804656
2 0.0325240524781328
2 0.0238520408163263
2 0.0347543731778415
2 0.0294825072886287
2 0.0291807580174909
2 0.0295830903790077
2 0.0322419825072864
2 0.0188360058309021
2 0.0302478134110785
2 0.0212587463556851
2 0.0291654518950427
2 0.0351632653061204
2 0.0224569970845473
2 0.0288527696792987
2 0.0245561224489789
2 0.030577988338191
2 0.0275561224489786
2 0.0314723032069956
2 0.0256472303206992
2 0.0290189504373166
2 0.0324868804664711
2 0.0250634110787157
2 0.0318899416909614
2 0.0324256559766752
2 0.031474489795917
2 0.0260954810495605
2 0.0201392128279875
2 0.0299307580174908
2 0.0322266763848383
2 0.0291741982507282
2 0.0282099125364422
2 0.0232594752186586
2 0.540782653061225
2 0.271780612244897
2 0.273808673469386
2 0.261846938775509
2 0.250772959183673
2 0.263966836734693
2 0.277964285714284
2 0.260102040816325
2 0.240816326530611
2 0.246272959183672
2 0.250107142857142
2 0.263369897959183
2 0.241329081632652
2 0.28155357142857
2 0.279173469387754
2 0.252265306122448
2 0.260760204081632
2 0.259260204081632
2 0.248155612244897
2 0.244145408163264
2 0.254201530612243
2 0.267204081632652
2 0.254744897959182
2 0.272002551020407
2 0.275660714285713
2 0.255923469387755
2 0.270127551020407
2 0.263676020408162
2 0.25852551020408
2 0.231977040816326
2 0.23979081632653
2 0.263247448979591
2 0.257086734693876
2 0.287362244897958
2 0.249426020408163
2 0.248461734693877
2 0.252602040816325
2 0.263729591836734
2 0.270081632653061
2 0.252632653061224
2 0.257369897959183
2 0.257025510204081
2 0.246533163265305
2 0.264931122448979
2 0.25710969387755
2 0.28508163265306
2 0.266017857142856
2 0.282869897959182
2 0.237311224489795
2 0.253994897959183
2 0.268091836734693
2 0.267892857142856
2 0.256612244897958
2 0.269163265306121
2 0.268926020408162
2 0.246655612244897
2 0.266308673469387
2 0.254201530612244
2 0.252227040816326
2 0.266232142857141
2 0.261272959183672
2 0.260721938775509
2 0.254056122448979
2 0.246311224489795
2 0.248262755102039
2 0.256374999999999
2 0.270020408163264
2 0.259183673469387
2 0.26310969387755
2 0.245278061224489
2 0.249051020408162
2 0.239607142857142
2 0.267502551020407
2 0.254454081632652
2 0.248094387755101
2 0.225448979591836
2 0.266025510204081
2 0.263905612244897
2 0.249846938775509
2 0.261617346938775
2 0.288272959183672
2 0.26460969387755
2 0.26812244897959
2 0.268573979591836
2 0.267510204081632
2 0.276280612244896
2 0.270357142857141
2 0.262535714285713
2 0.249808673469387
2 0.262872448979591
2 0.26445663265306
2 0.264464285714285
2 0.25710969387755
2 0.276403061224489
2 0.278316326530611
2 0.271941326530611
2 0.264517857142856
2 0.227645408163265
2 0.247811224489795
2 0.262635204081631
2 0.426892857142857
2 0.385817346938775
2 0.378892857142857
2 0.375196428571429
2 0.377714285714286
2 0.383449489795918
2 0.381795663265306
2 0.398796173469388
2 0.423927295918367
2 0.497097448979593
2 0.387024234693879
};
\addplot [line width=0.48pt, dimgray84, forget plot]
table {%
-0.15 0.993901785714286
0.15 0.993901785714286
};
\addplot [color4, mark=triangle*, mark size=3, mark options={solid}, only marks, forget plot]
table {%
0 0.958778647986649
};
\addplot [line width=0.48pt, dimgray84, forget plot]
table {%
0.85 0.990705782312925
1.15 0.990705782312925
};
\addplot [color4, mark=triangle*, mark size=3, mark options={solid}, only marks, forget plot]
table {%
1 0.918255042396152
};
\addplot [line width=0.48pt, dimgray84, forget plot]
table {%
1.85 0.972142857142857
2.15 0.972142857142857
};
\addplot [color4, mark=triangle*, mark size=3, mark options={solid}, only marks, forget plot]
table {%
2 0.813729104217316
};
\end{axis}

\end{tikzpicture}

%% file: extension-tikz-figures/shared_legend.tex
\begin{tikzpicture}
\begin{axis}[
    width=0,
    height=0,
    at={(0,0)},
    scale only axis,
    xmin=0,
    xmax=0,
    ybar,
    ybar legend,
    xtick={},
    ymin=0,
    ymax=0,
    ytick={},
    axis background/.style={fill=white},
    legend style={area legend,fill, legend cell align=center, align=center, draw=white!15!black, font=\scriptsize, at={(0, 0)}, anchor=center, /tikz/every even column/.append style={column sep=2em}},
    legend columns=10,
]
\addplot [thick, color0,fill=color0]
table {%
0 1
};
\addlegendentry{Out-of-band}
\addplot [thick, color1,fill=color1]
table {%
0 1
};
\addlegendentry{Dynamic}
\addplot [thick, color2,fill=color2]
table {%
0 1
};
\addlegendentry{TDMA ($T_{\ell=10}$)}
\addplot [thick, color3,fill=color3]
table {%
0 1
};
\addlegendentry{TDMA ($T_{\ell=100}$)}
\addplot [thick, color4,fill=color4]
table {%
0 1
};
\addlegendentry{FDMA (1/15)}

\end{axis}

\end{tikzpicture}

%% file: extension-tikz-figures/perf-task-00.tex
\begin{tikzpicture}

\pgfplotsset{every tick label/.append style={font=\scriptsize}}
\pgfplotsset{scaled x ticks=false}
\pgfkeys{/pgf/number format/.cd,1000 sep={\,}}

\begin{axis}[
width=\ffffwidth,
height=\fheight,
axis line style={lightgray204},
tick align=outside,
x grid style={lightgray204},
xmajorgrids,
xmajorticks=false,
xmin=-0.5, xmax=4.5,
xtick={0,1,2,3,4},
xticklabels={$0$,$10$,$20$,$40$,$60$,$70$},
xtick style={color=darkslategray38},
y grid style={lightgray204},
ylabel=\textcolor{darkslategray38}{Reward},
ymajorgrids,
ymajorticks=true,
ymin=0.5, ymax=1,
ytick style={color=darkslategray38}
]
\draw[draw=white,fill=color0,line width=0.32pt] (axis cs:-0.4,0) rectangle (axis cs:0.4,0.980206488095238);
\draw[draw=white,fill=color1,line width=0.32pt] (axis cs:0.6,0) rectangle (axis cs:1.4,0.959614261904763);
\draw[draw=white,fill=color2,line width=0.32pt] (axis cs:1.6,0) rectangle (axis cs:2.4,0.867360333333334);
\draw[draw=white,fill=color3,line width=0.32pt] (axis cs:2.6,0) rectangle (axis cs:3.4,0.842684523809524);
\draw[draw=white,fill=color4,line width=0.32pt] (axis cs:3.6,0) rectangle (axis cs:4.4,0.873084916666667);
\addplot [line width=0.864pt, darkslategray66]
table {%
0 0.962360143522321
0 0.992620797179345
};
\addplot [line width=0.864pt, darkslategray66]
table {%
1 0.942596404528512
1 0.973472883572857
};
\addplot [line width=0.864pt, darkslategray66]
table {%
2 0.829577026497262
2 0.901817212992976
};
\addplot [line width=0.864pt, darkslategray66]
table {%
3 0.796165180200833
3 0.877699907252857
};
\addplot [line width=0.864pt, darkslategray66]
table {%
4 0.835101611069345
4 0.903665730024286
};
\end{axis}

\end{tikzpicture}

%% file: extension-tikz-figures/perf-task-12.tex
\begin{tikzpicture}

\pgfplotsset{every tick label/.append style={font=\scriptsize}}
\pgfplotsset{scaled x ticks=false}
\pgfkeys{/pgf/number format/.cd,1000 sep={\,}}

\begin{axis}[
width=\ffffwidth,
height=\fheight,
axis line style={lightgray204},
tick align=outside,
x grid style={lightgray204},
xmajorgrids,
xmajorticks=false,
xmin=-0.5, xmax=4.5,
xtick={0,1,2,3,4},
xticklabels={$0$,$10$,$20$,$40$,$60$,$70$},
xtick style={color=darkslategray38},
y grid style={lightgray204},
ylabel=\textcolor{darkslategray38}{Reward},
ymajorgrids,
ymajorticks=true,
ymin=0.5, ymax=1,
ytick style={color=darkslategray38}
]
\draw[draw=white,fill=color0,line width=0.32pt] (axis cs:-0.4,0) rectangle (axis cs:0.4,0.981695865079365);
\draw[draw=white,fill=color1,line width=0.32pt] (axis cs:0.6,0) rectangle (axis cs:1.4,0.926482654761905);
\draw[draw=white,fill=color2,line width=0.32pt] (axis cs:1.6,0) rectangle (axis cs:2.4,0.830435936507937);
\draw[draw=white,fill=color3,line width=0.32pt] (axis cs:2.6,0) rectangle (axis cs:3.4,0.825890757936508);
\draw[draw=white,fill=color4,line width=0.32pt] (axis cs:3.6,0) rectangle (axis cs:4.4,0.838276408730159);
\addplot [line width=0.864pt, darkslategray66]
table {%
0 0.961606772632937
0 0.991466175539861
};
\addplot [line width=0.864pt, darkslategray66]
table {%
1 0.90803088646871
1 0.943510199855893
};
\addplot [line width=0.864pt, darkslategray66]
table {%
2 0.810941828291944
2 0.836298081494643
};
\addplot [line width=0.864pt, darkslategray66]
table {%
3 0.812352176905833
3 0.833090515956587
};
\addplot [line width=0.864pt, darkslategray66]
table {%
4 0.776648130505278
4 0.888341583933333
};
\end{axis}

\end{tikzpicture}

%% file: extension-tikz-figures/perf-task-102.tex
\begin{tikzpicture}

\pgfplotsset{every tick label/.append style={font=\scriptsize}}
\pgfplotsset{scaled x ticks=false}
\pgfkeys{/pgf/number format/.cd,1000 sep={\,}}

\begin{axis}[
width=\ffffwidth,
height=\fheight,
axis line style={lightgray204},
tick align=outside,
x grid style={lightgray204},
xmajorgrids,
xmajorticks=false,
xmin=-0.5, xmax=4.5,
xtick={0,1,2,3,4},
xticklabels={$0$,$10$,$20$,$40$,$60$,$70$},
xtick style={color=darkslategray38},
y grid style={lightgray204},
ylabel=\textcolor{darkslategray38}{Reward},
ymajorgrids,
ymajorticks=true,
ymin=0.5, ymax=1,
ytick style={color=darkslategray38}
]
\draw[draw=white,fill=color0,line width=0.32pt] (axis cs:-0.4,0) rectangle (axis cs:0.4,0.990107403184165);
\draw[draw=white,fill=color1,line width=0.32pt] (axis cs:0.6,0) rectangle (axis cs:1.4,0.985826894721744);
\draw[draw=white,fill=color2,line width=0.32pt] (axis cs:1.6,0) rectangle (axis cs:2.4,0.863258270940906);
\draw[draw=white,fill=color3,line width=0.32pt] (axis cs:2.6,0) rectangle (axis cs:3.4,0.861464319420539);
\draw[draw=white,fill=color4,line width=0.32pt] (axis cs:3.6,0) rectangle (axis cs:4.4,0.902713384251291);
\addplot [line width=0.864pt, darkslategray66]
table {%
0 0.982746583190451
0 0.996901984918865
};
\addplot [line width=0.864pt, darkslategray66]
table {%
1 0.978148319380683
1 0.992467894315645
};
\addplot [line width=0.864pt, darkslategray66]
table {%
2 0.858049398655246
2 0.868086907995335
};
\addplot [line width=0.864pt, darkslategray66]
table {%
3 0.859354173444865
3 0.864353880242603
};
\addplot [line width=0.864pt, darkslategray66]
table {%
4 0.878297189938088
4 0.925639527621073
};
\end{axis}

\end{tikzpicture}

%% file: extension-tikz-figures/perf-task-110.tex
\begin{tikzpicture}

\pgfplotsset{every tick label/.append style={font=\scriptsize}}
\pgfplotsset{scaled x ticks=false}
\pgfkeys{/pgf/number format/.cd,1000 sep={\,}}

\begin{axis}[
width=\ffffwidth,
height=\fheight,
axis line style={lightgray204},
tick align=outside,
x grid style={lightgray204},
xmajorgrids,
xmajorticks=false,
xmin=-0.5, xmax=4.5,
xtick={0,1,2,3,4},
xticklabels={$0$,$10$,$20$,$40$,$60$,$70$},
xtick style={color=darkslategray38},
y grid style={lightgray204},
ylabel=\textcolor{darkslategray38}{Reward},
ymajorgrids,
ymajorticks=true,
ymin=0.5, ymax=1,
ytick style={color=darkslategray38}
]
\draw[draw=white,fill=color0,line width=0.32pt] (axis cs:-0.4,0) rectangle (axis cs:0.4,0.986500165880166);
\draw[draw=white,fill=color1,line width=0.32pt] (axis cs:0.6,0) rectangle (axis cs:1.4,0.932400605963106);
\draw[draw=white,fill=color2,line width=0.32pt] (axis cs:1.6,0) rectangle (axis cs:2.4,0.622994858429858);
\draw[draw=white,fill=color3,line width=0.32pt] (axis cs:2.6,0) rectangle (axis cs:3.4,0.602940193050192);
\draw[draw=white,fill=color4,line width=0.32pt] (axis cs:3.6,0) rectangle (axis cs:4.4,0.878334453381953);
\addplot [line width=0.864pt, darkslategray66]
table {%
0 0.975498213078457
0 0.996160372941146
};
\addplot [line width=0.864pt, darkslategray66]
table {%
1 0.898221270535787
1 0.961034028471416
};
\addplot [line width=0.864pt, darkslategray66]
table {%
2 0.610436096785302
2 0.63718648631727
};
\addplot [line width=0.864pt, darkslategray66]
table {%
3 0.600070425633976
3 0.606780863449356
};
\addplot [line width=0.864pt, darkslategray66]
table {%
4 0.808177883862398
4 0.930955322322379
};
\end{axis}

\end{tikzpicture}

%% file: extension-tikz-figures/packet-loss-task-00.tex
\begin{tikzpicture}

	\pgfplotsset{every tick label/.append style={font=\scriptsize}}
	\pgfplotsset{scaled x ticks=false}
	\pgfkeys{/pgf/number format/.cd,1000 sep={\,}}
	
	\begin{axis}[
		width=\ffffwidth,
		height=\fheight,
		axis line style={lightgray204},
		tick align=outside,
		x grid style={lightgray204},
		xmajorgrids,
		xmajorticks=false,
		xmin=-0.5, xmax=4.5,
		xtick={0,1,2,3,4},
		xtick style={color=darkslategray38},
		y grid style={lightgray204},
		ylabel=\textcolor{darkslategray38}{Rejection (pkt/ms)},
		ymajorgrids,
		ymajorticks=true,
		ymin=0.0, ymax=2,
		ytick style={color=darkslategray38}
		]
\draw[draw=white,fill=color0,line width=0.32pt] (axis cs:-0.4,0) rectangle (axis cs:0.4,0.232409651248551);
\draw[draw=white,fill=color1,line width=0.32pt] (axis cs:0.6,0) rectangle (axis cs:1.4,0.547328163178583);
\draw[draw=white,fill=color2,line width=0.32pt] (axis cs:1.6,0) rectangle (axis cs:2.4,1.1302654109589);
\draw[draw=white,fill=color3,line width=0.32pt] (axis cs:2.6,0) rectangle (axis cs:3.4,1.86637294995573);
\draw[draw=white,fill=color4,line width=0.32pt] (axis cs:3.6,0) rectangle (axis cs:4.4,1.93608638241683);
\addplot [line width=0.864pt, darkslategray66]
table {%
	0 0.215854275313455
	0 0.253731875566326
};
\addplot [line width=0.864pt, darkslategray66]
table {%
	1 0.52463667215553
	1 0.567278445554021
};
\addplot [line width=0.864pt, darkslategray66]
table {%
	2 1.0986944562286
	2 1.15574861782962
};
\addplot [line width=0.864pt, darkslategray66]
table {%
	3 1.80733699871411
	3 1.9118960638096
};
\addplot [line width=0.864pt, darkslategray66]
table {%
	4 1.89626548856327
	4 1.98494573114112
};
\end{axis}

\end{tikzpicture}

%% file: extension-tikz-figures/packet-loss-task-12.tex
\begin{tikzpicture}

	\pgfplotsset{every tick label/.append style={font=\scriptsize}}
	\pgfplotsset{scaled x ticks=false}
	\pgfkeys{/pgf/number format/.cd,1000 sep={\,}}
	
	\begin{axis}[
		width=\ffffwidth,
		height=\fheight,
		axis line style={lightgray204},
		tick align=outside,
		x grid style={lightgray204},
		xmajorgrids,
		xmajorticks=false,
		xmin=-0.5, xmax=4.5,
		xtick={0,1,2,3,4},
		xticklabels={$0$,$10$,$20$,$40$,$60$,$70$},
		xtick style={color=darkslategray38},
		y grid style={lightgray204},
		ylabel=\textcolor{darkslategray38}{Rejection (pkt/ms)},
		ymajorgrids,
		ymajorticks=true,
		ymin=0.0, ymax=2,
		ytick style={color=darkslategray38}
		]
\draw[draw=white,fill=color0,line width=0.32pt] (axis cs:-0.4,0) rectangle (axis cs:0.4,0.0788460877633071);
\draw[draw=white,fill=color1,line width=0.32pt] (axis cs:0.6,0) rectangle (axis cs:1.4,0.189941246403281);
\draw[draw=white,fill=color2,line width=0.32pt] (axis cs:1.6,0) rectangle (axis cs:2.4,0.0656591706210027);
\draw[draw=white,fill=color3,line width=0.32pt] (axis cs:2.6,0) rectangle (axis cs:3.4,0.109801521377984);
\draw[draw=white,fill=color4,line width=0.32pt] (axis cs:3.6,0) rectangle (axis cs:4.4,0.817526094947463);
\addplot [line width=0.864pt, darkslategray66]
table {%
	0 0.0695394430603759
	0 0.0882443044053469
};
\addplot [line width=0.864pt, darkslategray66]
table {%
	1 0.176026802322576
	1 0.208031514326573
};
\addplot [line width=0.864pt, darkslategray66]
table {%
	2 0.0562579054913693
	2 0.0763076068530362
};
\addplot [line width=0.864pt, darkslategray66]
table {%
	3 0.0965210185363295
	3 0.12222934886771
};
\addplot [line width=0.864pt, darkslategray66]
table {%
	4 0.775899240298922
	4 0.85026599091445
};
\end{axis}

\end{tikzpicture}

%% file: extension-tikz-figures/packet-loss-task-102.tex
\begin{tikzpicture}

	\pgfplotsset{every tick label/.append style={font=\scriptsize}}
	\pgfplotsset{scaled x ticks=false}
	\pgfkeys{/pgf/number format/.cd,1000 sep={\,}}
	
	\begin{axis}[
		width=\ffffwidth,
		height=\fheight,
		axis line style={lightgray204},
		tick align=outside,
		x grid style={lightgray204},
		xmajorgrids,
		xmajorticks=false,
		xmin=-0.5, xmax=4.5,
		xtick={0,1,2,3,4},
		xticklabels={$0$,$10$,$20$,$40$,$60$,$70$},
		xtick style={color=darkslategray38},
		y grid style={lightgray204},
		ylabel=\textcolor{darkslategray38}{Rejection (pkt/ms)},
		ymajorgrids,
		ymajorticks=true,
		ymin=0.0, ymax=2,
		ytick style={color=darkslategray38}
		]
\draw[draw=white,fill=darkseagreen12518595,line width=0.32pt] (axis cs:-0.4,0) rectangle (axis cs:0.4,0.00564556207136468);
\draw[draw=white,fill=peru21513172,line width=0.32pt] (axis cs:0.6,0) rectangle (axis cs:1.4,0.0702333860759494);
\draw[draw=white,fill=cornflowerblue121150222,line width=0.32pt] (axis cs:1.6,0) rectangle (axis cs:2.4,0.0967476489028213);
\draw[draw=white,fill=sandybrown22319372,line width=0.32pt] (axis cs:2.6,0) rectangle (axis cs:3.4,0.0185049722227714);
\draw[draw=white,fill=color4,line width=0.32pt] (axis cs:3.6,0) rectangle (axis cs:4.4,0.448889062591601);
\addplot [line width=0.864pt, darkslategray66]
table {%
	0 0.00347897803399265
	0 0.00822791163900208
};
\addplot [line width=0.864pt, darkslategray66]
table {%
	1 0.0601105579509494
	1 0.0812593063686709
};
\addplot [line width=0.864pt, darkslategray66]
table {%
	2 0.0880117531347963
	2 0.106322302311912
};
\addplot [line width=0.864pt, darkslategray66]
table {%
	3 0.0131686782586347
	3 0.0244303254186354
};
\addplot [line width=0.864pt, darkslategray66]
table {%
	4 0.422928077306392
	4 0.471521424090829
};
\end{axis}

\end{tikzpicture}

%% file: extension-tikz-figures/packet-loss-task-110.tex
\begin{tikzpicture}
	
	\pgfplotsset{every tick label/.append style={font=\scriptsize}}
	\pgfplotsset{scaled x ticks=false}
	\pgfkeys{/pgf/number format/.cd,1000 sep={\,}}
	
	\begin{axis}[
		width=\ffffwidth,
		height=\fheight,
		axis line style={lightgray204},
		tick align=outside,
		x grid style={lightgray204},
		xmajorgrids,
		xmajorticks=false,
		xmin=-0.5, xmax=4.5,
		xtick={0,1,2,3,4},
		xticklabels={$0$,$10$,$20$,$40$,$60$,$70$},
		xtick style={color=darkslategray38},
		y grid style={lightgray204},
		ylabel=\textcolor{darkslategray38}{Rejection (pkt/ms)},
		ymajorgrids,
		ymajorticks=true,
		ymin=0.0, ymax=2,
		ytick style={color=darkslategray38}
		]
\draw[draw=white,fill=color0,line width=0.32pt] (axis cs:-0.4,0) rectangle (axis cs:0.4,0.0176790103058717);
\draw[draw=white,fill=color1,line width=0.32pt] (axis cs:0.6,0) rectangle (axis cs:1.4,0.0389702065815426);
\draw[draw=white,fill=color2,line width=0.32pt] (axis cs:1.6,0) rectangle (axis cs:2.4,0.090313050164923);
\draw[draw=white,fill=color3,line width=0.32pt] (axis cs:2.6,0) rectangle (axis cs:3.4,0);
\draw[draw=white,fill=color4,line width=0.32pt] (axis cs:3.6,0) rectangle (axis cs:4.4,0.281862444253508);
\addplot [line width=0.864pt, darkslategray66]
table {%
	0 0.012976511080361
	0 0.0213538088353745
};
\addplot [line width=0.864pt, darkslategray66]
table {%
	1 0.0329950128724177
	1 0.0449900353991488
};
\addplot [line width=0.864pt, darkslategray66]
table {%
	2 0.0839805114636664
	2 0.0983015001922415
};
\addplot [line width=0.864pt, darkslategray66]
table {%
	3 0
	3 0
};
\addplot [line width=0.864pt, darkslategray66]
table {%
	4 0.264598725195368
	4 0.299432820363324
};
\end{axis}

\end{tikzpicture}

%% file: extension-tikz-figures/packet-dead-task-00.tex
\begin{tikzpicture}

	\pgfplotsset{every tick label/.append style={font=\scriptsize}}
	\pgfplotsset{scaled x ticks=false}
	\pgfkeys{/pgf/number format/.cd,1000 sep={\,}}
	
	\begin{axis}[
		width=\ffffwidth,
		height=\fheight,
		axis line style={lightgray204},
		tick align=outside,
		x grid style={lightgray204},
		xmajorgrids,
		xmajorticks=false,
		xmin=-0.5, xmax=4.5,
		xtick={0,1,2,3,4},
		xticklabels={$0$,$10$,$20$,$40$,$60$,$70$},
		xtick style={color=darkslategray38},
		y grid style={lightgray204},
		ylabel=\textcolor{darkslategray38}{Drop rate (pkt/ms)},
		ymajorgrids,
		ymajorticks=true,
		ymin=0.0, ymax=4,
		ytick style={color=darkslategray38}
		]
		\draw[draw=white,fill=color0,line width=0.32pt] (axis cs:-0.4,0) rectangle (axis cs:0.4,0.00400020690725382);
		\draw[draw=white,fill=color1,line width=0.32pt] (axis cs:0.6,0) rectangle (axis cs:1.4,0.00706008881487649);
		\draw[draw=white,fill=color2,line width=0.32pt] (axis cs:1.6,0) rectangle (axis cs:2.4,0.0660289394619608);
		\draw[draw=white,fill=color3,line width=0.32pt] (axis cs:2.6,0) rectangle (axis cs:3.4,0.0341970367249491);
		\draw[draw=white,fill=color4,line width=0.32pt] (axis cs:3.6,0) rectangle (axis cs:4.4,0.0134840712974559);
		\addplot [line width=0.864pt, darkslategray66]
		table {%
			0 0.0025354751107816
			0 0.00553399391347828
		};
		\addplot [line width=0.864pt, darkslategray66]
		table {%
			1 0.00541215653621982
			1 0.00895035439217319
		};
		\addplot [line width=0.864pt, darkslategray66]
		table {%
			2 0.0605736081085748
			2 0.0718303004454881
		};
		\addplot [line width=0.864pt, darkslategray66]
		table {%
			3 0.0298455143599466
			3 0.0393204506354891
		};
		\addplot [line width=0.864pt, darkslategray66]
		table {%
			4 0.011203443756326
			4 0.0168955780480692
		};
		\end{axis}

\end{tikzpicture}

%% file: extension-tikz-figures/packet-dead-task-12.tex
\begin{tikzpicture}

	\pgfplotsset{every tick label/.append style={font=\scriptsize}}
	\pgfplotsset{scaled x ticks=false}
	\pgfkeys{/pgf/number format/.cd,1000 sep={\,}}
	
	\begin{axis}[
		width=\ffffwidth,
		height=\fheight,
		axis line style={lightgray204},
		tick align=outside,
		x grid style={lightgray204},
		xmajorgrids,
		xmajorticks=false,
		xmin=-0.5, xmax=4.5,
		xtick={0,1,2,3,4},
		xticklabels={$0$,$10$,$20$,$40$,$60$,$70$},
		xtick style={color=darkslategray38},
		y grid style={lightgray204},
		ylabel=\textcolor{darkslategray38}{Drop rate (pkt/ms)},
		ymajorgrids,
		ymajorticks=true,
		ymin=0.0, ymax=4,
		ytick style={color=darkslategray38}
		]
\draw[draw=white,fill=color0,line width=0.32pt] (axis cs:-0.4,0) rectangle (axis cs:0.4,0.0442844542899924);
\draw[draw=white,fill=color1,line width=0.32pt] (axis cs:0.6,0) rectangle (axis cs:1.4,0.279699745652309);
\draw[draw=white,fill=color2,line width=0.32pt] (axis cs:1.6,0) rectangle (axis cs:2.4,2.45866580258415);
\draw[draw=white,fill=color3,line width=0.32pt] (axis cs:2.6,0) rectangle (axis cs:3.4,2.53462915601023);
\draw[draw=white,fill=color4,line width=0.32pt] (axis cs:3.6,0) rectangle (axis cs:4.4,0.14601191340985);
\addplot [line width=0.864pt, darkslategray66]
table {%
	0 0.0388512478206469
	0 0.0496580727339856
};
\addplot [line width=0.864pt, darkslategray66]
table {%
	1 0.267129177609132
	1 0.294393090117662
};
\addplot [line width=0.864pt, darkslategray66]
table {%
	2 2.43466877809956
	2 2.48458716543805
};
\addplot [line width=0.864pt, darkslategray66]
table {%
	3 2.50219374485933
	3 2.56182855508951
};
\addplot [line width=0.864pt, darkslategray66]
table {%
	4 0.137647827774434
	4 0.155432537826114
};
\end{axis}

\end{tikzpicture}

%% file: extension-tikz-figures/packet-dead-task-102.tex
\begin{tikzpicture}
	
	\pgfplotsset{every tick label/.append style={font=\scriptsize}}
	\pgfplotsset{scaled x ticks=false}
	\pgfkeys{/pgf/number format/.cd,1000 sep={\,}}
	
	\begin{axis}[
		width=\ffffwidth,
		height=\fheight,
		axis line style={lightgray204},
		tick align=outside,
		x grid style={lightgray204},
		xmajorgrids,
		xmajorticks=false,
		xmin=-0.5, xmax=4.5,
		xtick={0,1,2,3,4},
		xticklabels={$0$,$10$,$20$,$40$,$60$,$70$},
		xtick style={color=darkslategray38},
		y grid style={lightgray204},
		ylabel=\textcolor{darkslategray38}{Drop rate (pkt/ms)},
		ymajorgrids,
		ymajorticks=true,
		ymin=0.0, ymax=4,
		ytick style={color=darkslategray38}
		]
\draw[draw=white,fill=color0,line width=0.32pt] (axis cs:-0.4,0) rectangle (axis cs:0.4,0.0196906063356056);
\draw[draw=white,fill=color1,line width=0.32pt] (axis cs:0.6,0) rectangle (axis cs:1.4,0.0406426607193123);
\draw[draw=white,fill=color2,line width=0.32pt] (axis cs:1.6,0) rectangle (axis cs:2.4,1.88046122923939);
\draw[draw=white,fill=color3,line width=0.32pt] (axis cs:2.6,0) rectangle (axis cs:3.4,3.72738900450919);
\draw[draw=white,fill=color4,line width=0.32pt] (axis cs:3.6,0) rectangle (axis cs:4.4,0.240816065762293);
\addplot [line width=0.864pt, darkslategray66]
table {%
	0 0.0152621158432349
	0 0.0238265839013992
};
\addplot [line width=0.864pt, darkslategray66]
table {%
	1 0.0326822561332092
	1 0.0488856357317628
};
\addplot [line width=0.864pt, darkslategray66]
table {%
	2 1.84114467988998
	2 1.91553264254734
};
\addplot [line width=0.864pt, darkslategray66]
table {%
	3 3.68693470505983
	3 3.76972621986646
};
\addplot [line width=0.864pt, darkslategray66]
table {%
	4 0.226953275443491
	4 0.258163435057606
};
\end{axis}

\end{tikzpicture}

%% file: extension-tikz-figures/packet-dead-task-110.tex
\begin{tikzpicture}

	\pgfplotsset{every tick label/.append style={font=\scriptsize}}
	\pgfplotsset{scaled x ticks=false}
	\pgfkeys{/pgf/number format/.cd,1000 sep={\,}}
	
	\begin{axis}[
		width=\ffffwidth,
		height=\fheight,
		axis line style={lightgray204},
		tick align=outside,
		x grid style={lightgray204},
		xmajorgrids,
		xmajorticks=false,
		xmin=-0.5, xmax=4.5,
		xtick={0,1,2,3,4},
		xticklabels={$0$,$10$,$20$,$40$,$60$,$70$},
		xtick style={color=darkslategray38},
		y grid style={lightgray204},
		ylabel=\textcolor{darkslategray38}{Drop rate (pkt/ms)},
		ymajorgrids,
		ymajorticks=true,
		ymin=0.0, ymax=4,
		ytick style={color=darkslategray38}
		]
\draw[draw=white,fill=color0,line width=0.32pt] (axis cs:-0.4,0) rectangle (axis cs:0.4,0.0324147037526089);
\draw[draw=white,fill=color1,line width=0.32pt] (axis cs:0.6,0) rectangle (axis cs:1.4,0.306992681585515);
\draw[draw=white,fill=color2,line width=0.32pt] (axis cs:1.6,0) rectangle (axis cs:2.4,2.05857121410342);
\draw[draw=white,fill=color3,line width=0.32pt] (axis cs:2.6,0) rectangle (axis cs:3.4,3.4216706868097);
\draw[draw=white,fill=color4,line width=0.32pt] (axis cs:3.6,0) rectangle (axis cs:4.4,0.125259149566321);
\addplot [line width=0.864pt, darkslategray66]
table {%
	0 0.0255592684329344
	0 0.0422849076755974
};
\addplot [line width=0.864pt, darkslategray66]
table {%
	1 0.286173997736791
	1 0.32919012119802
};
\addplot [line width=0.864pt, darkslategray66]
table {%
	2 2.02411027537693
	2 2.09888734381367
};
\addplot [line width=0.864pt, darkslategray66]
table {%
	3 3.37339362849647
	3 3.47286584179679
};
\addplot [line width=0.864pt, darkslategray66]
table {%
	4 0.110479154855088
	4 0.140458547175799
};
\end{axis}

\end{tikzpicture}

%% file: extension-tikz-figures/shared_legend-line.tex
\begin{tikzpicture}
\begin{axis}[
    width=0,
    height=0,
    at={(0,0)},
    scale only axis,
    xmin=0,
    xmax=0,
    xtick={},
    ymin=0,
    ymax=0,
    ytick={},
    axis background/.style={fill=white},
    legend style={legend cell align=center, align=center, draw=white!15!black, font=\scriptsize, at={(0, 0)}, anchor=center, /tikz/every even column/.append style={column sep=2em}},
    legend columns=10,
]
\addplot [semithick, color0,mark=o]
table {%
0 1
};
\addlegendentry{Out-of-band}
\addplot [semithick, color1,mark=x]
table {%
0 1
};
\addlegendentry{Dynamic}
\addplot [semithick, color2,mark=triangle]
table {%
0 1
};
\addlegendentry{TDMA ($T_{\ell=10}$)}
\addplot [semithick, color3,mark=+]
table {%
0 1
};
\addlegendentry{TDMA ($T_{\ell=100}$)}
\addplot [semithick, color4,mark=diamond]
table {%
0 1
};
\addlegendentry{FDMA (1/15)}

\end{axis}

\end{tikzpicture}

%% file: extension-tikz-figures/latency1-task-12.tex
\begin{tikzpicture}
\pgfplotsset{every tick label/.append style={font=\scriptsize}}

	\begin{axis}[
		width=\ffffwidth,
		height=\fheight,
		axis line style={lightgray204},
		tick align=outside,
		x grid style={lightgray204},
		xlabel=\textcolor{darkslategray38}{Latency (ms)},
		xmajorgrids,
		xmajorticks=true,
		xmin=-0, xmax=70,
		xtick style={color=darkslategray38},
		y grid style={lightgray204},
		ylabel=\textcolor{darkslategray38}{Empirical CDF},
		ymajorgrids,
		ymajorticks=true,
		ymin=0, ymax=1,
		ytick style={color=darkslategray38}
		]
\addplot[dashed, semithick,color=lightgray204]
table{%
50	0
50	1
};
	\addplot [semithick, color0, mark = o, mark repeat=25]
table {%
-inf 0
0 0.00347222222222222
0 0.00694444444444444
0 0.0104166666666667
0 0.0138888888888889
0 0.0173611111111111
0 0.0208333333333333
0 0.0243055555555556
0 0.0277777777777778
0 0.03125
0 0.0347222222222222
0 0.0381944444444444
0 0.0416666666666667
0 0.0451388888888889
0 0.0486111111111111
0 0.0520833333333333
0 0.0555555555555556
0 0.0590277777777778
0 0.0625
0 0.0659722222222222
0 0.0694444444444444
0 0.0729166666666667
0 0.0763888888888889
0 0.0798611111111111
0 0.0833333333333333
0 0.0868055555555556
0 0.0902777777777778
0 0.09375
0 0.0972222222222222
0 0.100694444444444
0 0.104166666666667
0 0.107638888888889
0 0.111111111111111
0 0.114583333333333
0 0.118055555555556
0 0.121527777777778
0 0.125
0 0.128472222222222
0 0.131944444444444
0 0.135416666666667
0 0.138888888888889
0 0.142361111111111
0 0.145833333333333
0 0.149305555555556
0 0.152777777777778
0 0.15625
0 0.159722222222222
0 0.163194444444444
0.25 0.166666666666667
0.333333333333333 0.170138888888889
0.333333333333333 0.173611111111111
0.333333333333333 0.177083333333333
0.333333333333333 0.180555555555556
0.333333333333333 0.184027777777778
0.333333333333333 0.1875
0.333333333333333 0.190972222222222
0.333333333333333 0.194444444444444
0.5 0.197916666666667
0.5 0.201388888888889
0.5 0.204861111111111
0.666666666666667 0.208333333333333
0.666666666666667 0.211805555555556
0.666666666666667 0.215277777777778
0.666666666666667 0.21875
0.666666666666667 0.222222222222222
0.666666666666667 0.225694444444444
1 0.229166666666667
1 0.232638888888889
1 0.236111111111111
1 0.239583333333333
1 0.243055555555556
1 0.246527777777778
1 0.25
1 0.253472222222222
1 0.256944444444444
1 0.260416666666667
1 0.263888888888889
1.25 0.267361111111111
1.33333333333333 0.270833333333333
1.33333333333333 0.274305555555556
1.33333333333333 0.277777777777778
1.5 0.28125
1.5 0.284722222222222
1.5 0.288194444444444
1.66666666666667 0.291666666666667
1.66666666666667 0.295138888888889
1.66666666666667 0.298611111111111
1.66666666666667 0.302083333333333
1.66666666666667 0.305555555555556
1.75 0.309027777777778
1.75 0.3125
2 0.315972222222222
2 0.319444444444444
2 0.322916666666667
2 0.326388888888889
2 0.329861111111111
2 0.333333333333333
2.25 0.336805555555556
2.25 0.340277777777778
2.25 0.34375
2.33333333333333 0.347222222222222
2.33333333333333 0.350694444444444
2.33333333333333 0.354166666666667
2.33333333333333 0.357638888888889
2.33333333333333 0.361111111111111
2.33333333333333 0.364583333333333
2.33333333333333 0.368055555555556
2.66666666666667 0.371527777777778
2.66666666666667 0.375
2.66666666666667 0.378472222222222
2.66666666666667 0.381944444444444
2.66666666666667 0.385416666666667
2.66666666666667 0.388888888888889
2.66666666666667 0.392361111111111
2.75 0.395833333333333
3 0.399305555555556
3 0.402777777777778
3 0.40625
3 0.409722222222222
3 0.413194444444444
3 0.416666666666667
3 0.420138888888889
3 0.423611111111111
3 0.427083333333333
3 0.430555555555556
3.33333333333333 0.434027777777778
3.33333333333333 0.4375
3.33333333333333 0.440972222222222
3.33333333333333 0.444444444444444
3.33333333333333 0.447916666666667
3.66666666666667 0.451388888888889
3.66666666666667 0.454861111111111
4 0.458333333333333
4 0.461805555555556
4 0.465277777777778
4.33333333333333 0.46875
4.33333333333333 0.472222222222222
4.5 0.475694444444444
4.66666666666667 0.479166666666667
4.66666666666667 0.482638888888889
4.66666666666667 0.486111111111111
4.66666666666667 0.489583333333333
4.66666666666667 0.493055555555556
4.75 0.496527777777778
5 0.5
5 0.503472222222222
5 0.506944444444444
5 0.510416666666667
5 0.513888888888889
5.25 0.517361111111111
5.33333333333333 0.520833333333333
5.33333333333333 0.524305555555556
5.33333333333333 0.527777777777778
5.33333333333333 0.53125
5.5 0.534722222222222
5.66666666666667 0.538194444444444
6 0.541666666666667
6 0.545138888888889
6 0.548611111111111
6.33333333333333 0.552083333333333
6.33333333333333 0.555555555555556
6.33333333333333 0.559027777777778
6.33333333333333 0.5625
6.33333333333333 0.565972222222222
6.5 0.569444444444444
6.5 0.572916666666667
6.66666666666667 0.576388888888889
6.66666666666667 0.579861111111111
7 0.583333333333333
7 0.586805555555556
7 0.590277777777778
7.33333333333333 0.59375
7.33333333333333 0.597222222222222
7.33333333333333 0.600694444444444
7.33333333333333 0.604166666666667
7.66666666666667 0.607638888888889
8 0.611111111111111
8 0.614583333333333
8 0.618055555555556
8 0.621527777777778
8 0.625
8.33333333333333 0.628472222222222
8.33333333333333 0.631944444444444
8.33333333333333 0.635416666666667
8.33333333333333 0.638888888888889
8.66666666666667 0.642361111111111
8.66666666666667 0.645833333333333
9 0.649305555555556
9 0.652777777777778
9 0.65625
9 0.659722222222222
9 0.663194444444444
9 0.666666666666667
9 0.670138888888889
9 0.673611111111111
9 0.677083333333333
9 0.680555555555556
9.25 0.684027777777778
9.33333333333333 0.6875
9.66666666666667 0.690972222222222
10 0.694444444444444
10 0.697916666666667
10 0.701388888888889
10 0.704861111111111
10 0.708333333333333
10 0.711805555555556
10.3333333333333 0.715277777777778
10.3333333333333 0.71875
10.6666666666667 0.722222222222222
10.6666666666667 0.725694444444444
11 0.729166666666667
11 0.732638888888889
11 0.736111111111111
11.25 0.739583333333333
11.3333333333333 0.743055555555556
11.3333333333333 0.746527777777778
11.6666666666667 0.75
12 0.753472222222222
12 0.756944444444444
12 0.760416666666667
12 0.763888888888889
13 0.767361111111111
14 0.770833333333333
14 0.774305555555556
14.6666666666667 0.777777777777778
14.6666666666667 0.78125
15 0.784722222222222
15 0.788194444444444
15 0.791666666666667
15.6666666666667 0.795138888888889
15.6666666666667 0.798611111111111
15.75 0.802083333333333
16 0.805555555555556
16 0.809027777777778
16.3333333333333 0.8125
16.3333333333333 0.815972222222222
17 0.819444444444444
17 0.822916666666667
17 0.826388888888889
17.3333333333333 0.829861111111111
18.5 0.833333333333333
18.5 0.836805555555556
18.6666666666667 0.840277777777778
18.6666666666667 0.84375
19 0.847222222222222
19 0.850694444444444
19.3333333333333 0.854166666666667
20 0.857638888888889
20 0.861111111111111
20.3333333333333 0.864583333333333
20.3333333333333 0.868055555555556
20.6 0.871527777777778
20.75 0.875
21 0.878472222222222
21.6666666666667 0.881944444444444
22 0.885416666666667
22.3333333333333 0.888888888888889
22.3333333333333 0.892361111111111
23 0.895833333333333
24.3333333333333 0.899305555555556
25 0.902777777777778
27.25 0.90625
29 0.909722222222222
30.3333333333333 0.913194444444444
31.6666666666667 0.916666666666667
32.6666666666667 0.920138888888889
33 0.923611111111111
33.6666666666667 0.927083333333333
34 0.930555555555556
37 0.934027777777778
38.3333333333333 0.9375
42.3333333333333 0.940972222222222
44.3333333333333 0.944444444444444
46.6666666666667 0.947916666666667
48.6 0.951388888888889
51 0.954861111111111
51 0.958333333333333
52 0.961805555555556
54 0.965277777777778
54 0.96875
56.5 0.972222222222222
58.3333333333333 0.975694444444444
59.2 0.979166666666667
63 0.982638888888889
64 0.986111111111111
64.6666666666667 0.989583333333333
65.75 0.993055555555556
69.6666666666667 0.996527777777778
70 1
};
\addplot [semithick, color1, mark = x, mark repeat=25, mark phase=5]
table {%
0 0
0 0.00395256916996047
0 0.00790513833992095
0 0.0118577075098814
0 0.0158102766798419
0 0.0197628458498024
0 0.0237154150197628
0 0.0276679841897233
0 0.0316205533596838
0 0.0355731225296443
0 0.0395256916996047
0 0.0434782608695652
0 0.0474308300395257
0 0.0513833992094862
0 0.0553359683794466
0 0.0592885375494071
0 0.0632411067193676
0 0.0671936758893281
0 0.0711462450592885
0 0.075098814229249
0 0.0790513833992095
0 0.08300395256917
0 0.0869565217391304
0 0.0909090909090909
0 0.0948616600790514
0 0.0988142292490119
0 0.102766798418972
0 0.106719367588933
0 0.110671936758893
0 0.114624505928854
0 0.118577075098814
0 0.122529644268775
0 0.126482213438735
0 0.130434782608696
0 0.134387351778656
0 0.138339920948617
0 0.142292490118577
0 0.146245059288538
0 0.150197628458498
0 0.154150197628458
0 0.158102766798419
0 0.162055335968379
0 0.16600790513834
0 0.1699604743083
0 0.173913043478261
0 0.177865612648221
0 0.181818181818182
0 0.185770750988142
0 0.189723320158103
0 0.193675889328063
0 0.197628458498024
0 0.201581027667984
0 0.205533596837945
0 0.209486166007905
0 0.213438735177866
0 0.217391304347826
0 0.221343873517787
0 0.225296442687747
0 0.229249011857708
0 0.233201581027668
0 0.237154150197628
0 0.241106719367589
0 0.245059288537549
0 0.24901185770751
0 0.25296442687747
0 0.256916996047431
0 0.260869565217391
0 0.264822134387352
0 0.268774703557312
0 0.272727272727273
0 0.276679841897233
0 0.280632411067194
0 0.284584980237154
0 0.288537549407115
0 0.292490118577075
0 0.296442687747036
0 0.300395256916996
0 0.304347826086957
0 0.308300395256917
0 0.312252964426877
0 0.316205533596838
0 0.320158102766798
0 0.324110671936759
0 0.328063241106719
0 0.33201581027668
0 0.33596837944664
0 0.339920948616601
0 0.343873517786561
0 0.347826086956522
0 0.351778656126482
0 0.355731225296443
0 0.359683794466403
0 0.363636363636364
0 0.367588932806324
0 0.371541501976285
0 0.375494071146245
0 0.379446640316206
0 0.383399209486166
0 0.387351778656126
0 0.391304347826087
0 0.395256916996047
0 0.399209486166008
0 0.403162055335968
0 0.407114624505929
0 0.411067193675889
0 0.41501976284585
0 0.41897233201581
0 0.422924901185771
0 0.426877470355731
0.333333333333333 0.430830039525692
1 0.434782608695652
1.5 0.438735177865613
1.5 0.442687747035573
1.75 0.446640316205534
1.75 0.450592885375494
2 0.454545454545455
2.6 0.458498023715415
2.75 0.462450592885375
3 0.466403162055336
3.4 0.470355731225296
3.8 0.474308300395257
4 0.478260869565217
5.2 0.482213438735178
5.8 0.486166007905138
7 0.490118577075099
7 0.494071146245059
7.4 0.49802371541502
7.4 0.50197628458498
10.5 0.505928853754941
11 0.509881422924901
11 0.513833992094862
11.6666666666667 0.517786561264822
12.6 0.521739130434783
14.3333333333333 0.525691699604743
14.3333333333333 0.529644268774704
14.3333333333333 0.533596837944664
14.5 0.537549407114625
15 0.541501976284585
15.3333333333333 0.545454545454545
15.4 0.549407114624506
15.4 0.553359683794466
15.75 0.557312252964427
16.5 0.561264822134387
16.5 0.565217391304348
16.6 0.569169960474308
17.6 0.573122529644269
17.8 0.577075098814229
18.5 0.58102766798419
18.5 0.58498023715415
19.1666666666667 0.588932806324111
19.6666666666667 0.592885375494071
19.8333333333333 0.596837944664032
20 0.600790513833992
20 0.604743083003953
20.3333333333333 0.608695652173913
20.4 0.612648221343874
20.6 0.616600790513834
21.25 0.620553359683794
21.8 0.624505928853755
22.6 0.628458498023715
23.3333333333333 0.632411067193676
24 0.636363636363636
24.2 0.640316205533597
24.3333333333333 0.644268774703557
24.5 0.648221343873518
24.6666666666667 0.652173913043478
25 0.656126482213439
25 0.660079051383399
26 0.66403162055336
26 0.66798418972332
26.25 0.671936758893281
27.25 0.675889328063241
28 0.679841897233202
29 0.683794466403162
29.25 0.687747035573123
29.5 0.691699604743083
29.6666666666667 0.695652173913043
30.1666666666667 0.699604743083004
30.2 0.703557312252964
30.3333333333333 0.707509881422925
30.4 0.711462450592885
31 0.715415019762846
31 0.719367588932806
31 0.723320158102767
31.5 0.727272727272727
32 0.731225296442688
32.1666666666667 0.735177865612648
32.4 0.739130434782609
33.5 0.743083003952569
34 0.74703557312253
34.8 0.75098814229249
34.8 0.754940711462451
35 0.758893280632411
36 0.762845849802372
36.5 0.766798418972332
38 0.770750988142292
38.3333333333333 0.774703557312253
38.5 0.778656126482213
39 0.782608695652174
39.25 0.786561264822134
39.6 0.790513833992095
40.25 0.794466403162055
41.3333333333333 0.798418972332016
42 0.802371541501976
42.3333333333333 0.806324110671937
43.2 0.810276679841897
43.2 0.814229249011858
44.1666666666667 0.818181818181818
45 0.822134387351779
45.5 0.826086956521739
45.6666666666667 0.8300395256917
46.4 0.83399209486166
47.8 0.837944664031621
47.8 0.841897233201581
49.8333333333333 0.845849802371542
52 0.849802371541502
53.5 0.853754940711462
54.8333333333333 0.857707509881423
56.5 0.861660079051383
56.6666666666667 0.865612648221344
58.4 0.869565217391304
58.6 0.873517786561265
58.6666666666667 0.877470355731225
58.75 0.881422924901186
59.2 0.885375494071146
59.4 0.889328063241107
60.4285714285714 0.893280632411067
61.5 0.897233201581028
61.5714285714286 0.901185770750988
61.6666666666667 0.905138339920949
62 0.909090909090909
62.3333333333333 0.91304347826087
62.6666666666667 0.91699604743083
63.5 0.920948616600791
63.75 0.924901185770751
64 0.928853754940711
64 0.932806324110672
64.2 0.936758893280632
64.3333333333333 0.940711462450593
65 0.944664031620553
67 0.948616600790514
67.8 0.952569169960474
68.75 0.956521739130435
69 0.960474308300395
69.1666666666667 0.964426877470356
69.3333333333333 0.968379446640316
69.3333333333333 0.972332015810277
69.3333333333333 0.976284584980237
69.5 0.980237154150198
69.6666666666667 0.984189723320158
70 0.988142292490119
70 0.992094861660079
70 0.99604743083004
70 1
};

\addplot [semithick, color2, mark = triangle, mark repeat=25, mark phase=10]
table {%
-inf 0
0 0.00374531835205993
0 0.00749063670411985
0 0.0112359550561798
0 0.0149812734082397
0 0.0187265917602996
0 0.0224719101123595
0 0.0262172284644195
0 0.0299625468164794
0 0.0337078651685393
0 0.0374531835205993
0 0.0411985018726592
0 0.0449438202247191
0 0.048689138576779
0 0.052434456928839
0 0.0561797752808989
0 0.0599250936329588
0 0.0636704119850187
0 0.0674157303370786
0 0.0711610486891386
0 0.0749063670411985
0 0.0786516853932584
0 0.0823970037453183
0 0.0861423220973783
0 0.0898876404494382
0 0.0936329588014981
0 0.0973782771535581
0 0.101123595505618
0 0.104868913857678
0 0.108614232209738
0 0.112359550561798
0 0.116104868913858
0 0.119850187265918
0 0.123595505617978
0 0.127340823970037
0 0.131086142322097
0 0.134831460674157
0 0.138576779026217
0 0.142322097378277
0 0.146067415730337
0 0.149812734082397
0 0.153558052434457
0 0.157303370786517
0 0.161048689138577
0 0.164794007490637
0 0.168539325842697
0 0.172284644194757
0 0.176029962546816
0 0.179775280898876
0 0.183520599250936
0 0.187265917602996
0 0.191011235955056
0 0.194756554307116
0 0.198501872659176
0 0.202247191011236
0 0.205992509363296
0 0.209737827715356
0 0.213483146067416
0 0.217228464419476
0 0.220973782771536
0 0.224719101123595
0 0.228464419475655
0 0.232209737827715
0 0.235955056179775
0 0.239700374531835
0 0.243445692883895
0 0.247191011235955
0 0.250936329588015
0 0.254681647940075
0 0.258426966292135
0 0.262172284644195
0 0.265917602996255
0 0.269662921348315
0 0.273408239700375
0 0.277153558052434
0 0.280898876404494
0 0.284644194756554
0 0.288389513108614
0 0.292134831460674
0 0.295880149812734
0 0.299625468164794
0 0.303370786516854
0 0.307116104868914
0 0.310861423220974
0 0.314606741573034
0 0.318352059925094
0 0.322097378277154
0 0.325842696629214
0 0.329588014981273
0 0.333333333333333
0 0.337078651685393
0 0.340823970037453
0 0.344569288389513
0 0.348314606741573
0 0.352059925093633
0 0.355805243445693
0 0.359550561797753
0 0.363295880149813
0 0.367041198501873
0 0.370786516853933
0 0.374531835205993
0 0.378277153558052
0 0.382022471910112
0 0.385767790262172
0 0.389513108614232
0 0.393258426966292
0 0.397003745318352
0 0.400749063670412
0 0.404494382022472
0 0.408239700374532
0 0.411985018726592
0 0.415730337078652
0 0.419475655430712
0 0.423220973782772
0 0.426966292134831
0 0.430711610486891
0 0.434456928838951
0 0.438202247191011
0 0.441947565543071
0 0.445692883895131
0 0.449438202247191
0 0.453183520599251
0 0.456928838951311
0 0.460674157303371
0 0.464419475655431
0 0.468164794007491
0 0.471910112359551
0 0.47565543071161
0 0.47940074906367
0 0.48314606741573
0 0.48689138576779
0 0.49063670411985
0 0.49438202247191
0 0.49812734082397
0 0.50187265917603
0 0.50561797752809
0 0.50936329588015
0 0.51310861423221
0 0.51685393258427
0 0.52059925093633
0 0.524344569288389
0 0.528089887640449
0 0.531835205992509
0 0.535580524344569
0 0.539325842696629
0 0.543071161048689
0 0.546816479400749
0 0.550561797752809
0 0.554307116104869
0 0.558052434456929
0 0.561797752808989
0 0.565543071161049
0 0.569288389513109
0 0.573033707865168
0 0.576779026217228
0 0.580524344569288
0 0.584269662921348
0 0.588014981273408
0 0.591760299625468
0 0.595505617977528
0 0.599250936329588
0 0.602996254681648
0 0.606741573033708
0 0.610486891385768
0 0.614232209737828
0 0.617977528089888
0 0.621722846441948
0 0.625468164794007
0 0.629213483146067
0 0.632958801498127
0 0.636704119850187
0 0.640449438202247
0 0.644194756554307
0 0.647940074906367
0 0.651685393258427
0 0.655430711610487
0 0.659176029962547
0 0.662921348314607
0 0.666666666666667
0 0.670411985018727
0 0.674157303370786
0 0.677902621722846
0 0.681647940074906
0 0.685393258426966
0 0.689138576779026
0 0.692883895131086
0 0.696629213483146
0 0.700374531835206
0 0.704119850187266
0 0.707865168539326
0 0.711610486891386
0 0.715355805243446
0 0.719101123595506
0 0.722846441947566
0 0.726591760299625
0 0.730337078651685
0 0.734082397003745
0 0.737827715355805
0 0.741573033707865
0 0.745318352059925
0 0.749063670411985
0 0.752808988764045
0 0.756554307116105
0 0.760299625468165
0 0.764044943820225
0 0.767790262172285
0 0.771535580524345
0 0.775280898876405
0 0.779026217228464
0 0.782771535580524
0 0.786516853932584
0 0.790262172284644
0 0.794007490636704
0 0.797752808988764
0 0.801498127340824
0 0.805243445692884
0 0.808988764044944
0 0.812734082397004
0 0.816479400749064
0 0.820224719101124
0 0.823970037453184
30.75 0.827715355805243
58 0.831460674157303
60 0.835205992509363
60.25 0.838951310861423
64 0.842696629213483
66.75 0.846441947565543
68 0.850187265917603
68.5 0.853932584269663
68.5 0.857677902621723
68.5 0.861423220973783
69 0.865168539325843
69 0.868913857677903
69 0.872659176029963
69.25 0.876404494382023
69.3333333333333 0.880149812734082
69.5 0.883895131086142
69.5 0.887640449438202
69.6666666666667 0.891385767790262
69.6666666666667 0.895131086142322
70 0.898876404494382
70 0.902621722846442
70 0.906367041198502
70 0.910112359550562
70 0.913857677902622
70 0.917602996254682
70 0.921348314606742
70 0.925093632958802
70 0.928838951310861
70 0.932584269662921
70 0.936329588014981
70 0.940074906367041
70 0.943820224719101
70 0.947565543071161
70 0.951310861423221
70 0.955056179775281
70 0.958801498127341
70 0.962546816479401
70 0.966292134831461
70 0.970037453183521
70 0.973782771535581
70 0.97752808988764
70 0.9812734082397
70 0.98501872659176
70 0.98876404494382
70 0.99250936329588
70 0.99625468164794
70 1
};

	\addplot [semithick, color3, mark=+, mark repeat=25, mark phase = 15]
table {%
-inf 0
0 0.00403225806451613
0 0.00806451612903226
0 0.0120967741935484
0 0.0161290322580645
0 0.0201612903225806
0 0.0241935483870968
0 0.0282258064516129
0 0.032258064516129
0 0.0362903225806452
0 0.0403225806451613
0 0.0443548387096774
0 0.0483870967741935
0 0.0524193548387097
0 0.0564516129032258
0 0.0604838709677419
0 0.0645161290322581
0 0.0685483870967742
0 0.0725806451612903
0 0.0766129032258065
0 0.0806451612903226
0 0.0846774193548387
0 0.0887096774193548
0 0.092741935483871
0 0.0967741935483871
0 0.100806451612903
0 0.104838709677419
0 0.108870967741935
0 0.112903225806452
0 0.116935483870968
0 0.120967741935484
0 0.125
0 0.129032258064516
0 0.133064516129032
0 0.137096774193548
0 0.141129032258065
0 0.145161290322581
0 0.149193548387097
0 0.153225806451613
0 0.157258064516129
0 0.161290322580645
0 0.165322580645161
0 0.169354838709677
0 0.173387096774194
0 0.17741935483871
0 0.181451612903226
0 0.185483870967742
0 0.189516129032258
0 0.193548387096774
0 0.19758064516129
0 0.201612903225806
0 0.205645161290323
0 0.209677419354839
0 0.213709677419355
0 0.217741935483871
0 0.221774193548387
0 0.225806451612903
0 0.229838709677419
0 0.233870967741935
0 0.237903225806452
0 0.241935483870968
0 0.245967741935484
0 0.25
0 0.254032258064516
0 0.258064516129032
0 0.262096774193548
0 0.266129032258065
0 0.270161290322581
0 0.274193548387097
0 0.278225806451613
0 0.282258064516129
0 0.286290322580645
0 0.290322580645161
0 0.294354838709677
0 0.298387096774194
0 0.30241935483871
0 0.306451612903226
0 0.310483870967742
0 0.314516129032258
0 0.318548387096774
0 0.32258064516129
0 0.326612903225806
0 0.330645161290323
0 0.334677419354839
0 0.338709677419355
0 0.342741935483871
0 0.346774193548387
0 0.350806451612903
0 0.354838709677419
0 0.358870967741935
0 0.362903225806452
0 0.366935483870968
0 0.370967741935484
0 0.375
0 0.379032258064516
0 0.383064516129032
0 0.387096774193548
0 0.391129032258065
0 0.395161290322581
0 0.399193548387097
0 0.403225806451613
0 0.407258064516129
0 0.411290322580645
0 0.415322580645161
0 0.419354838709677
0 0.423387096774194
0 0.42741935483871
0 0.431451612903226
0 0.435483870967742
0 0.439516129032258
0 0.443548387096774
0 0.44758064516129
0 0.451612903225806
0 0.455645161290323
0 0.459677419354839
0 0.463709677419355
0 0.467741935483871
0 0.471774193548387
0 0.475806451612903
0 0.479838709677419
0 0.483870967741935
0 0.487903225806452
0 0.491935483870968
0 0.495967741935484
0 0.5
0 0.504032258064516
0 0.508064516129032
0 0.512096774193548
0 0.516129032258065
0 0.520161290322581
0 0.524193548387097
0 0.528225806451613
0 0.532258064516129
0 0.536290322580645
0 0.540322580645161
0 0.544354838709677
0 0.548387096774194
0 0.55241935483871
0 0.556451612903226
0 0.560483870967742
0 0.564516129032258
0 0.568548387096774
0 0.57258064516129
0 0.576612903225806
0 0.580645161290323
0 0.584677419354839
0 0.588709677419355
0 0.592741935483871
0 0.596774193548387
0 0.600806451612903
0 0.604838709677419
0 0.608870967741935
0 0.612903225806452
0 0.616935483870968
0 0.620967741935484
0 0.625
0 0.629032258064516
0 0.633064516129032
0 0.637096774193548
0 0.641129032258065
0 0.645161290322581
0 0.649193548387097
0 0.653225806451613
0 0.657258064516129
0 0.661290322580645
0 0.665322580645161
0 0.669354838709677
0 0.673387096774194
0 0.67741935483871
0 0.681451612903226
0 0.685483870967742
0 0.689516129032258
0 0.693548387096774
0 0.69758064516129
0 0.701612903225806
0 0.705645161290323
0 0.709677419354839
0 0.713709677419355
0 0.717741935483871
0 0.721774193548387
0 0.725806451612903
0 0.729838709677419
0 0.733870967741935
0 0.737903225806452
0 0.741935483870968
0 0.745967741935484
0 0.75
0 0.754032258064516
0 0.758064516129032
0 0.762096774193548
0 0.766129032258065
0 0.770161290322581
0 0.774193548387097
0 0.778225806451613
0 0.782258064516129
0 0.786290322580645
0 0.790322580645161
0 0.794354838709677
0 0.798387096774194
0 0.80241935483871
0 0.806451612903226
0 0.810483870967742
0 0.814516129032258
0 0.818548387096774
0 0.82258064516129
0 0.826612903225806
0 0.830645161290323
0 0.834677419354839
0 0.838709677419355
0 0.842741935483871
0 0.846774193548387
0 0.850806451612903
0 0.854838709677419
0 0.858870967741935
0 0.862903225806452
17 0.866935483870968
52.5 0.870967741935484
52.6 0.875
56.8571428571428 0.879032258064516
58.6666666666667 0.883064516129032
60.6666666666667 0.887096774193548
67.3333333333333 0.891129032258065
67.5 0.895161290322581
67.75 0.899193548387097
68.5 0.903225806451613
68.75 0.907258064516129
69 0.911290322580645
69 0.915322580645161
69.25 0.919354838709677
69.5 0.923387096774194
69.6666666666667 0.92741935483871
69.6666666666667 0.931451612903226
69.75 0.935483870967742
70 0.939516129032258
70 0.943548387096774
70 0.94758064516129
70 0.951612903225806
70 0.955645161290323
70 0.959677419354839
70 0.963709677419355
70 0.967741935483871
70 0.971774193548387
70 0.975806451612903
70 0.979838709677419
70 0.983870967741935
70 0.987903225806452
70 0.991935483870968
70 0.995967741935484
70 1
};

	\addplot [semithick, color4, mark=diamond, mark repeat=25, mark phase = 20]
table {%
-inf 0
0 0.00338983050847458
0 0.00677966101694915
0 0.0101694915254237
0 0.0135593220338983
0 0.0169491525423729
0 0.0203389830508475
0 0.023728813559322
0 0.0271186440677966
0 0.0305084745762712
0 0.0338983050847458
0 0.0372881355932203
0 0.0406779661016949
0 0.0440677966101695
0 0.0474576271186441
0 0.0508474576271186
0 0.0542372881355932
0 0.0576271186440678
0 0.0610169491525424
0 0.0644067796610169
0 0.0677966101694915
0 0.0711864406779661
0 0.0745762711864407
0 0.0779661016949153
0 0.0813559322033898
0 0.0847457627118644
0 0.088135593220339
0 0.0915254237288136
0 0.0949152542372881
0 0.0983050847457627
0 0.101694915254237
0 0.105084745762712
0 0.108474576271186
0 0.111864406779661
0 0.115254237288136
0 0.11864406779661
0 0.122033898305085
0 0.125423728813559
0 0.128813559322034
0 0.132203389830508
0 0.135593220338983
0 0.138983050847458
0 0.142372881355932
0 0.145762711864407
0 0.149152542372881
0 0.152542372881356
0 0.155932203389831
0.2 0.159322033898305
0.2 0.16271186440678
0.25 0.166101694915254
0.666666666666667 0.169491525423729
0.666666666666667 0.172881355932203
1 0.176271186440678
1 0.179661016949153
1.25 0.183050847457627
1.33333333333333 0.186440677966102
2 0.189830508474576
2.57142857142857 0.193220338983051
3.33333333333333 0.196610169491525
4 0.2
4.5 0.203389830508475
4.5 0.206779661016949
4.66666666666667 0.210169491525424
4.66666666666667 0.213559322033898
5 0.216949152542373
7 0.220338983050847
7.66666666666667 0.223728813559322
8.33333333333333 0.227118644067797
8.66666666666667 0.230508474576271
10.3333333333333 0.233898305084746
10.6666666666667 0.23728813559322
10.6666666666667 0.240677966101695
11 0.24406779661017
11.3333333333333 0.247457627118644
11.6666666666667 0.250847457627119
12 0.254237288135593
12 0.257627118644068
13.6666666666667 0.261016949152542
14 0.264406779661017
14.3333333333333 0.267796610169492
15 0.271186440677966
15.3333333333333 0.274576271186441
15.3333333333333 0.277966101694915
15.75 0.28135593220339
18.3333333333333 0.284745762711864
19 0.288135593220339
19 0.291525423728814
19.3333333333333 0.294915254237288
19.5 0.298305084745763
20 0.301694915254237
20.75 0.305084745762712
21 0.308474576271186
21 0.311864406779661
23 0.315254237288136
23.3333333333333 0.31864406779661
23.5 0.322033898305085
25.6666666666667 0.325423728813559
26 0.328813559322034
26.3333333333333 0.332203389830508
27 0.335593220338983
28 0.338983050847458
28 0.342372881355932
28 0.345762711864407
28.25 0.349152542372881
28.3333333333333 0.352542372881356
29.6666666666667 0.355932203389831
30 0.359322033898305
31 0.36271186440678
32 0.366101694915254
32.5 0.369491525423729
32.75 0.372881355932203
33 0.376271186440678
33 0.379661016949153
34 0.383050847457627
34.3333333333333 0.386440677966102
34.3333333333333 0.389830508474576
34.6666666666667 0.393220338983051
35 0.396610169491525
36 0.4
36.3333333333333 0.403389830508475
36.3333333333333 0.406779661016949
36.6666666666667 0.410169491525424
37 0.413559322033898
37 0.416949152542373
37 0.420338983050847
37.6666666666667 0.423728813559322
39 0.427118644067797
39 0.430508474576271
39 0.433898305084746
39.6666666666667 0.43728813559322
40.3333333333333 0.440677966101695
41 0.444067796610169
42 0.447457627118644
42 0.450847457627119
42 0.454237288135593
42 0.457627118644068
43 0.461016949152542
43.3333333333333 0.464406779661017
43.6666666666667 0.467796610169492
43.6666666666667 0.471186440677966
44 0.474576271186441
44 0.477966101694915
44.3333333333333 0.48135593220339
45 0.484745762711864
45 0.488135593220339
45.25 0.491525423728814
45.75 0.494915254237288
46 0.498305084745763
46 0.501694915254237
46.5 0.505084745762712
47.5 0.508474576271186
47.75 0.511864406779661
48 0.515254237288136
48 0.51864406779661
48.6666666666667 0.522033898305085
49.3333333333333 0.525423728813559
50 0.528813559322034
50 0.532203389830509
50.3333333333333 0.535593220338983
50.3333333333333 0.538983050847458
52 0.542372881355932
52 0.545762711864407
52.3333333333333 0.549152542372881
52.3333333333333 0.552542372881356
52.6666666666667 0.555932203389831
53.3333333333333 0.559322033898305
53.6666666666667 0.56271186440678
53.75 0.566101694915254
53.75 0.569491525423729
54 0.572881355932203
54.3333333333333 0.576271186440678
54.6666666666667 0.579661016949153
54.6666666666667 0.583050847457627
55 0.586440677966102
55 0.589830508474576
55 0.593220338983051
55.5 0.596610169491525
55.6666666666667 0.6
55.6666666666667 0.603389830508475
56 0.606779661016949
56 0.610169491525424
56.5 0.613559322033898
56.6666666666667 0.616949152542373
56.6666666666667 0.620338983050847
57 0.623728813559322
57 0.627118644067797
57.25 0.630508474576271
57.3333333333333 0.633898305084746
57.6666666666667 0.63728813559322
57.6666666666667 0.640677966101695
57.75 0.644067796610169
58.3333333333333 0.647457627118644
58.3333333333333 0.650847457627119
58.3333333333333 0.654237288135593
58.6666666666667 0.657627118644068
59 0.661016949152542
59 0.664406779661017
59 0.667796610169492
59.5 0.671186440677966
59.6666666666667 0.674576271186441
60 0.677966101694915
60 0.68135593220339
61 0.684745762711864
61 0.688135593220339
61 0.691525423728814
61.3333333333333 0.694915254237288
61.3333333333333 0.698305084745763
62 0.701694915254237
62.25 0.705084745762712
62.3333333333333 0.708474576271186
62.5 0.711864406779661
62.5 0.715254237288136
62.5 0.71864406779661
62.75 0.722033898305085
63 0.725423728813559
63 0.728813559322034
63 0.732203389830508
63 0.735593220338983
63 0.738983050847458
63.3333333333333 0.742372881355932
63.3333333333333 0.745762711864407
63.5 0.749152542372881
63.6666666666667 0.752542372881356
63.6666666666667 0.75593220338983
63.6666666666667 0.759322033898305
64.3333333333333 0.76271186440678
64.5 0.766101694915254
64.6666666666667 0.769491525423729
65 0.772881355932203
65 0.776271186440678
65 0.779661016949153
65 0.783050847457627
65 0.786440677966102
65 0.789830508474576
65.3333333333333 0.793220338983051
65.6666666666667 0.796610169491525
65.6666666666667 0.8
65.6666666666667 0.803389830508475
65.6666666666667 0.806779661016949
65.75 0.810169491525424
66.3333333333333 0.813559322033898
66.3333333333333 0.816949152542373
66.3333333333333 0.820338983050847
66.5 0.823728813559322
66.6666666666667 0.827118644067797
66.6666666666667 0.830508474576271
66.6666666666667 0.833898305084746
66.6666666666667 0.83728813559322
66.6666666666667 0.840677966101695
67 0.84406779661017
67 0.847457627118644
67 0.850847457627119
67 0.854237288135593
67 0.857627118644068
67 0.861016949152542
67.3333333333333 0.864406779661017
67.3333333333333 0.867796610169492
67.3333333333333 0.871186440677966
67.3333333333333 0.874576271186441
67.5 0.877966101694915
67.6666666666667 0.88135593220339
67.6666666666667 0.884745762711864
68 0.888135593220339
68 0.891525423728814
68 0.894915254237288
68 0.898305084745763
68.3333333333333 0.901694915254237
68.6666666666667 0.905084745762712
68.6666666666667 0.908474576271186
68.6666666666667 0.911864406779661
68.6666666666667 0.915254237288136
69 0.91864406779661
69 0.922033898305085
69 0.925423728813559
69 0.928813559322034
69.25 0.932203389830508
69.3333333333333 0.935593220338983
69.5 0.938983050847458
69.6666666666667 0.942372881355932
69.6666666666667 0.945762711864407
69.6666666666667 0.949152542372881
69.6666666666667 0.952542372881356
69.6666666666667 0.955932203389831
70 0.959322033898305
70 0.96271186440678
70 0.966101694915254
70 0.969491525423729
70 0.972881355932203
70 0.976271186440678
70 0.979661016949153
70 0.983050847457627
70 0.986440677966102
70 0.989830508474576
70 0.993220338983051
70 0.996610169491525
70 1
};

\end{axis}

\end{tikzpicture}